\documentclass[useAMS,usenatbib]{mnras}
\usepackage{natbib}
\usepackage{amsmath}
\usepackage{graphicx}
\usepackage{multirow}
\usepackage{url}
\numberwithin{equation}{section}
\title[Molecules in AGN winds]{The origin of fast molecular outflows in quasars: molecule formation in AGN-driven galactic winds}
\author[A. J. Richings and C.-A. Faucher-Gigu\`{e}re]{Alexander J. Richings$^{1}$\thanks{Email: a.j.richings@northwestern.edu} and Claude-Andr\'{e} Faucher-Gigu\`{e}re$^{1}$\\
$^{1}$Center for Interdisciplinary Exploration and Research in Astrophysics (CIERA) and Department of Physics and Astronomy,\\ 
Northwestern University, 2145 Sheridan Road, Evanston, IL 60208, USA}

\begin{document}

\date{\today}

\pagerange{\pageref{firstpage}--\pageref{lastpage}} \pubyear{2017}

\maketitle

\label{firstpage}

\begin{abstract} 
We explore the origin of fast molecular outflows that have been observed in Active Galactic Nuclei (AGN). Previous numerical studies have shown that it is difficult to create such an outflow by accelerating existing molecular clouds in the host galaxy, as the clouds will be destroyed before they can reach the high velocities that are observed. In this work, we consider an alternative scenario where molecules form in-situ within the AGN outflow. We present a series of hydro-chemical simulations of an isotropic AGN wind interacting with a uniform medium. We follow the time-dependent chemistry of 157 species, including 20 molecules, to determine whether molecules can form rapidly enough to produce the observed molecular outflows. We find H$_{2}$ outflow rates up to $140 \, \rm{M}_{\odot} \, \rm{yr}^{-1}$, which is sensitive to density, AGN luminosity, and metallicity. We compute emission and absorption lines of CO, OH and warm (a few hundred K) H$_{2}$ from the simulations in post-processing. The CO-derived outflow rates and OH absorption strengths at solar metallicity agree with observations, although the maximum line of sight velocities from the model CO spectra are a factor $\approx 2$ lower than is observed. We derive a CO (1$-$0) to H$_{2}$ conversion factor of $\alpha_{\rm{CO} \, (1-0)} = 0.13 \, \rm{M}_{\odot} \, (\rm{K} \, \rm{km} \, \rm{s}^{-1} \, \rm{pc}^{2})^{-1}$, 6 times lower than is commonly assumed in observations of such systems. We find strong emission from the mid-infrared lines of H$_{2}$. The mass of H$_{2}$ traced by this infrared emission is within a few per cent of the total H$_{2}$ mass. This H$_{2}$ emission may be observable by JWST. 
\end{abstract}


\section{Introduction}

Recent observations of galaxies hosting luminous active galactic nuclei (AGN) have found fast ouflows of molecular gas. They have been detected in several molecular lines, including CO \citep{feruglio10, feruglio15, cicone12, sun14}, OH \citep{fischer10, veilleux13, spoon13, stone16}, warm H$_{2}$ emission \citep{rupke13a}, and dense tracers such as HCN, HNC and HCO$^{+}$ \citep{aalto12, aalto15}. 

These outflows are spatially extended on kpc scales, with typical outflow rates of $\sim 100 - 1000 \, \rm{M}_{\odot} \, \rm{yr}^{-1}$ and velocities up to $\sim 1000 \, \rm{km} \, \rm{s}^{-1}$. While molecular outflow rates from starburst-dominated galaxies are typically comparable to the star formation rate, those in AGN-dominated systems can significantly exceed the star formation rate of the host galaxy \citep[e.g.][]{sturm11, cicone14, fiore17}. Furthermore, the velocity of the outflow is found to correlate with the AGN luminosity, with the fastest outflows only found in the most luminous quasars. This suggests that the fast molecular outflows in these systems are being driven by the AGN. Observations of molecular outflows have also derived gas depletion time-scales of $\sim 10^{6} - 10^{8} \, \rm{yr}$, which has been taken as evidence for negative feedback from AGN rapidly quenching star formation in the host galaxy \citep{sturm11, cicone14}. 

Galaxy-scale outflows driven by AGN have also been observed in both the neutral and ionized phases \citep[e.g.][]{moe09, rupke11, greene12, harrison12, harrison14, liu13}. Interestingly, the mass outflow rates in the different phases of AGN-driven galaxy-scale outflows appear roughly comparable \citep[e.g.][]{fauchergiguereetal12, fiore17}. Furthermore, luminous quasars appear to drive such galaxy-scale, cool outflows at all redshifts, from the local Universe out to $z > 6$ \citep{maiolino12, cicone15}. 

Measurements of molecular outflows can be used to deduce the energetics of the AGN wind. \mbox{\citet{gonzalezalfonso17}} measured OH lines in 14 local ULIRGs with clear evidence for outflows. By modelling these lines with radiative transfer models, they calculated the energetics of the outflows. In most sources, they found momentum rates of $\approx 2-5 L_{\rm{IR}} / c$. However, some showed higher momentum boosts, up to $\approx 20$, which may imply an energy-conserving phase of expansion (\citealt{fauchergiguere12}, FGQ12 hereafter; \citealt{zubovas12}; \citealt{costa14}; \citealt{tombesi15}; \citealt{stern16}). \citet{cicone14} also found momentum boosts of $\approx 20$ in their AGN-dominated systems. 

Understanding the energetics of AGN winds is important for determining how AGN interact with their host galaxy. It has been shown that feedback from AGN can play a role in quenching star formation in the most massive galaxies, and is needed by galaxy formation models to reproduce the massive end of the observed galaxy stellar mass function \citep{springel05, hopkins08, booth09, bower12, bower17, crain15, tremmel16}, and to reproduce the colours of massive galaxies \citep{gabor12, henriques15, feldmann16, trayford16}. Note that the AGN feedback that operates in the most massive haloes (i.e. galaxy clusters) is likely to be driven by relativistic jets \citep[e.g.][]{croton06}, rather than the type of AGN outflows that we study in this paper. 

It has also been suggested that AGN feedback can shape the observed scaling relations between supermassive black holes (SMBHs) and their host galaxies \citep{silk98, wyithe03, dimatteo05, murray05}, although \citet{anglesalcazar13, anglesalcazar15, anglesalcazar17} found that the SMBH scaling relations in their cosmological simulations arose primarily due to feeding of gas onto the SMBH via gravitational torques, with only a weak dependence on feedback. 

While there is growing evidence for fast molecular outflows driven by AGN, it is still not understood how these molecular outflows are formed. \citet{narayanan06} and \citet{narayanan08} explored the effects of galactic winds (both star formation- and AGN-driven) on molecular gas emission from CO using hydrodynamic simulations of galaxy mergers, and they showed that the resulting molecular outflows in their simulations produced detectable CO emission. However, their simulations simply assumed that half of the cold neutral gas was molecular, and did not model the non-equilibrium chemistry of CO. 

The origin of the observed fast molecular outflows thus remains unclear. One possibility is that the hot, high-velocity wind launched from regions close to the AGN accelerates existing molecular clouds in the host galaxy. For example, \citet{gaspari17} considered the entrainment of hot ($\sim 10^{7} \, \rm{K}$), warm ($\sim 10^{4} - 10^{5} \, \rm{K}$) and cold ($\la 100 \, \rm{K}$) gas from the host galaxy by an AGN wind in their model for AGN feeding and feedback. They argued that such entrainment would produce a multiphase outflow, with velocities and outflow rates in the hot/warm/cold phases of a few $\times 10^{3}$/$10^{3}$/$500 \, \rm{km} \, \rm{s}^{-1}$ and $10$/$100$/several $100 \, \rm{M}_{\odot} \, \rm{yr}^{-1}$, respectively, in good agreement with observed multiphase outflows. However, numerical studies of the interaction of a hot, fast wind with a cold cloud find that the cloud will tend to be rapidly destroyed by the hot wind, before it can be accelerated to the high velocities that have been observed \citep{klein94, scannapieco15, bruggen16, schneider17, zhang17}. Furthermore, even if a dense molecular cloud is long-lived, it is difficult to accelerate it significantly due to its small cross section. 

Alternatively, the molecules may have formed in-situ, within the material swept up by the AGN-driven outflow. It has been shown that galactic winds can efficiently cool under certain conditions \citep{wang95, silich03, silich04, tenoriotagle07, thompson16, scannapieco17}, which may explain the presence of cool gas in observed galactic outflows \citep[e.g.][]{martin15}. If this cooling gas can form molecules rapidly enough, it could also produce the observed fast molecular outflows. \citet{neufeld89} also demonstrated that molecules can form in the cooling gas behind a dissociative shock, although they explored a very different physical regime, with velocities $60 - 100 \, \rm{km} \, \rm{s}^{-1}$ and pre-shock densities $10^{4-6} \, \rm{cm}^{-3}$. 

\citet{ferrara16} recently explored this alternative scario for forming molecular gas within AGN outflows. They ran a series of hydrodynamic simulations of a $1 \, \rm{kpc}$ patch of the shell of gas swept up from the ambient ISM by an AGN outflow, including a treatment for dust destruction via sputtering. They found that, while the gas can cool to $\sim 10^{4} \, \rm{K}$, it remains as a single phase, rather than forming a two-phase medium, and clumps of dense gas are smoothed out by pressure gradients. They also found that dust grains are rapidly destroyed, within $\sim 10^{4} \, \rm{yr}$. They therefore concluded that clumps of molecular gas cannot condense out of the AGN outflow material. However, their study only focussed on the formation of cold clumps and the survivability of dust grains, and did not explicitly follow the time-dependent molecular chemistry. Furthermore, their simulations only followed a patch of the swept up shell in the AGN outflow, rather than the full AGN wind structure and its interaction with the ambient ISM, which will miss the effects of the bulk motions of the outflowing material. Additionally, they only considered one ambient medium density and one forward shock velocity, so it remains unclear whether different conditions may be more conducive to the formation of molecular clumps. Therefore, this scenario requires further investigation before we can discount it altogether. 

In this paper we investigate in-situ molecule formation in AGN winds using a series of hydro-chemical simulations of an isotropic AGN wind interacting with a uniform ambient ISM. These simulations include a treatment for the time-dependent chemistry of the gas, which allows us to follow the formation and destruction of molecules, including H$_{2}$ and commonly observed molecules such as CO, OH and HCO$^{+}$. The chemical model is valid over a wide range of physical conditions, from cold, molecular gas to hot, highly ionized gas. This enables us to self-consistently follow the chemistry (along with associated radiative cooling and heating processes) in all phases of the AGN wind. 

We consider an idealised setup, with a uniform ambient ISM, as this matches the setup studied analytically by FGQ12, providing us with a useful framework in which to interpret our results. The simplified geometry also allows us to focus on the chemistry in this paper, which is the most novel aspect of our simulations. We explore a range of ambient ISM densities ($1 - 10 \, \rm{cm}^{-3}$), AGN luminosities ($10^{45} - 10^{46} \, \rm{erg} \, \rm{s}^{-1}$) and metallicities ($0.1 - 1 \, \rm{Z}_{\odot}$\footnote{Throughout this paper we use solar abundances taken from table 1 of \citet{wiersma09}, with a solar metallicity $\rm{Z}_{\odot} = 0.0129$.}), to determine under what conditions molecular outflows are likely to form. 

We also perform radiative transfer calculations on our simulations in post-processing to compute observable molecular lines of CO, OH and IR-traced warm H$_{2}$. These allow us to rigorously compare our simulations to observations, and to compute the conversion factors between CO emission and H$_{2}$ mass. 

The remainder of this paper is divided into the following sections. We describe the AGN wind model, the chemistry solver module and the simulation suite in Section~\ref{simulations_sect}. In Section~\ref{parameter_vars_sect} we explore the effects of the ambient medium density, AGN luminosity and metallicity on the molecular content of AGN winds. We compare variations of the model in Section~\ref{model_vars_sect} to quantify how uncertainties in the model parameters affect our results. In Section~\ref{molecular_lines_sect} we compute emission and absorption lines of CO, OH and IR-traced H$_{2}$ from the simulations in post-processing, which we compare to observations. We summarise our main results in Section~\ref{conclusions_sect}. In Appendix~\ref{resolution_appendix} we present resolution tests, and we explore the uncertainties in the local shielding approximation used in the simulations in Appendix~\ref{shielding_appendix}. 

\section{Simulations}\label{simulations_sect} 

We ran a series of 3D hydro-chemical simulations of an idealised, isotropic AGN wind interacting with an initially uniform ambient ISM. We used the hydrodynamics+gravity code \textsc{gizmo} \citep{hopkins15} with the Meshless Finite Mass (MFM) hydro solver option. MFM is a Lagrangian hydrodynamics method that combines many of the advantages of grid-based methods (such as the accurate capture of shocks and fluid instabilities) and Smoothed Particle Hydrodynamics (SPH) methods (such as exact conservation of mass, energy and momentum, and better angular momentum conservation); see \citet{hopkins15} for more details. 

To follow the chemical evolution of molecules and ions in the simulations, we have coupled \textsc{gizmo} to the \textsc{chimes} chemistry solver module \citep{richings14a, richings14b}. The chemical and thermal processes that are included in \textsc{chimes} are summarised in Section~\ref{chemistry_sect} below. 

We simulate a periodic box $1.6 - 5.0 \, \rm{kpc}$ across. Each simulation is run for $1 \, \rm{Myr}$, which is approximately the time it would take an outflow of $1000 \, \rm{km} \, \rm{s}^{-1}$ to reach $1 \, \rm{kpc}$. For comparison, \citet{gonzalezalfonso17} modelled the molecular outflows observed in OH lines from a sample of 14 ULIRGs that host AGN, and they found flow times ($r/v$) of $0.1 - 2 \, \rm{Myr}$. The box size for each simulation was chosen to ensure that the outflow does not reach the edge of the box within $1 \, \rm{Myr}$. The gas is initially in thermal and chemical equilibrium in the presence of the redshift zero extragalactic UV background of \citet{haardt01}, i.e. before the radiation from the AGN is switched on. 

To reduce the computational cost, we only simulate one octant of the box at high resolution, with the remainder of the box at a factor 8 lower mass resolution. We use a fiducial resolution of $30 \, \rm{M}_{\odot}$ per gas particle, with 32 kernel neighbours, in the high-resolution region. Some runs use a lower resolution of $240 \, \rm{M}_{\odot}$ per particle, and we include one high-resolution run with $10 \, \rm{M}_{\odot}$ per particle (see Section~\ref{suite_sect}). 

We include gravity from a central black hole using a single collisionless particle with a mass $M_{\rm{BH}} = 10^{8} \, \rm{M}_{\odot}$. We also include a static potential from an isothermal sphere to represent the galaxy potential, with an enclosed mass at radius $R$ of: 

\begin{equation} 
M_{\rm{gal}}(<R) = \frac{2 \sigma^{2} R}{G}. 
\end{equation} 
We use a velocity dispersion $\sigma = 200 \, \rm{km} \, \rm{s}^{-1}$, which corresponds to a black hole mass of $10^{8} \, \rm{M}_{\odot}$ on the $M_{\rm{BH}} - \sigma$ relation \citep[e.g.][]{gultekin09}. 

We also include self-gravity of the gas. For gas particles, we use an adaptive gravitational softening that is set to be equal to the inter-particle spacing, which is defined as $h_{\rm{inter}} = (m_{\rm{gas}} / \rho)^{1/3}$ for a gas particle with mass $m_{\rm{gas}}$ and density $\rho$. We impose a minimum gravitational softening of $0.1 \, \rm{pc}$ at the fiducial mass resolution. The gravitational softening of the black hole particle is fixed at $1 \, \rm{pc}$. 

Note that, since we consider a uniform ambient medium, these simulations may represent an outflow from a buried quasar nucleus. Once the outflow breaks out of the galactic disc, the density of the ambient medium will decrease rapidly and the wind bubble may lose its internal pressure support when the hot gas vents out \citep[e.g.][]{stern16}.

\subsection{AGN wind model}\label{agn_wind_sect} 

We inject an isotropic AGN wind at the centre of the simulation box by spawning new gas particles within the central $1 \, \rm{pc}$ with an outward radial velocity $v_{\rm{in}} = 30 \, 000 \, \rm{km} \, \rm{s}^{-1}$. Such velocities have been observed in broad absorption line (BAL) quasars \citep[e.g.][]{weymann81, gibson09}, and in ultra-fast outflows from quasars observed in X-rays \citep[e.g.][]{feruglio15, nardini15, tombesi15}, and could be launched, for example, from the accretion disc around the black hole \citep{murray95}, although we do not explicitly model the launching mechanism in this work. Following FGQ12, we parameterise the momentum injection rate of the AGN wind as: 

\begin{equation}\label{momentum_eqn} 
\dot{M}_{\rm{in}} v_{\rm{in}} = \tau_{\rm{in}} \frac{L_{\rm{AGN}}}{c}, 
\end{equation} 
where $\dot{M}_{\rm{in}}$ is the mass injection rate of the wind and $L_{\rm{AGN}}$ is the AGN luminosity. $\tau_{\rm{in}}$ is a parameter that is related to the mass loading of the wind (see equation 3 of FGQ12); we use $\tau_{\rm{in}} = 1$. 

For our choice of $v_{\rm{in}}$ and $\tau_{\rm{in}}$, and a given $L_{\rm{AGN}}$, we can calculate $\dot{M}_{\rm{in}}$ from equation~\ref{momentum_eqn}. For each hydrodynamic time-step we can then calculate the mass of gas to inject. We spawn wind particles in pairs oriented in opposite directions, so that their net momentum vector is zero. If the injected mass in a time-step exceeds twice the gas particle mass in the high-resolution region, we spawn pairs of gas particles with random orientations at random distances between 0 and 1 pc from the centre of the box. All wind particles have a fixed mass equal to that of the high-resolution particles, even if they are injected into a low-resolution region. Wind particles are initially cool ($10^{4} \, \rm{K}$), and will later be shock-heated by the reverse shock. Any injected mass that is insufficient to create a pair of gas particles is carried over to later time-steps. For a more detailed discussion of the different components of the wind bubble that results, we refer readers to FGQ12. 

\subsection{Chemistry and cooling}\label{chemistry_sect} 

We use the \textsc{chimes} chemistry solver module \citep{richings14a, richings14b} to follow the evolution of 157 chemical species, including all ionization states of 11 elements\footnote{H, He, C, N, O, Ne, Mg, Si, S, Ca, Fe.} and 20 molecular species\footnote{H$_{2}$, H$_{2}^{+}$, H$_{3}^{+}$, OH, H${_2}$O, C$_{2}$, O$_{2}$, HCO$^{+}$, CH, CH$_{2}$, CH$_{3}^{+}$, CO, CH$^{+}$, CH$_{2}^{+}$, OH$^{+}$, H$_{2}$O$^{+}$, H$_{3}$O$^{+}$, CO$^{+}$, HOC$^{+}$, O$_{2}^{+}$.}. \textsc{chimes} also calculates the radiative cooling and heating rates using the non-equilibrium chemical abundances. This chemical model covers a wide range of physical conditions relevant to AGN winds, from hot, highly ionized gas to the cold, CO-bright molecular gas that is the focus of this study. The chemical model gives us a system of 158 coupled differential equations (157 chemical rate equations and the thermal energy equation). Each time-step \textsc{chimes} integrates these equations for every active gas particle using the \textsc{cvode} non-linear differential equation solver from the \textsc{sundials}\footnote{\url{https://computation.llnl.gov/casc/sundials/main.html}} suite. For this integration, we use relative and absolute tolerances of $10^{-4}$ and $10^{-10}$, respectively, for most particles. However, for particles with temperatures $T > 10^{6} \, \rm{K}$ or densities $n_{\rm{H}} < 50 \, \rm{cm}^{-3}$ we found it necessary to use a relative tolerance of $10^{-6}$. We impose a cooling floor at $T = 10 \, \rm{K}$. 

A full description of the chemical reactions and thermal processes included in \textsc{chimes} can be found in \citet{richings14a}. For this work, we have updated the photoelectric heating rate from dust grains used in \textsc{chimes} to that given in equations 19 and 20 of \citet{wolfire03}, which better models the effects of PAHs. Below we summarise the most important thermo-chemical processes that are relevant for this study. \\ 

\textbf{H$_{2}$ formation.} We include the formation of H$_{2}$ in the gas phase (via H$^{-}$) and on dust grains. For the formation rate of H$_{2}$ on dust grains, we use equation 18 of \citet{cazaux02}, where we scale the dust abundance linearly with metallicity. This assumes that the dust-to-metals ratio is constant. However, it is unclear whether dust grains can survive the strong shocks and high gas temperatures found in AGN winds, and some studies have suggested that the dust grains will be rapidly destroyed in such conditions \citep[e.g.][]{ferrara16}. As we will discuss further in Section~\ref{model_vars_sect}, the formation of molecules in an AGN wind is sensitive to the assumed dust-to-metals ratio, and this is therefore a major uncertainty in our model. 

\textbf{Photoionization and photoheating.} We include the photoionization of atoms and ions and the photodissociation of molecules, along with the associated heating, due to the AGN radiation field. To calculate the photochemical rates, we use the spectrum of an average quasar from \citet{sazonov04}, scaled to a given bolometric AGN luminosity $L_{\rm{AGN}}$. At a radius $R$ from the black hole, the flux of hydrogen-ionizing photons (with energy $h \nu > 1 \, \rm{Ryd}$) is then: 

\begin{equation} 
f_{\rm{ion}} = 1.62 \times 10^{11} \, \rm{cm}^{-2} \, \rm{s}^{-1} \left( \frac{L_{\rm{AGN}}}{10^{46} \, \rm{erg} \, \rm{s}^{-1}} \right) \left( \frac{R}{\rm{kpc}} \right)^{-2}. 
\end{equation} 

To model shielding of gas from the AGN radiation, we use a local shielding approximation that assumes that a gas particle is shielded by material within a shielding length $L_{\rm{sh}}$ of the particle. The column density of species $i$, with volume density $n_{i}$, is then: 

\begin{equation}\label{shield_eqn} 
N_{i} = n_{i} L_{\rm{sh}}. 
\end{equation} 
This also assumes that $n_{i}$ is uniform over the shielding length. 

For the shielding length, we use a Sobolev-like approximation based on the local gradient of the density, $\rho$: 

\begin{equation}\label{sobolev_eqn} 
L_{\rm{sh}} = \frac{1}{2} \left( h_{\rm{inter}} + \frac{\rho}{|{\nabla \rho}|} \right), 
\end{equation} 
where the inter-particle spacing $h_{\rm{inter}}$ for a gas particle of mass $m_{\rm{gas}}$ is defined such that $\rho = m_{\rm{gas}} / h_{\rm{inter}}^{3}$, and is related to the kernel smoothing length $h_{\rm{sml}}$ and the number of neighbours $N_{\rm{ngb}}$ as $h_{\rm{inter}} = h_{\rm{sml}} / N_{\rm{ngb}}^{1/3}$. This approximation has previously been used in cosmological simulations \citep[e.g.][]{gnedin09, hopkins17}, and simulations of isolated disc galaxies that use the \textsc{chimes} chemical model \citep{richings16}, although not always with the $h_{\rm{inter}}$ term. 

\citet{safranekshrader17} recently compared several local approximations for radiative shielding, including the density gradient-based approach that we use, in galactic disc simulations in post-processing. They found that, of the methods that depend only on local quantities, using the Jeans length, with a temperature capped at $40 \, \rm{K}$, gave the most accurate results for H$_{2}$ and CO fractions, as compared to a full radiative transfer calculation. However, in Appendix~\ref{shielding_appendix} we compare these two local approximations in our simulations to the true column densities computed from a simple ray tracing algorithm in post-processing, and we find that the density gradient-based approach is the most accurate. This is likely because the AGN winds that we simulate are a very different environment to the turbulent galactic discs that \citet{safranekshrader17} studied, so it is possible that the Jeans length assumption is not valid for cold clumps in an AGN wind. Nevertheless, there is still a large scatter of an order of magnitude between the density gradient-based approximation and the true ray tracing-based column densities. We explore the impact of these uncertainties on our results in Section~\ref{model_vars_sect}. 

Once we have calculated the column densities for a given particle from equations~\ref{shield_eqn} and \ref{sobolev_eqn}, we then attenuate the photochemical and photoheating rates as a function of the column densities of H\textsc{i}, H$_{2}$, He\textsc{i}, He\textsc{ii}, CO and the dust extinction, $A_{\rm{v}}$, as described in \citet{richings14b}. We use $A_{\rm{v}} = 4.0 \times 10^{-22} N_{\rm{H_{tot}}} Z / \rm{Z}_{\odot}$, where $N_{\rm{H_{tot}}}$ is the total hydrogen column density. This is an average of the dust models of \citet{weingartner01} for the Milky Way and the Large and Small Magellanic Clouds (see appendix A of \citealt{krumholz11}). This also assumes a constant dust-to-metals ratio, which is a major uncertainty in this work, as discussed above. 

We use a temperature-dependent H$_{2}$ self-shielding function that was fit to a series of photodissociation region (PDR) models created with version 13.01 of \textsc{cloudy}\footnote{\url{http://nublado.org/}} \citep[last described by][]{ferland13}, as described in appendix B of \citet{richings14b}. Doppler broadening of the Lyman-Werner lines, for example by turbulence, tends to weaken the self-shielding of H$_{2}$ \citep[e.g.][]{draine96, ahn07}. We assume a constant turbulent broadening parameter $b_{\rm{turb}} = 7.1 \, \rm{km} \, \rm{s}^{-1}$, which is typical for molecular clouds in the Milky Way. This is added to the thermal broadening in quadrature to give the total broadening parameter, which is used in the self-shielding function. 

\textbf{Compton cooling and heating from AGN radiation.} The original \textsc{chimes} model only included Compton cooling from the cosmic microwave background. For this work, we have added non-relativistic Compton cooling and heating from the AGN radiation field, with a Compton temperature $T_{\rm{C}} = 2 \times 10^{7} \, \rm{K}$ \citep{sazonov04}. The cooling rate (where a negative value corresponds to heating) for an AGN luminosity $L_{\rm{AGN}}$ at a distance $R$ is then: 

\begin{equation}\label{compton_eqn} 
\Lambda_{\rm{Compton}} = \frac{\sigma_{\rm{T}}}{m_{\rm{e}} c^{2}} \left( \frac{L_{\rm{AGN}}}{\pi R^{2}} \right) k_{\rm{B}} (T - T_{\rm{C}}) n_{\rm{e}} \, \rm{erg} \, \rm{s}^{-1} \, \rm{cm}^{-3}, 
\end{equation} 
where $\sigma_{\rm{T}}$, $m_{\rm{e}}$ and $k_{\rm{B}}$ are the Thomson cross section, electron mass and Boltzmann constant, respectively, and $T$ and $n_{\rm{e}}$ are the temperature and electron density of the gas. 

In cold gas ($T < T_{\rm{C}}$), this will be a source of heating. At the high post-shock temperatures corresponding to the wind velocity, $v_{\rm{in}} = 30 \, 000 \, \rm{km} \, \rm{s}^{-1}$, used in our simulations ($\sim 10^{10} \, \rm{K}$), the non-relativistic assumption can break down. However, FGQ12 argued that, in this regime, the temperatures of electrons and protons become thermally decoupled, which leads to two-temperature effects that cause the shocked electrons to effectively remain non-relativistic and the cooling time of the protons to become limited by Coulomb interactions with the electrons. This generally increases the cooling time of the shocked wind by a large factor relative to relativistic Compton cooling. Since we do not model two-temperature effects in our simulations, we would overestimate the cooling in the hot AGN wind if we used the relativistic Compton cooling rate. We therefore use the non-relativistic rate given above. Since we find that the shocked wind does not significantly cool in our simulations, our results are not sensitive to the exact details of how Compton cooling is modelled.

\textbf{Cosmic rays.} We include ionization, dissociation of molecules, and heating due to cosmic rays. The strength of the cosmic rays is parameterised by the primary ionization rate of H\textsc{i} by cosmic rays, $\zeta_{\rm{HI}}$. The ionization and dissociation rates of other species are then scaled to the H\textsc{i} rate according to their relative rates in the \textsc{umist} database\footnote{\url{http://www.udfa.net/}} \citep{mcelroy13} where available, or using the equations in \citet{lotz67}, \citet{silk70} and \citet{langer78} otherwise. Secondary ionizations of H\textsc{i} and He\textsc{i} from cosmic rays are calculated using the tables of \citet{furlanetto10}, assuming a primary electron energy of $35 \, \rm{eV}$. 

The cosmic ray ionization rate of H\textsc{i} in the Milky Way, inferred from observations of H$_{3}^{+}$, is $\zeta_{\rm{HI}}^{\rm{MW}} = 1.8 \times 10^{-16} \, \rm{s}^{-1}$ \citep{indriolo12}. However, the quasar host galaxies where fast molecular outflows have been observed are typically ULIRGs, with higher star formation rates than the Milky Way. We therefore expect that the cosmic ray density will also be higher in these ULIRGs than in the Milky Way. \citet{lacki10} present 1D models of cosmic ray diffusion in star-forming galaxies. Their fig. 15 shows the cosmic ray energy density, $U_{\rm{CR}}$, as a function of gas surface density, $\Sigma_{\rm{g}}$, in their models. The gas surface density in the Milky Way is $\Sigma_{\rm{g}}^{\rm{MW}} \approx 10 \, \rm{M}_{\odot} \, \rm{pc}^{-2} = 0.002 \, \rm{g} \, \rm{cm}^{-2}$, while ULIRGs have surface densities up to $\Sigma_{\rm{g}}^{\rm{ULIRG}} \approx 10^{4} \, \rm{M}_{\odot} \, \rm{pc}^{-2} = 2 \, \rm{g} \, \rm{cm}^{-2}$ \citep[see fig. 11a of][]{kennicutt11}. If we assume that $\zeta_{\rm{HI}} \propto U_{\rm{CR}}$ then, using fig. 15 of \citet{lacki10} to get $U_{\rm{CR}}$, and with $\zeta_{\rm{HI}}^{\rm{MW}}$ from \citet{indriolo12}, we find $\zeta_{\rm{HI}}^{\rm{ULIRG}} = 5.7 \times 10^{-15} \, \rm{s}^{-1}$. We use this value in most of our simulations, although we include one run with the Milky Way value to explore how uncertainties in this rate affect our results. 

\subsection{Simulation suite}\label{suite_sect} 

Table~\ref{sims_table} summarises the physical parameters of our simulations. For our fiducial model, we considered an AGN with a bolometric luminosity $L_{\rm{AGN}} = 10^{46} \, \rm{erg} \, \rm{s}^{-1}$, embedded in an ambient ISM with a uniform initial density $n_{\rm{H}} = 10 \, \rm{cm}^{-3}$ and solar metallicity. These conditions are typical for the ULIRG host galaxies in which fast molecular outflows have been observed. Note that, while mean densities in the central $\approx 500 \, \rm{pc}$ of ULIRGs are typically much higher than this ($\sim 10^{3} - 10^{4} \, \rm{cm}^{-3}$, e.g. \citealt{downes98}), measured outflow parameters suggest that, on average, the outflows must have expanded along under-dense paths (see FGQ12). Our spherically-symmetric initial conditions may be viewed as an approximation to AGN winds expanding in the nuclei of buried quasars, in which all directions from the central AGN are significantly covered by ambient gas. Observations offer some evidence that this setup is appropriate for understanding the generation of quasar-driven molecular outflows. For example, \citet{gonzalezalfonso17} note that the highest molecular outflow velocities are found in buried sources.

This fiducial model is labelled nH10\_L46\_Z1. We then ran a further three simulations where the ambient ISM density, AGN luminosity and metallicity were each reduced in turn by a factor of 10. This set of four simulations, run at our fiducial resolution of $30 \, \rm{M}_{\odot}$ per gas particle, allows us to explore how molecule formation is affected by the physical parameters of the system. These parameter variations are presented in Section~\ref{parameter_vars_sect}. 

\begin{table*}
\centering
\begin{minipage}{168mm}
\caption{Simulation runs and parameters: initial ambient ISM density $n_{\rm{H}}$, bolometric AGN luminosity $L_{\rm{AGN}}$, gas metallicity $Z$, gas particle mass $m_{\rm{gas}}$ in the high-resolution octant, minimum gravitational softening $\epsilon_{\rm{min}}$ for gas particles, and the side length of the high-resolution octant $l_{\rm{hiRes}}$. The model variations are discussed further in the text.}
\label{sims_table}
\begin{tabular*}{\linewidth}{@{}l@{\extracolsep{\fill}} cccccc}
\hline 
Name & $n_{\rm{H}} \, (\rm{cm}^{-3})$ & $L_{\rm{AGN}} \, (\rm{erg} \, \rm{s}^{-1})$ & $Z / \rm{Z}_{\odot}$ & $m_{\rm{gas}} \, (\rm{M}_{\odot})$ & $\epsilon_{\rm{min}} \, (\rm{pc})$ & $l_{\rm{hiRes}} \, (\rm{kpc})$ \\ 
\hline
\multicolumn{7}{c}{Parameter variations} \\ 
\hline
nH10\_L46\_Z1 & 10 & $10^{46}$ & 1 & 30 & 0.1 & 1.2 \\ 
nH1\_L46\_Z1 & 1 & $10^{46}$ & 1 & 30 & 0.1 & 2.5 \\ 
nH10\_L45\_Z1 & 10 & $10^{45}$ & 1 & 30 & 0.1 & 0.8 \\
nH10\_L46\_Z0.1 & 10 & $10^{46}$ & 0.1 & 30 & 0.1 & 1.2 \\ 
\hline
\multicolumn{7}{c}{Resolution variations} \\ 
\hline
nH10\_L46\_Z1\_lowRes & 10 & $10^{46}$ & 1 & 240 & 0.2 & 1.2 \\ 
nH1\_L46\_Z1\_lowRes & 1 & $10^{46}$ & 1 & 240 & 0.2 & 2.5 \\ 
nH10\_L45\_Z1\_lowRes & 10 & $10^{45}$ & 1 & 240 & 0.2 & 0.8 \\ 
nH10\_L45\_Z1\_hiRes & 10 & $10^{45}$ & 1 & 10 & 0.07 & 0.8 \\ 
nH10\_L46\_Z0.1\_lowRes & 10 & $10^{46}$ & 0.1 & 240 & 0.2 & 1.2 \\ 
\hline
\multicolumn{7}{c}{Model variations} \\ 
\hline
lowCR & 10 & $10^{46}$ & 1 & 240 & 0.2 & 1.2 \\ 
jeansLimiter & 10 & $10^{46}$ & 1 & 240 & 0.2 & 1.2 \\ 
noAGNflux & 10 & $10^{46}$ & 1 & 240 & 0.2 & 1.2 \\ 
lowShield & 10 & $10^{46}$ & 1 & 240 & 0.2 & 1.2 \\ 
lowDust10 & 10 & $10^{46}$ & 1 & 240 & 0.2 & 1.2 \\ 
lowDust100 & 10 & $10^{46}$ & 1 & 240 & 0.2 & 1.2 \\ 
\hline
\end{tabular*}
\end{minipage}
\end{table*}

To test the effects of numerical resolution on our results, we repeated these four simulations with a lower resolution of $240 \, \rm{M}_{\odot}$ per gas particle. We also performed one high-resolution run, nH10\_L45\_Z1\_hiRes, with a resolution of $10 \, \rm{M}_{\odot}$ per gas particle. We used the low-luminosity model for this high-resolution run, rather than our fiducial model, as it was computationally cheaper. These resolution tests are presented in Appendix~\ref{resolution_appendix}. 

Finally, we performed a further six runs of nH10\_L46\_Z1\_lowRes in which we varied different aspects of our AGN and chemistry modelling, to explore uncertainties in the models. These runs all used the lower resolution of $240 \, \rm{M}_{\odot}$ per particle, and are described in more detail below. \\ 

\textbf{lowCR:} In our fiducial model we assume a primary ionization rate of H\textsc{i} due to cosmic rays of $\zeta_{\rm{HI}} = 5.7 \times 10^{-15} \, \rm{s}^{-1}$, which we expect to be appropriate for ULIRGs. However, this rate is uncertain as it is based on scaling the observed Milky Way rate up to typical ULIRG conditions using the 1D cosmic ray diffusion models of \citet{lacki10}. To quantify how uncertainties in this rate might affect molecule formation, we performed one run with the Milky Way rate, $1.8 \times 10^{-16} \, \rm{s}^{-1}$ \citep{indriolo12}, which is a factor $\approx 30$ lower than our fiducial value. 

\textbf{jeansLimiter:} \citet{bate97} and \citet{truelove97} demonstrated that hydrodynamic simulations that include self-gravity may experience artificial fragmentation if they do not resolve the Jeans scale. Our simulations do not include an explicit Jeans limiter to ensure that the Jeans scale is always resolved. At our fiducial resolution, the thermal Jeans scale becomes unresolved at densities $n_{\rm{H}} \ga 10^{4} \, \rm{cm}^{-3}$ at $100 \, \rm{K}$. Beyond this density, the Jeans mass is smaller than the particle mass, $30 \, \rm{M}_{\odot}$. Note that, in the MFM hydro solver, a gas particle is roughly equivalent to a gas cell in a grid-based code. Thus the effective spatial and mass resolution in MFM is the inter-particle spacing and the particle mass, respectively, rather than the kernel smoothing length and the kernel mass as in SPH codes (see section 4.2.1 of \citealt{hopkins17} for a more detailed discussion of the resolution requirements of MFM). 

To investigate whether artificial fragmentation may affect our results, we ran one simulation in which we included a density-dependent pressure floor to ensure that the thermal Jeans scale is always resolved. This method to prevent artificial fragmentation has previously been used in galaxy simulations \citep[e.g.][]{robertson08, hopkins11, richings16}. The pressure floor, $P_{\rm{floor}}$, was chosen such that the Jeans length is always at least a factor $N_{\rm{J,} \, \rm{l}}$ times the smoothing length, $h_{\rm{sml}}$. Then, from equation 2.14 of \citet{richings16}: 

\begin{equation}\label{floor_eqn}
P_{\rm{floor}} = \frac{G}{\pi \gamma} N_{\rm{J, \, l}}^{2} h_{\rm{sml}}^{2} \rho^{2}, 
\end{equation}
where $\rho$ is the gas density, and the ratio of specific heats is $\gamma = 5/3$. We used $N_{\rm{J,} \, \rm{l}} = 3.2$, which is equivalent to resolving the Jeans mass by 4 times the kernel mass, as used by \citet{richings16}. \\ 

\textbf{noAGNflux:} The radiation field of the AGN may have an important effect on the formation of molecular outflows, as it can destroy molecules via photodissociation, and it can heat the cold gas via photoheating and photoelectric heating from dust grains. It can also affect how the AGN wind evolves, as Compton cooling due to the AGN radiation can enable the hot wind bubble to cool under certain conditions, which will make the wind momentum-driven rather than energy-driven (e.g. \citealt{king03}; FGQ12; \citealt{costa14}). We therefore performed one run with the AGN radiation switched off, to investigate the impact that it has on the chemistry and the AGN wind structure.

\textbf{lowShield:} As described in Section~\ref{chemistry_sect}, we use a local approximation for the shielding column density of each gas particle. However, in Appendix~\ref{shielding_appendix} we show that, while this approximation does on average follow the true column density, as computed from ray tracing in post-processing, there is a scatter of a factor of 10 between the two. We therefore ran one simulation in which the local column density was reduced by a factor of 10, to quantify how these uncertainties might affect our results. 

\textbf{lowDust10/100:} Dust grains are important for molecule formation as they catalyse H$_{2}$ formation and shield molecules from photodissociating radiation. In our chemical model we assume that the dust abundance scales linearly with metallicity. However, it is unclear whether dust grains will survive in AGN winds, as some studies have suggesed that dust grains will be rapidly destroyed by shocks and sputtering \citep[e.g.][]{ferrara16}. We therefore ran two simulations with dust abundances reduced by a factor of 10 and 100, to explore how uncertainties in the dust abundance might affect our results. 

The results of these model variations are presented in Section~\ref{model_vars_sect}. 

\section{Parameter variations}\label{parameter_vars_sect} 

\begin{figure*}
\centering
\mbox{
	\includegraphics[width=168mm]{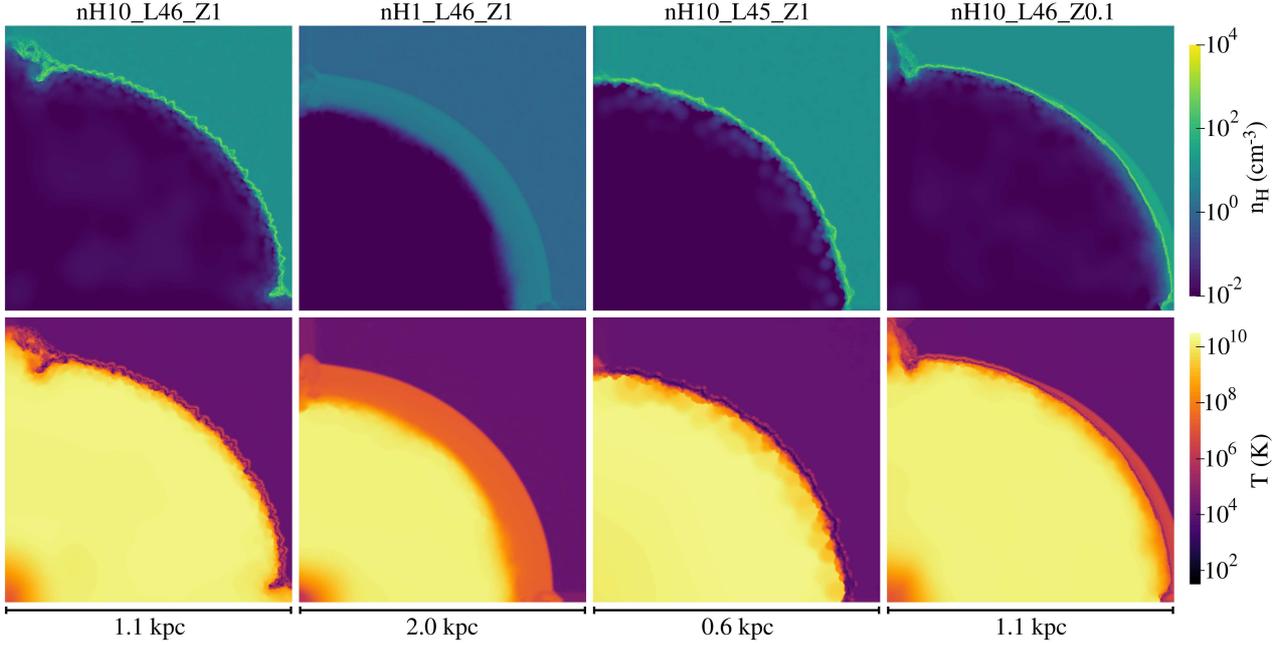}}
\caption{Snapshots from the four `parameter variations' runs at the end of each simulation, after $1 \, \rm{Myr}$, showing 2D slices of the gas density (top row) and temperature (bottom row). These 2D slices are taken along the $x=y$ plane through the high-resolution octant. We then project the mean density and mass-weighted mean temperature in a slice $0.03 \, \rm{kpc}$ thick onto this plane. The size of each panel is indicated at the bottom of the figure. The outflow is energy-driven in all cases, with a hot bubble that maintains a high temperature of $\sim 10^{10} \, \rm{K}$, surrounded by a thin shell of material swept up from the ambient ISM. This swept up material is able to radiatively cool within $1 \, \rm{Myr}$ in all runs except the low ambient density run, nH1\_L46\_Z1.}
\label{gasFig}
\end{figure*}

In this section we focus on the four fiducial-resolution simulations in which we vary the ambient ISM density, AGN luminosity, and metallicity. Fig.~\ref{gasFig} shows snapshots of the high-resolution octant at the end of each simulation, after $1 \, \rm{Myr}$. The top and bottom rows show 2D slices of the gas density and temperature, respectively, along the $x=y$ plane, i.e. a vertical plane diagonally through the high-resolution octant. The spatial extent of these images ranges from $0.6$ to $2.0 \, \rm{kpc}$, as indicated at the bottom of the figure. 

\begin{figure}
\centering
\mbox{
	\includegraphics[width=84mm]{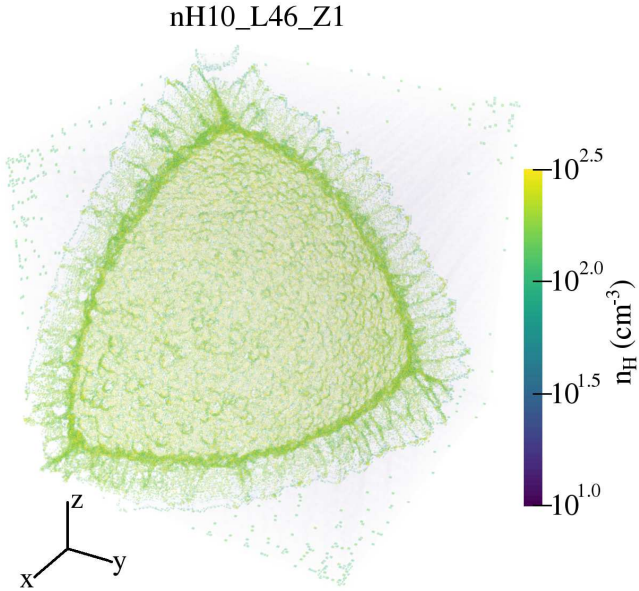}}
\caption{A 3D volume rendering of the gas density in the high-resolution octant of run nH10\_L46\_Z1. We see density inhomogeneities in the shell of swept up gas, which are created when this material radiatively cools.}
\label{gasFig3D}
\end{figure}

We see artifacts near the boundaries between the high- and low-resolution regions (i.e. the left and bottom edges of each panel), which break the spherical symmetry of the outflow. These are particularly noticeable in the nH10\_L46\_Z1 run (left-hand column). They occur because the kernel used by the MFM hydro solver smoothes over gas particles with different masses near these boundaries, which can lead to inaccuracies in the hydro solver. However, we find that these artifacts are only near the boundaries with the low-resolution regions, and most of the volume is unaffected. For example, in Fig.~\ref{gasFig3D} we show a 3D volume rendering of the gas density in the high-resolution octant of run nH10\_L46\_Z1. We see in this 3D view that the artifacts are only found along the boundaries with the low-resolution regions. 

To ensure that these artifacts do not affect our results, we restrict our analysis throughout this paper to particles in the high-resolution region with a polar and azimuthal angle (in spherical polar coordinates) between $15^{\circ}$ and $75^{\circ}$. This defines a wedge through the high-resolution octant that avoids regions near the boundaries with the low-resolution regions. We will refer to this as the high-resolution wedge for the remainder of this paper. For quantities such as molecular masses and outflow rates, we scale up our results for the wedge to what we would get in a full sphere by multiplying by the ratio of the solid angle of a sphere to the solid angle subtended by the wedge (a factor of 17). 

In all four simulations we see a bubble of hot ($\sim 10^{10} \, \rm{K}$), low density ($n_{\rm{H}} \sim 10^{-2} \, \rm{cm}^{-3}$) gas. This bubble has maintained its high temperature throughout the simulation, so radiative cooling is inefficient here and the outflows are in the energy-driven regime. Except for in the nH10\_L45\_Z1 run, the central $\sim 100 \, \rm{pc}$ is cooler. This corresponds to the position of the reverse shock (which we determine from the radial velocity of wind particles as a function of radius; not shown), and particles within this radius have not yet been shock-heated. In the low-luminosity run, we expect that the inner, pre-shock wind should also remain cool, but this is not apparent in Fig.~\ref{gasFig} because there the 10$\times$ lower $\dot{M}_{\rm in}$ implies that there are very few resolution elements in the wind at any given time. 

Close to the AGN, Compton cooling can potentially cool the hot bubble. This would lead to a momentum-driven wind \citep[e.g.][]{king03, costa14, ferrara16}, provided that this occurs beyond the free expansion radius, i.e. the radius at which the total mass injected by the nuclear wind is equal to the swept up mass (FGQ12). Using the non-relativistic Compton cooling rate given in equation~\ref{compton_eqn}, the cooling time at a radius $R$ from an AGN with luminosity $L_{\rm{AGN}}$ is: 

\begin{equation} 
t_{\rm{cool}} = \frac{3}{2} \frac{n m_{\rm{e}} c^{2}}{n_{\rm{e}} \sigma_{\rm{T}}} \left( \frac{\pi R^{2}}{L_{\rm{AGN}}} \right), 
\end{equation}
where $n$ and $n_{\rm{e}}$ are the total (including electrons, H, He and metals) and electron number densities, respectively. By equating $t_{\rm{cool}}$ to the flow time, $t_{\rm{flow}} = R / v_{\rm{s}}$, for the forward shock velocity $v_{\rm{s}}$, we can calculate the radius $R_{\rm{C}}$ within which we expect Compton cooling to be efficient: 

\begin{equation}\label{Rc_eqn} 
R_{\rm{C}} = 9.3 \left( \frac{L_{\rm{AGN}}}{10^{46} \, \rm{erg} \, \rm{s}^{-1}} \right) \left( \frac{v_{\rm{s}}}{300 \, \rm{km} \, \rm{s}^{-1}} \right)^{-1} \, \rm{pc}, 
\end{equation} 
where we assume $n/n_{\rm{e}} \approx 2$ for a fully ionized plasma. For forward shock velocities $v_{\rm{s}} \approx 300 - 1000 \, \rm{km} \, \rm{s}^{-1}$, as we find in our simulations on scales $\sim 1 \, \rm{kpc}$, we therefore expect Compton cooling to be inefficient beyond $\sim 10 \, \rm{pc}$. At smaller radii, $v_{\rm{s}}$ can be higher than the fiducial $300 \, \rm{km} \, \rm{s}^{-1}$ in equation~\ref{Rc_eqn}, so Compton cooling will often also be negligible at smaller radii. This is consistent with FGQ12's result that AGN-driven wind bubbles should be generically energy-conserving, provided that they are inflated by nuclear winds with high initial velocities $v_{\rm{in}} \ga 10 \, 000 \, \rm{km} \, \rm{s}^{-1}$. 

The model of \citet{king03} suggests that Compton cooling should be efficient out to radii up to $\sim 1 \, \rm{kpc}$, which would result in a momentum-driven wind. This discrepancy arises partly because \citet{king03} uses the relativistic Compton cooling time, which, for an initial wind velocity $v_{\rm{in}} = 30 \, 000 \, \rm{km} \, \rm{s}^{-1}$, is nearly two orders of magnitude shorter than for non-relativistic Compton cooling, as we use. However, FGQ12 argued that, in the hot shocked AGN wind, the electron and proton temperatures become thermally decoupled, and two-temperature effects result in cooling times that are much longer than the relativistic Compton cooling time used in \citet{king03}. Additionally, the model of \citet{king03} assumes an isothermal density profile ($n_{\rm{H}} \propto r^{-2}$), whereas our simulations assume a uniform ambient medium.

While the inner wind shock typically does not cool, the outer shock with the ambient medium has a lower velocity ($\sim v_{\rm{s}}$) and can efficiently cool at moderately high ambient densities \citep[e.g.][]{zubovas14, nims15}. At the outer edge of the hot bubble there is a thin shell of gas swept up from the ambient ISM. In the three runs with an ambient density of $10 \, \rm{cm}^{-3}$ (first, third and fourth columns in Fig.~\ref{gasFig}), this shell has cooled to $\sim 10^{4} \, \rm{K}$, and has been compressed, with densities up to $\sim 10^{4} \, \rm{cm}^{-3}$ (we explore the density and temperature distributions in more detail in Section~\ref{gas_properties_sect}). However, in run nH1\_L46\_Z1 the swept up ISM material is still at $\sim 10^{7} \, \rm{K}$ (which corresponds to the post-shock temperature of the forward shock velocity, $v_{\rm{s}} \sim 1000 \, \rm{km} \, \rm{s}^{-1}$), with only a moderate enhancement in density. This is understandable as the cooling time is longer at lower densities, so in the low-density run the swept up gas has had insufficient time to cool after $1 \, \rm{Myr}$. 

In the nH1\_L46\_Z1 run, the shell of swept up gas remains smooth. However, in the other three runs, radiative cooling in the swept up material leads to compression and collapse of the gas, which creates density inhomogeneities, so the shell is no longer smooth. These inhomogeneities can be seen in both the 2D (Fig.~\ref{gasFig}) and 3D (Fig.~\ref{gasFig3D}) views. It is in this dense, radiatively cooling shell that we might expect molecules to form. 

\subsection{Molecular masses and outflow rates}\label{molecule_sect} 

\begin{figure}
\centering
\mbox{
	\includegraphics[width=84mm]{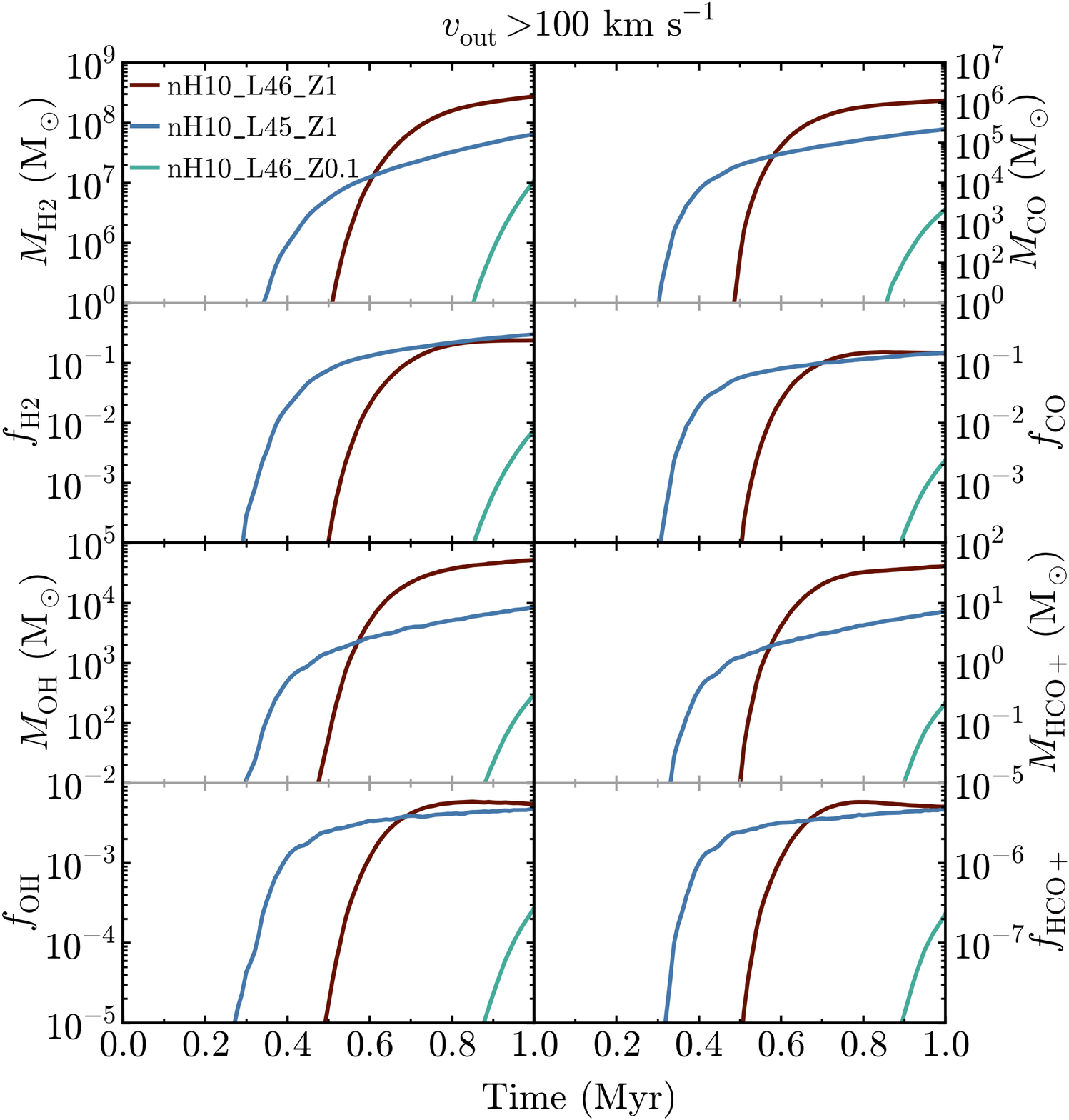}}
\caption{The molecular content of gas outflowing with a velocity $v_{\rm{out}} > 100 \, \rm{km} \, \rm{s}^{-1}$. Each pair of panels shows the mass (top) and molecular fraction (bottom) of a given species as a function of time, for H$_{2}$ (top left pair of panels), CO (top right), OH (bottom left) and HCO$^{+}$ (bottom right). Molecular masses are calculated in the high-resolution wedge and then multiplied by 17 to represent the mass in the full spherical shell. Molecular fractions are defined as the fraction of the total mass of hydrogen (H$_{2}$), carbon (CO and HCO$^{+}$), or oxygen (OH) in the given molecular species. Our fiducial run (nH10\_L46\_Z1; red curves) has an H$_{2}$ mass of $2.7 \times 10^{8} \, \rm{M}_{\odot}$ in the outflow after $1 \, \rm{Myr}$, which decreases with decreasing AGN luminosity (blue curves) and metallicity (green curves). Our low-density run (nH1\_L46\_Z1) is not shown here, as it did not form any molecules.} 
\label{molecularMassFig}
\end{figure}

Fig.~\ref{molecularMassFig} shows the total mass and molecular fractions of H$_{2}$, CO, OH and HCO$^{+}$ (top left, top right, bottom left and bottom right, respectively) in gas that is outflowing with a radial velocity $v_{\rm{out}} > 100 \, \rm{km} \, \rm{s}^{-1}$. The masses are calculated in the high-resolution wedge only, then multiplied by a factor of 17 (the ratio of the solid angle of a sphere to the solid angle subtended by the wedge) to represent the mass we would find in the full spherical shell. The molecular fractions are defined as the fraction of the total hydrogen mass in H$_{2}$, the fraction of the total carbon mass in CO or HCO$^{+}$, or the fraction of the total oxygen mass in OH. These fractions thus go to unity when all of the given element is in the given molecular species. 

In our fiducial run, nH10\_L46\_Z1, the total mass of H$_{2}$ in outflowing gas reaches $2.7 \times 10^{8} \, \rm{M}_{\odot}$ after $1 \, \rm{Myr}$ (top left panel of Fig.2; red curve). For comparison, observations of ULIRGs that host luminous quasars have found H$_{2}$ outflows with masses up to a few times $10^{8} \, \rm{M}_{\odot}$, as measured from e.g. CO emission \citep{feruglio10, cicone14} or OH emission and absorption \citep{gonzalezalfonso17}. The molecular H$_{2}$ fraction of outflowing gas in this run is 0.24 after $1 \, \rm{Myr}$. \citet{rupke13b} measure galactic winds in the neutral atomic phase (via Na \textsc{i} D) and the ionized phase (via O\textsc{i}, H$\alpha$ and N\textsc{ii}) in six ULIRGs, of which three are powered by AGN (F08572+3915, Mrk 231 and Mrk 273). Molecular outflows have also been observed in these three systems from CO emission \citep{cicone12, cicone14}. From these measurements, the molecular H$_{2}$ fractions of these observed outflows are $0.26 - 0.83$. 

When we decrease the AGN luminosity by a factor of 10 (nH10\_L45\_Z1; blue curves), the outflowing H$_{2}$ mass after $1 \, \rm{Myr}$ decreases by a factor of 4. However, the H$_{2}$ fraction of outflowing gas is similar to the fiducial run, so the H$_{2}$ mass is lower because the total mass swept up by the outflow is lower at lower AGN luminosities, which in our simulations correspond to less energetic winds. 

The H$_{2}$ mass in the low-metallicity run (nH10\_L46\_Z0.1; green curves) is a factor 26 lower after $1 \, \rm{Myr}$ compared to the fiducial run. The H$_{2}$ fraction is also correspondingly lower. This is partly due to the lower dust abundance, which is needed to catalyse H$_{2}$ formation and to shield H$_{2}$ from UV radiation. We will also see in Section~\ref{gas_properties_sect} that there is less cold, dense gas in the nH10\_L46\_Z0.1 run, which is likely at least partly because H$_{2}$ is an important coolant below $10^{4} \, \rm{K}$ (in Section~\ref{H2_lines_sect} we will show that warm H$_{2}$ emits strongly in infrared lines in our simulations). Metal line cooling is also reduced in the low-metallicity run. 

Note that, while the H$_{2}$ fraction in outflowing gas has reached a steady state in the fiducial run after $1 \, \rm{Myr}$, in the low-luminosity and low-metallicity runs it is still increasing at the end of the simulation. In these two runs, the H$_{2}$ fraction that we measure therefore depends on the time that we choose to measure it at. 

The low density run (nH1\_L46\_Z1; not shown) did not form any molecules after $1 \, \rm{Myr}$. We saw in Fig.~\ref{gasFig} that the swept up material in this run has had insufficient time to cool by the end of the simulation, so there is no cold, dense gas, which is needed for molecules to form. 

\begin{table}
\begin{minipage}{84mm}
\centering
\caption{OH abundances relative to H$_{2}$ after $1 \, \rm{Myr}$ in the three parameter variations runs that form molecules. For comparison, observations typically assume $\chi (\rm{OH}) = 5 \times 10^{-6}$ when inferring H$_{2}$ masses from OH emission and absorption.}
\label{OH_table}
\begin{tabular}{lc}
\hline 
Simulation & $\chi (\rm{OH}) = \mathit{n}_{\rm{OH}} / \mathit{n}_{\rm{H}_{2}}$ \\ 
\hline
nH10\_L46\_Z1 & $2.3 \times 10^{-5}$ \\ 
nH10\_L45\_Z1 & $1.6 \times 10^{-5}$ \\
nH10\_L46\_Z0.1 & $3.3 \times 10^{-6}$ \\ 
\hline
\end{tabular}
\end{minipage}
\end{table}

The masses and fractions of CO, OH and HCO$^{+}$ in outflowing gas show similar trends between the four parameter variations runs. After $1 \, \rm{Myr}$, we find OH fractions relative to the total oxygen mass ($f_{\rm{OH}} = (16/17) M_{\rm{OH}} / M_{\rm{O}_{\rm{tot}}}$) of $2.7 \times 10^{-4} - 5.4 \times 10^{-3}$ (excluding the low-density run, which does not form OH). Observational studies of OH absorption and emission typically fit radiative transfer models to the OH spectra to determine the OH column density, then convert this to the H$_{2}$ column density using an assumed OH fraction relative to H$_{2}$ ($\chi (\rm{OH}) = \mathit{n}_{\rm{OH}} / \mathit{n}_{\rm{H}_{2}}$). This OH conversion factor is a ratio of OH to H$_{2}$ number densities, unlike the CO conversion factor ($\alpha_{\rm{CO}}$, which we explore further in section~\ref{CO_lines_sect}), which is a ratio of H$_{2}$ mass to CO luminosity. Observational studies typically assume $\chi (\rm{OH}) = 5 \times 10^{-6}$ \citep[e.g.][]{sturm11}, which is based on observations of multiple OH lines in the Sagitarius B2 molecular cloud in the Galactic Centre \citep{goicoechea02}. 

Using the OH and H$_{2}$ fractions from Fig.~\ref{molecularMassFig}, along with the total oxygen abundance, we calculate $\chi (\rm{OH})$ in outflowing gas ($v_{\rm{out}} > 100 \, \rm{km} \, \rm{s}^{-1}$) after $1 \, \rm{Myr}$ in each simulation as follows: 

\begin{equation} 
\chi (\rm{OH}) = \frac{\mathit{n}_{\rm{OH}}}{\mathit{n}_{\rm{H}_{2}}} = \frac{2}{17} \frac{\mathit{M}_{\rm{OH}}}{\mathit{M}_{\rm{H}_{2}}} = \frac{\mathit{f}_{\rm{OH}}}{\mathit{f}_{\rm{H}_{2}}} \frac{\mathit{M}_{\rm{O_{tot}}} / 16}{\mathit{M}_{\rm{H_{tot}}} / 2}, 
\end{equation}  
where $M_{\rm{O_{tot}}}$ and $M_{\rm{H_{tot}}}$ are the total masses of oxygen and hydrogen, respectively. The values of $\chi (\rm{OH})$ from the simulations are shown in Table~\ref{OH_table}. Based on the fiducial run, which produces the strongest molecular outflows, observations of molecular outflows based on OH may overestimate H$_{2}$ masses by up to a factor $\approx 4$. However, we note that \citet{gonzalezalfonso17} assume an OH abundance relative to all H nuclei (not just H$_{2}$) of $2.5 \times 10^{-6}$, which is in good agreement with our simulations. 

The value of $\chi (\rm{OH})$ that we calculate from the simulations depends on the time at which we calculate it, although $1 \, \rm{Myr}$ corresponds to the typical flow times ($r / v_{\rm{out}}$) that are seen in observed AGN-driven molecular outflows \citep[e.g.][]{gonzalezalfonso17}. 

We can also use our simulations to calculate the mass outflow rate in molecular gas: 

\begin{align}\label{outflow_eqn}  
\frac{\rm{d}\mathit{M}_{\rm{H_{2}}}}{\rm{d}\mathit{t}} & = 17 \sum_{\rm{hiResWedge}} \left(\frac{m_{\rm{H_{2}}}}{t_{\rm{flow}}} \right) \nonumber \\
 & = 17 \sum_{\rm{hiResWedge}} \left(\frac{m_{\rm{H_{2}}} v_{\rm{out}}}{r} \right), 
\end{align} 
where $m_{\rm{H_{2}}}$ is the H$_{2}$ mass of a given particle, $t_{\rm{flow}} = r / v_{\rm{out}}$ is the flow time of the particle, $v_{\rm{out}}$ is its radial velocity, and $r$ is its radius from the AGN. The summation is over particles in the high-resolution wedge. We include only particles outflowing with a velocity $v_{\rm{out}} > 100 \, \rm{km} \, \rm{s}^{-1}$, for consistency with Fig.~\ref{molecularMassFig} (although the outflow rates are unchanged if we include all outflowing particles with $v_{\rm{out}} > 0 \, \rm{km} \, \rm{s}^{-1}$, since in our setup the outflowing gas is concentrated in a thin shell). The factor of $17$ then gives us the outflow rate from the full spherical shell. 

\begin{figure}
\centering
\mbox{
	\includegraphics[width=84mm]{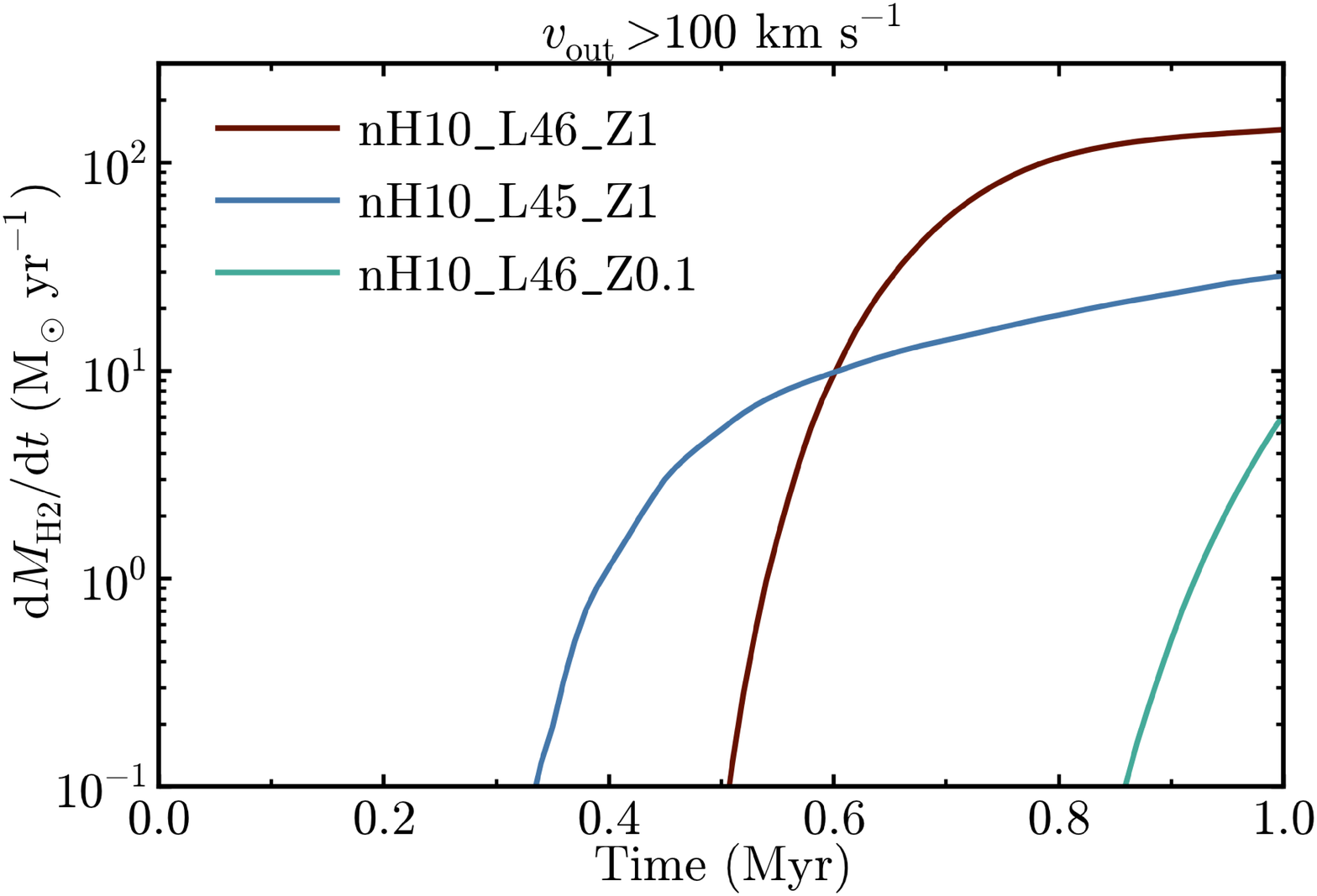}}
\caption{Mass outflow rates of H$_{2}$ in the fiducial (nH10\_L46\_Z1; red curve), low-luminosity (nH10\_L45\_Z1; blue curve), and low-metallicity (nH10\_L46\_Z0.1; green curve) runs, plotted versus time. These outflow rates are calculated in the high-resolution wedge and then multiplied by 17 to give the outflow rate in the full spherical shell. We include only particles outflowing with a velocity $v_{\rm{out}} > 100 \, \rm{km} \, \rm{s}^{-1}$. Our fiducial run reaches an H$_{2}$ outflow rate of $140 \, \rm{M}_{\odot} \, \rm{yr}^{-1}$ after $1 \, \rm{Myr}$.} 
\label{outflowRateFig}
\end{figure}

Fig.~\ref{outflowRateFig} shows the H$_{2}$ mass outflow rate versus time in the three parameter variations runs that form molecules. In our fiducial run (nH10\_L46\_Z1; red curve), the outflow rate reaches $140 \, \rm{M}_{\odot} \, \rm{yr}^{-1}$. For comparison, observations of luminous quasar host galaxies find H$_{2}$ outflow rates up to $\sim 100 - 1000 \, \rm{M}_{\odot} \, \rm{yr}^{-1}$ \citep[e.g.][]{cicone14, gonzalezalfonso17}. However, these observations infer H$_{2}$ masses from other molecules (e.g. CO or OH). As we saw above (and we will see in Section~\ref{CO_lines_sect}), the conversion factors between observed molecular lines and H$_{2}$ can introduce uncertainties of up to a factor $\approx 4-5$. Furthermore, observations typically estimate an outflow rate from the H$_{2}$ mass, size and velocity of the outflow as a whole, whereas we sum over individual particles in the simulations. In Section~\ref{CO_lines_sect} we explore a more rigorous comparison between the observations and our simulations, using radiative transfer calculations of CO line emission performed on our simulations in post-processing to estimate H$_{2}$ outflow rates in the same way as the observations. 

When the AGN luminosity is reduced by a factor of 10 (nH10\_L45\_Z1; magenta curve), the H$_{2}$ outflow rate is lower, reaching $29 \, \rm{M}_{\odot} \, \rm{yr}^{-1}$ after $1 \, \rm{Myr}$. As we saw in Fig.~\ref{molecularMassFig}, this is because less mass has been swept up from the ISM than in the fiducial run (the H$_{2}$ mass fraction in the low-luminosity run is roughly the same as in the reference run at 1 Myr). 

The low-metallicity run (nH10\_L46\_Z0.1; green curve) produces an even lower H$_{2}$ outflow rate, reaching $6 \, \rm{M}_{\odot} \, \rm{yr}^{-1}$ after $1 \, \rm{Myr}$. This reflects the lower H$_{2}$ mass in this run, primarily due to the lower dust abundance, as discussed above. 

In Appendix~\ref{resolution_appendix}, we explore the effects of numerical resolution on the H$_{2}$ outflow rate in the simulations. We find that the molecular outflow rate is not well converged, as it increases by a factor $\approx 4$ as the mass resolution is increased from 240 to $10 \,\rm{M}_{\odot}$ per gas particle (see Fig.~\ref{resFig}). Therefore, our predicted molecular fractions can be interpreted as lower limits. 

\subsection{Physical properties of molecular gas}\label{gas_properties_sect} 

\begin{figure*}
\centering
\mbox{
	\includegraphics[width=153mm]{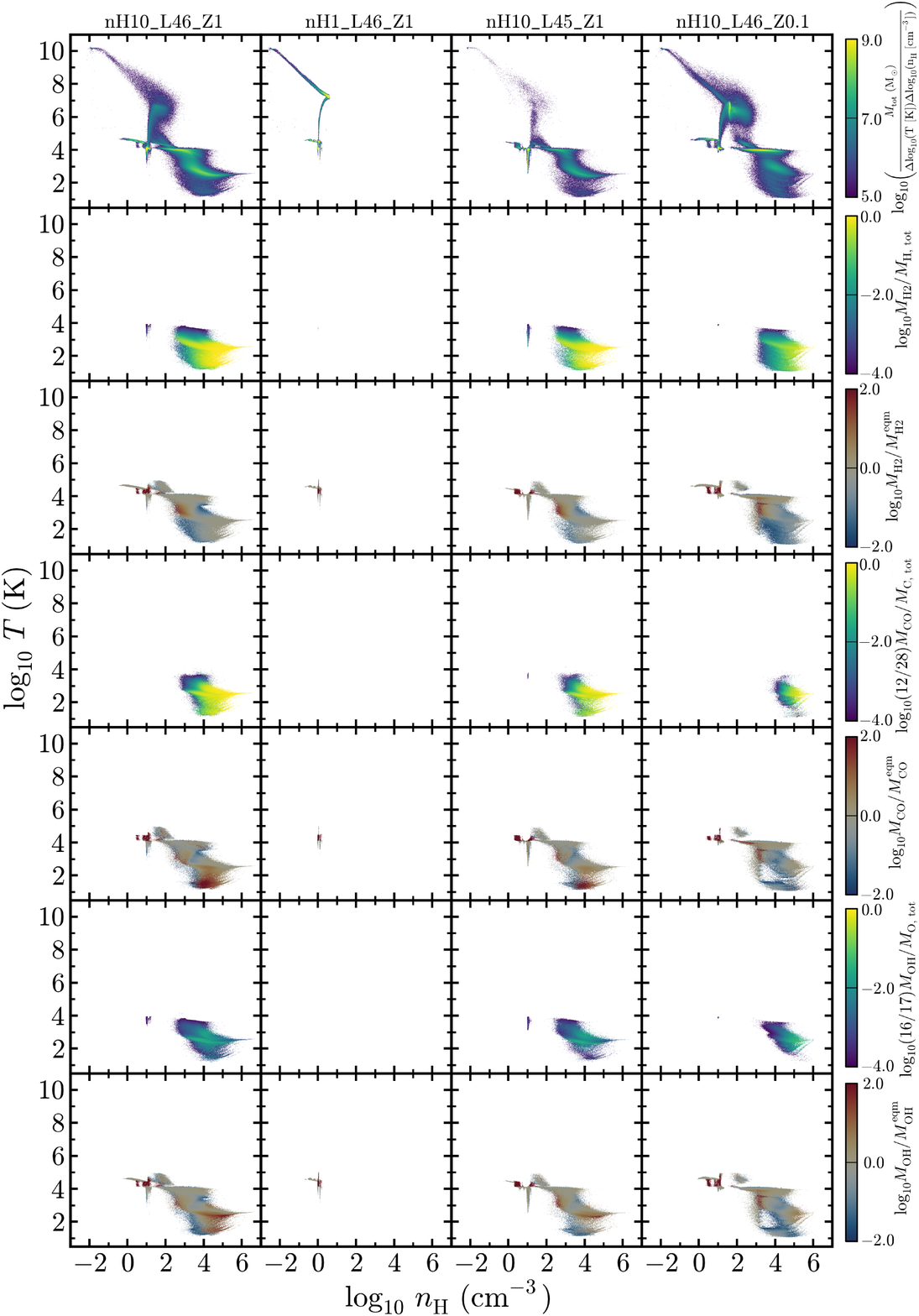}}
\caption{Temperature-density distributions of gas in the high-resolution wedge after $1 \, \rm{Myr}$ in the four parameter variations runs (from left to right: nH10\_L46\_Z1, nH1\_L46\_Z1, nH10\_L45\_Z1, and nH10\_L46\_Z0.1). From top to bottom, the colour scale indicates (in each temperature-density bin): gas mass, H$_{2}$ fraction, ratio of H$_{2}$ masses using non-equilibrium and equilibrium abundances, CO fraction, ratio of non-equilibrium to equilibrium CO masses, OH fraction, and ratio of non-equilibrium to equilibrium OH masses. The factors 12/28 and 16/17 in the colour scales of the fourth and sixth rows, respectively, are the ratios of the atomic weights of C to CO, and of O to OH, which ensures that these fractions go to unity when all carbon is in CO, or when all oxygen is in OH.} 
\label{TrhoFig}
\end{figure*}

Fig.~\ref{TrhoFig} shows the temperature-density distributions of gas particles in the high-resolution wedge after $1 \, \rm{Myr}$, in the four parameter variations runs (from left to right). 

In the top row, the colour scale indicates the mass of gas in each temperature-density bin. In the fiducial run (nH10\_L46\_Z1; top left panel), the injected wind particles have been shock-heated to $T \sim 10^{10} \, \rm{K}$, with a density $n_{\rm{H}} \sim 10^{-2} \, \rm{cm}^{-3}$. The ambient ISM is initially at $n_{\rm{H}} = 10 \, \rm{cm}^{-3}$, $T \sim 10^{4} \, \rm{K}$. The forward shock then heats this to $\sim 10^{7} \, \rm{K}$. The shock-heated ambient ISM is approximately in pressure equilibrium with the shocked wind material, albeit with a large scatter. Once the swept up material is able to radiatively cool, its density increases. At intermediate densities ($\sim 10^{2} - 10^{3} \, \rm{cm}^{-3}$), it cools to $\sim 10^{4} \, \rm{K}$, while at higher densities it cools below $10^{3} \, \rm{K}$. At the highest densities reached in this simulation ($\sim 10^{6} \, \rm{cm}^{-3}$), the thermal equilibrium temperature is $\approx 400 \, \rm{K}$. 

In the low-density run (nH1\_L46\_Z1; second column), almost all of the swept up material is still at the post-shock temperature and density of $\sim 10^{7} \, \rm{K}$ and a few $\rm{cm}^{-3}$, respectively, from passing through the forward shock. This material has not had sufficient time to cool by the end of the simulation after $1 \, \rm{Myr}$, so there is no cold, high-density gas. 

The low-luminosity run (nH10\_L45\_Z1; third column) has a similar temperature-density structure as the fiducial run, although there is more swept up material in the latter. We also saw in Fig.~\ref{gasFig} that the outflow reaches a larger radius after $1 \, \rm{Myr}$ in nH10\_L46\_Z1 than in nH10\_L45\_Z1, and the mass of swept up gas at a given time is simply the mass of the ambient medium that was initially within the radius that the outflow has reached at that time. 

The low-metallicity run (nH10\_L46\_Z0.1; fourth column) also has a similar temperature-density structure to the fiducial run, and with a similar mass of swept up gas. However, less gas has cooled below $10^{4} \, \rm{K}$ in nH10\_L46\_Z0.1, which is likely at least partly due to the reduced molecular H$_{2}$ cooling (we will see in Section~\ref{H2_lines_sect} that warm H$_{2}$ emits strongly in the infrared, and thus H$_{2}$ can be an important coolant). Metal line cooling is also lower in this run. 

In the second, fourth and sixth rows of Fig.~\ref{TrhoFig}, the colour scales indicate the H$_{2}$, CO and OH fractions, respectively, in each temperature-density bin. As above, these fractions have been normalised such that they are equal to one when all hydrogen is in H$_{2}$, all carbon is in CO, or all oxygen is in OH. In the fiducial run, the H$_{2}$ fraction reaches close to unity at densities $n_{\rm{H}} \ga 10^{3} \, \rm{cm}^{-3}$, even at temperatures up to $\sim 10^{3} \, \rm{K}$. In particular, comparing the first and second panels in the left-hand column, for the fiducial run, we see there is a significant amount of warm H$_{2}$ (at a few hundred K), which is likely to exhibit bright emission from the near- and mid-infrared H$_{2}$ lines, as has been observed, for example, in shocked molecular gas in galaxy mergers and galaxy clusters \citep[e.g.][]{vanderwerf93, sugai97, egami06, appleton06, appleton17, zakamska10, hill14}. We will explore the IR-traced H$_{2}$ emission further in Section~\ref{H2_lines_sect}. 

The CO fraction in the fiducial run approaches unity at a higher density ($n_{\rm{H}} \ga 10^{4} \, \rm{cm}^{-3}$) than for the H$_{2}$ fraction. This is the same behaviour as in typical PDR models, as CO requires a higher density than H$_{2}$ to become shielded from dissociating radiation, and results in `CO-dark' molecular gas at intermediate densities, which is rich in H$_{2}$ but not CO \citep[e.g.][]{tielens85, vandishoeck88, wolfire10, smith14}. Similarly to H$_{2}$, we also see significant amounts of CO at high temperatures, up to $\sim 10^{3} \, \rm{K}$. It would be interesting to look for warm CO in AGN-driven winds using high-J transitions of CO. Such high-J lines have been detected by \textit{Herschel} in several ULIRGs, including the quasar Mrk 231 \citep[e.g.][]{vanderwerf10}, although not specifically in the wind. 

In regions of the temperature-density diagram where CO is close to unity, approximately half of the oxygen is in CO (for the metal abundance ratios that we assume here, which are taken from table 1 of \citealt{wiersma09}). The OH fraction in the fiducial run is highest at densities $\ga 10^{4} \, \rm{cm}^{-3}$. However, the OH fraction never exceeds 0.3, and is typically much lower than this. In molecular gas with a CO fraction close to unity, most of the remaining oxygen is in either H$_{2}$O or O\textsc{i}, rather than in OH. 

The low-density run shows no significant H$_{2}$, CO or OH, as the swept up material has had insufficient time to cool, whilst in the low-luminosity run the H$_{2}$, CO and OH fractions show similar behaviour as in the fiducial run. 

At low metallicity, the transition to high molecular fractions in H$_{2}$, CO and OH occurs at a higher density than in the fiducial model (a few times $10^{4} \, \rm{cm}^{-3}$ for H$_{2}$, and $\sim 10^{5} \, \rm{cm}^{-3}$ for CO and OH). This is partly due to the lower formation rate of H$_{2}$ on dust grains (which we assume scales linearly with metallicity), and partly due to reduced shielding by dust, self-shielding, and cross shielding of CO by H$_{2}$. This trend of the atomic to molecular transition with metallicity has also been demonstrated in PDR models \citep[e.g.][]{mckee10, krumholz13, sternberg14}. 

We see in Fig.~\ref{TrhoFig} that our simulations (except the low-density run) produce dense ($n_{\rm{H}} > 10^{4} \, \rm{cm}^{-3}$) molecular gas in the outflow. Such dense gas has been obesrved, for example, from HCN in the wind of Mrk 231 \citep{aalto15}. While we do not track HCN in our chemical network, the bottom right panels of Fig.~\ref{molecularMassFig} show that our simulations do form a significant mass of HCO$^{+}$, which also traces dense gas. 

Finally, the colour scales in the third, fifth and seventh rows in Fig.~\ref{TrhoFig} show, for H$_{2}$, CO and OH respectively, the ratio of the mass of the given species in each temperature-density bin calculated using the non-equilibrium abundances to the corresponding mass assuming chemical equilibrium, in the high-resolution wedge after $1 \, \rm{Myr}$. To compute the equilibrium abundances, we evolved the chemistry of all gas particles in the snapshot at $1 \, \rm{Myr}$ at fixed temperature and fixed particle positions for a further $10 \, \rm{Gyr}$. 

At $T \la 300 \, \rm{K}$ and densities just below the transition to fully molecular, H$_{2}$, CO and OH are lower in non-equilibrium, by $1-2$ orders of magnitude, i.e. the transition to fully molecular would occur at a slightly lower density if we assumed equilibrium abundances. Particles here have not had enough time to fully form molecules, because the formation time-scale of molecules increases with decreasing density. At higher temperatures of $\sim 10^{3} \, \rm{K}$ at these densities, there is an enhancement in the non-equilibrium molecular fractions. 

At $T \la 100 \, \rm{K}$ and moderate densities ($10^{3} \la n_{\rm{H}} \la 10^{4} \, \rm{cm}^{-3}$), the CO mass fraction is higher in non-equilibrium than in equilibrium, by $\sim 2$ orders of magnitude. We find that particles in this region are out of thermal equilibrium, and are evolving rapidly through the density-temperature space. This explains why the CO abundances of these particles can be so far out of chemical equilibrium, as the chemistry cannot keep up with the rapid evolution through temperature-density space. 

The low-luminosity run shows similar trends in the non-equilibrium to equilibrium ratios as the fiducial run. However, the low-metallicity run shows a much stronger reduction of H$_{2}$, CO and OH abundances in non-equilibrium, compared to equilibrium, than we see in the fiducial run. This is unsurprising, as the formation rates of these molecules are lower at low metallicity, due to the lower dust abundance (for H$_{2}$) and the lower carbon and oxygen abundances (for CO and OH). It thus takes longer for them to reach equilibrium. 

At solar metallicity, the total outflowing mass of H$_{2}$ is 24 per cent lower in non-equilibrium than in equilibrium, in both the fiducial and low-luminosity runs. The non-equilibrium effects are stronger for CO and OH, where the fiducial and low-luminosity runs have 89 per cent and 65 per cent, respectively, more CO mass in non-equilibrium than equilibrium, and 96 per cent and 68 per cent, respectively, more OH mass. In the low-metallicity run, the non-equilibrium H$_{2}$, CO and OH masses are 78 per cent, 30 per cent and 46 per cent lower, respectively, than in equilibrium. 

\subsection{Dominant CO formation channels} 

Following the non-equilibrium evolution of CO requires a complex chemical network, with several different formation channels that create CO. In the \textsc{chimes} model, there are 18 reactions that form CO, in addition to reactions that form intermediate species between atomic/ionized carbon and CO (see appendix B of \citealt{richings14a} for a full list of reactions included in the chemical network). To explore whether one formation channel dominates the CO chemistry, we calculate the total formation rate of CO from each of these 18 reactions in simulation nH10\_L46\_Z1 after $1 \, \rm{Myr}$. We include only particles with a CO fraction of at least 10 per cent, to exclude particles where both the formation and destruction rates are high that do not contribute significantly to the total CO mass. 

\begin{table}
\begin{minipage}{84mm}
\centering
\caption{Dominant chemical reactions that form CO, in order of decreasing total rate, calculated from simulation nH10\_L46\_Z1 after $1 \, \rm{Myr}$.}
\label{coFormation}
\begin{tabular}{lc}
\hline 
Reaction & Total CO formation rate \\ 
 & ($\rm{M}_{\odot} \, \rm{yr}^{-1}$) \\
\hline
HCO$^{+}$ + e$^{-}$ $\rightarrow$ CO + H & $1.9 \times 10^{7}$ \\ 
CO$^{+}$ + H $\rightarrow$ CO + H$^{+}$ & $1.8 \times 10^{7}$ \\ 
OH + C$^{+}$ $\rightarrow$ CO + H$^{+}$ & $1.2 \times 10^{7}$ \\ 
OH + C $\rightarrow$ CO + H & $8.0 \times 10^{6}$ \\ 
O$_{2}$ + C $\rightarrow$ CO + O & $1.5 \times 10^{6}$ \\ 
\hline
\end{tabular}
\end{minipage}
\end{table}

We find that there is no single reaction that dominates the formation of CO. Table~\ref{coFormation} lists the five most dominant reactions, in order of decreasing rate. These reactions are also important in diffuse clouds in the Milky Way (see, for example, figure 4 in \citealt{vandishoeck86}). \citet{nelson97} suggested a simplified CO model in which C$^{+}$ first forms CH$_{2}^{+}$, which is converted rapidly to CH and CH$_{2}$. These can then react with O to form CO, or they can be photodissociated. \citet{nelson97} do not explicitly follow these intermediate species, instead treating the whole formation channel via CH and CH$_{2}$ together, according to the relevant rate coefficients (see their equations $18-21$). However, we find that the reactions that form CO from CH or CH$_{2}$ contribute $<0.1$ per cent to the total formation rate in our reference simulation. Therefore, the simplified CO model of \citet{nelson97} cannot be applied to the physical regimes that we study here. 

\section{Model variations}\label{model_vars_sect} 

In this section we consider the six model variations runs, which we compare to our standard model implementation. These runs all use the fiducial setup, with an initial ambient ISM density of $10 \, \rm{cm}^{-3}$, an AGN luminosity of $10^{46} \, \rm{erg} \, \rm{s}^{-1}$, and solar metallicity. We then vary different aspects of the modelling that are uncertain, to quantify how these uncertainties impact our results. These runs all used a resolution of $240 \, \rm{M}_{\odot}$ per gas particle, which is a factor 8 lower mass resolution than was used in the main parameter variations runs in the previous section. We present a comparison of the different resolutions in Appendix~\ref{resolution_appendix}. 

\begin{figure}
\centering
\mbox{
	\includegraphics[width=84mm]{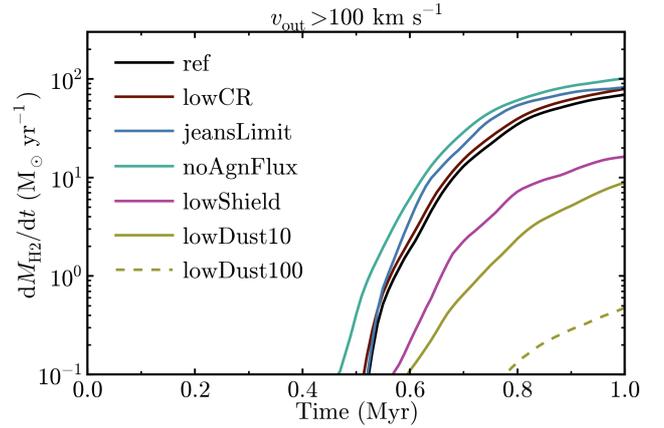}}
\caption{As Fig.~\ref{outflowRateFig}, but for the six model variations runs (lowCR, red curve; jeansLimit, blue curve; noAGNflux, green curve; lowShield, magenta curve; lowDust10, yellow solid curve; lowDust100, yellow dashed curve), and the reference model (ref, i.e. nH10\_L46\_Z1\_lowRes in Table~\ref{sims_table}; black curve). These runs are at a resolution of $240 \, \rm{M}_{\odot}$ per gas particle. The largest effect on the H$_{2}$ outflow rate is seen when we reduce the shielding column density by a factor of 10 (lowShield), or we reduce the dust abundance by a factor of 10 (lowDust10) or 100 (lowDust100).}
\label{modelVarsFig}
\end{figure}

Fig.~\ref{modelVarsFig} shows the mass outflow rate of H$_{2}$ in gas outflowing with $v_{\rm{out}} > 100 \, \rm{km} \, \rm{s}^{-1}$ in the reference model (ref, i.e. nH10\_L46\_Z1\_lowRes in Table~\ref{sims_table}; black curve) and in the six model variations runs. As in Fig.~\ref{outflowRateFig}, the H$_{2}$ mass outflow rate is calculated in the high-resolution wedge and then multiplied by 17 to get the outflow rate in a full sphere, using equation~\ref{outflow_eqn}. 

The H$_{2}$ outflow rate in the reference model reaches $69 \, \rm{M}_{\odot} \, \rm{yr}^{-1}$ after $1 \, \rm{Myr}$. This is a factor two lower than we found at higher resolution in run nH10\_L46\_Z1 (Fig.~\ref{outflowRateFig}; see also Appendix~\ref{resolution_appendix}). In the lowCR run (red curve), we use a cosmic ray ionization rate of $\zeta_{\rm{HI}} = 1.8 \times 10^{-16} \, \rm{s}^{-1}$, as measured in the Milky Way \citep{indriolo12} and a factor $\approx 30$ lower than the fiducial value we use in the reference model for ULIRGs. This run produces a similar H$_{2}$ outflow rate as the reference model, reaching $79 \, \rm{M}_{\odot} \, \rm{yr}^{-1}$ after $1 \, \rm{Myr}$. This suggests that uncertainties in the cosmic ray ionization rate do not have a strong impact on the H$_{2}$ outflow rate. 

In the jeansLimit run (blue curve), we include a pressure floor such that the Jeans length is always resolved by at least 3.2 smoothing lengths. The reference model does not include this Jeans limiter, and so it might be susceptible to artificial fragmentation \citep[e.g.][]{bate97, truelove97}. In Fig.~\ref{modelVarsFig}, we see that the H$_{2}$ outflow rate in the jeansLimit run is also close to the reference model, reaching $82 \, \rm{M}_{\odot} \, \rm{yr}^{-1}$ after $1 \, \rm{Myr}$. Therefore, if artificial fragmentation is occuring in the reference model, it does not significantly affect the H$_{2}$ outflow rate. 

To test how the AGN radiation affects the formation and destruction of molecules in the AGN wind, we performed one run without the AGN radiation (noAGNflux; green curve). Unsurprisingly, the H$_{2}$ outflow rate is higher in this run, reaching $100 \, \rm{M}_{\odot} \, \rm{yr}^{-1}$ after $1 \, \rm{Myr}$. 

In the reference model, we calculate the shielding column density using a local approximation based on the density gradient (see Section~\ref{chemistry_sect}). In Appendix~\ref{shielding_appendix} we show that, while this local approximation on average reproduces the true column density (computed via ray tracing in post-processing), there is a large scatter between the two of an order of magnitude. To test how these uncertainties might affect our results, we performed one run with column densities reduced by a factor of 10 (lowShield; magenta curve). The H$_{2}$ outflow rate is noticeably lower in this run, reaching just $16 \, \rm{M}_{\odot} \, \rm{yr}^{-1}$ after $1 \, \rm{Myr}$, a factor of 4.3 less than in the reference model. This highlights that uncertainties in the shielding column density can have a large impact on molecule formation, although the lowShield run is a worst-case scenario, as for most particles the local shielding approximation is closer to (or even less than) the true value (see Fig.~\ref{shieldFig}). 

In the reference model we assume that the dust abundance scales linearly with metallicity. However, it is not clear whether dust grains can survive the strong shocks and high gas temperatures found in AGN winds. To investigate how sensitive our results are to uncertainties in the dust abundance, we performed two runs at solar metallicity but with dust abundances reduced by a factor of 10 and 100 (lowDust10 and lowDust100, respectively; solid and dashed yellow curves in Fig.~\ref{modelVarsFig}). We see that the dust abundance has the largest impact of all the model variations runs, reducing the H$_{2}$ outflow rate after $1 \, \rm{Myr}$ by a factor of $\approx 8$ and $\approx 150$ when the dust abundance is lowered by a factor of 10 and 100, respectively. Therefore, to form a significant amount of H$_{2}$ in the AGN wind that can explain the observed fast molecular outflows, the swept up material requires a dust-to-metals ratio that is close to the Milky Way value, or higher. 

\citet{ferrara16} performed a series of hydrodynamic simulations of the swept up ISM region of an AGN outflow. They followed the evolution of dust grains in their simulations, including destruction by sputtering in hot gas and accounting for the initial destruction of dust grains by the forward shock, both modelled based on the results of \citet{dwek96}. \citet{ferrara16} found that dust grains are rapidly destroyed, within $\sim 10^{4} \, \rm{yr}$, in the swept up ISM. We can estimate the sputtering time-scale in our fiducial model (nH10\_L46\_Z1) using equation 14 in \citet{hirashita15} \citep[see also][]{nozawa06}: 

\begin{equation} 
\tau_{\rm{sput}} = 7.1 \times 10^{5} \left( \frac{a}{1 \, \mu\rm{m}} \right) \left(\frac{n_{\rm{H}}}{1 \, \rm{cm}^{-3}} \right)^{-1} \, \rm{yr}, 
\end{equation} 
where $a$ is the grain radius. This equation is valid at gas temperatures $\ga 10^{6} \, \rm{K}$, but at lower temperatures sputtering is inefficient. The post-shock density in the swept up ISM is 4 times the ambient density, i.e. $40 \, \rm{cm}^{-3}$. So, for $a = 0.1 \, \mu\rm{m}$ \citep[as assumed by][]{ferrara16}, $\tau_{\rm{sput}} = 1800 \, \rm{yr}$. In our fiducial model, it takes $0.5 \, \rm{Myr}$ for the swept up ISM to cool and start forming molecules (see Fig.~\ref{molecularMassFig}). This suggests that dust grains in the swept up ISM will be destroyed by sputtering before that gas can cool. 

However, \citet{ferrara16} only included dust destruction, and they did not consider any dust formation mechanisms. For example, \citet{hirashita17} highlighted the importance of dust growth via accretion of metals from the gas phase onto grains in their models of dust evolution in elliptical galaxies. We can estimate the accretion time-scale using equation 20 of \citet{asano13}: 

\begin{align} 
\tau_{\rm{acc}} = 2.0 \times 10^{7} & \left( \frac{a}{0.1 \, \mu\rm{m}} \right) \left( \frac{n_{\rm{H}}}{100 \, \rm{cm}^{-3}} \right)^{-1} \nonumber \\  
 & \times \left( \frac{T}{50 \, \rm{K}} \right)^{-1/2} \left( \frac{Z}{0.02} \right)^{-1} \, \rm{yr}, 
\end{align} 
where the metallicity $Z = Z_{\odot} = 0.0129$ in our fiducial model. From the top left panel of Fig.~\ref{TrhoFig}, we see that the swept up ISM covers a large range of densities and temperatures. As an example, consider $n_{\rm{H}} = 10^{4} \, \rm{cm}^{-3}$, $T = 1000 \, \rm{K}$. Then the accretion time-scale for $a = 0.1 \, \mu\rm{m}$ grains is $\tau_{\rm{acc}} = 0.07 \, \rm{Myr}$. Thus it is feasible for dust grains to re-form after the swept up ISM has cooled and before the end of the simulation at $1 \, \rm{Myr}$. 

Dust growth from ISM accretion requires seed grains to be present. There are several possible sources of such seed dust in the post-shock cooling layers of AGN-driven galactic winds. For example, it is observed that star formation-driven galactic winds carry large dust masses \citep[e.g.][]{hoopes05, roussel10, hutton14, melendez15, hodgeskluck16}. As the AGN wind interacts with star formation-driven outflows, some dust is likely to mix. Furthermore, while the main $\sim 1000 \, \rm{km} \, \rm{s}^{-1}$ shock fronts in AGN-driven outflows can easily destroy dust, shocks driven in ambient over-densities will have lower velocities \citep{klein94}. Such over-densities can therefore protect dust from destruction by shocks. The dust protected by over-densities can subsequently mix with the cooling layers of AGN-driven outflows.

Additionally, \citet{boschman15} explored another formation channel for H$_{2}$ on polycyclic aromatic hydrocarbons (PAHs). They showed that this mechanism for H$_{2}$ formation can have an important impact on the structure of PDRs at high temperatures ($T > 200 \, \rm{K}$), which is particularly relevant for our AGN wind models, where we find a lot of dense gas at such temperatures. H$_{2}$ formation on PAHs is not included in our chemical model. 

To conclusively determine whether sufficient dust can survive or re-form to produce the observed fast molecular outflows, we will need to model the evolution of dust grains, including destruction from shocks and sputtering and formation from accretion. The resulting dust abundances, and the resulting grain size distributions (as $\tau_{\rm{sput}}$ and $\tau_{\rm{acc}}$ both depend on the grain radius), along with a treatment for the PAH chemistry, will then need to be coupled to H$_{2}$ formation and shielding in the chemical model. 

\section{Observable molecular lines}\label{molecular_lines_sect} 

Our chemical model includes several molecular species that are commonly seen in observations, such as CO and OH. We can therefore use our simulations to make predictions for observable emission and absorption lines from these molecules, which we can compare to observations. We model the molecular lines using radiative transfer calculations performed on snapshots from our simulations in post-processing, using the publicly available Monte Carlo radiative transfer code \textsc{radmc-3d}\footnote{\url{http://www.ita.uni-heidelberg.de/~dullemond/software/radmc-3d/}}, version 0.40 \citep{dullemond12}. \textsc{Radmc-3d} follows the line radiative transfer, along with thermal dust emission and absorption. We include anisotropic scattering by dust grains, and we include the radiation from the AGN, using the average quasar spectrum from \citet{sazonov04}. 

\textsc{Radmc-3d} includes several approximations to calculate the level populations of a given molecular species. The simplest approach is to assume Local Thermodynamic Equilibrium (LTE). However, \citet{duartecabral15} showed that assuming LTE can overestimate CO emission in their hydrodynamic simulations of spiral galaxies. We therefore use an approximate non-LTE treatment using the Local Velocity Gradient (LVG) method \citep{sobolev57}. See \citet{shetty11} for a full description of the LVG method implemented in \textsc{radmc-3d}. 

\textsc{Radmc-3d} operates on a grid, rather than on gas particles, so we first need to project the gas particles from the simulation snapshot onto a 3D grid. We construct an Adaptive Mesh Refinement (AMR) grid, with a base grid of $16^{3}$ cells over the high-resolution octant. We then refine any cells that contain more than 8 particles by splitting the cell into 8 smaller cells, and we continue refining the grid in this way until all cells contain no more than 8 particles. This requires $8-10$ levels of refinement, which results in a minimum cell size of $0.07-0.3 \, \rm{pc}$, depending on the simulation. 

Once we have constructed the AMR grid, we then interpolate the particle densities, temperatures and velocities onto the grid. We do this using a cubic spline kernel with a smoothing length that encloses 32 neighbouring particles, as used in the simulations. We include two species of dust grains, graphite and silicate, using opacities from \citet{martin90} (based on the grain size distribution of \citealt{mathis77}). We use dust-to-gas ratios of $2.4 \times 10^{-3} \, Z / \rm{Z}_{\odot}$ and $4.0 \times 10^{-3} \, Z / \rm{Z}_{\odot}$ for graphite and silicate respectively, which we assume scale linearly with metallciity, $Z$. At solar metallicity, these are taken from \textsc{cloudy} version 13.01 \citep{ferland13}, using the `ISM' grain abundances. In cells with a gas temperature $T > 10^{5} \, \rm{K}$, we set the dust abundance to zero. This temperature is chosen for consistency with the chemical model, as we switch off the molecular chemistry and dust processes above this temperature. At high gas temperatures, we expect dust grains to be rapidly destroyed by sputtering. 

We also include a microturbulent Doppler broadening parameter $b_{\rm{turb}} = 7.1 \, \rm{km} \, \rm{s}^{-1}$ (in addition to thermal broadening), for consistency with the H$_{2}$ self-shielding function used in the simulations. However, we will see that the resolved gas motions produce lines that are much broader than this. 

There is a possible inconsistency between the chemistry, which is calculated on the particles, and the radiative transfer, which is calculated on the grid cells. For example, suppose there are two particles that contribute equally to a given cell. One of these particles is cold ($100 \, \rm{K}$) with a CO fraction of 1, while the other is hot ($10^{7} \, \rm{K}$) with a CO fraction of 0. If we weight both particles equally, the cell will have a temperature $\approx 5 \times 10^{6} \, \rm{K}$ and a CO fraction of 0.5, i.e. half the CO will be unrealistically hot. However, from the point of view of the chemistry solver, all of the CO was cold. To avoid such inconsistencies, when interpolating particle temperatures and velocities onto the grid we weight each particle by its mass of the molecular species (CO in the above example), rather than the total particle mass. 

Finally, we mirror the high-resolution octant in the line of sight direction, so that we include both the approaching (blueshifted) and receding (redshifted) sides of the outflow. We then run \textsc{radmc-3d} to produce maps of CO (Section~\ref{CO_lines_sect}), OH (Section~\ref{OH_lines_sect}), and IR-traced H$_{2}$ (Section~\ref{H2_lines_sect}) emission and absorption. These maps contain 1024 $\times$ 1024 spatial pixels covering the high-resolution octant, and 2000 equally spaced velocity bins extending from $-1000$ to $+1000 \, \rm{km} \, \rm{s}^{-1}$. For each line, we also create a map of the continuum emission only, which we then subtract from the full line emission map. 

\begin{figure*}
\centering
\mbox{
	\includegraphics[width=168mm]{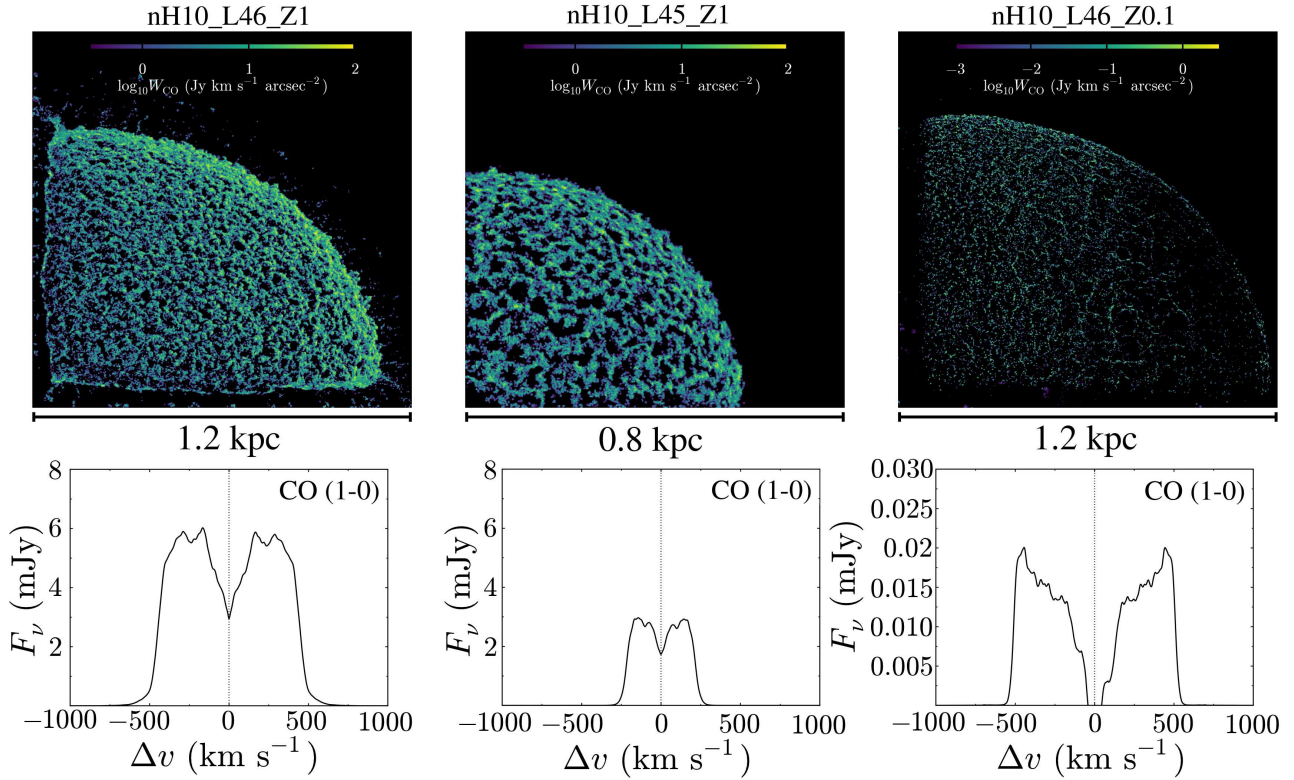}}
\caption{Velocity-integrated maps of CO (1$-$0) emission (top row) and CO (1$-$0) spectra (bottom row) from runs nH10\_L46\_Z1 (left), nH10\_L45\_Z1 (centre), and nH10\_L46\_Z0.1 (right) after $1 \, \rm{Myr}$. These were calculated in the high-resolution octant, which was mirrored in the line of sight direction to include both the receding and approaching sides of the outflow. The spectra have been normalised to a distance of $184 \, \rm{Mpc}$, which corresponds to the distance of Mrk 231. Note that the dips in the spectra at $\Delta v = 0$ are the result of artifacts along the boundary between the high- and low-resolution regions, and are not physical. These do not affect our comparisons with observed CO luminosities, for which we only integrate over the broad wings of the line, and also likely do not significantly affect our estimates of the CO to H$_{2}$ conversion factors (see text).} 
\label{coEmissionFig}
\end{figure*}

\subsection{CO emission}\label{CO_lines_sect} 

We calculate the emission from the three lowest rotational transitions of CO, J=1$-$0, 2$-$1, and 3$-$2, with rest frame wavelengths of $2.6 \, \rm{mm}$, $1.3 \, \rm{mm}$ and $0.9 \, \rm{mm}$, respectively. These lines have been observed at high velocities (up to $\sim 1000 \, \rm{km} \, \rm{s}^{-1}$) in several AGN host galaxies \citep[e.g.][]{feruglio10, feruglio15, cicone12, cicone14}. We use molecular data for CO from the \textsc{lamda} database\footnote{\url{http://home.strw.leidenuniv.nl/~moldata/}} \citep{schoier05}. To calculate the non-LTE level populations of CO, we include collisions with ortho- and para-H$_{2}$, using collisional rate coefficients from \citet{yang10} and assuming an ortho-to-para ratio of 3:1. The collisional rate coefficients are only tabulated up to $3000 \, \rm{K}$, so we assume that they are constant above this temperature. We will primarily focus on the (1$-$0) line, but we will use the higher transitions to explore the excitation of CO. 

The top row of Fig.~\ref{coEmissionFig} shows velocity-integrated maps of the continuum-subtracted CO (1$-$0) line emission in nH10\_L46\_Z1 (left), nH10\_L45\_Z1 (centre), and nH10\_L46\_Z0.1 (right), from the final snapshot after $1 \, \rm{Myr}$, while the bottom row shows the continuum-subtracted CO(1$-$0) spectra from these three runs. These spectra only show the flux from the high-resolution octant (which has also been mirrored in the line of sight direction, i.e. to create one quadrant), in other words they have not been multiplied by 4 to get the spectrum from the full spherical outflow. The fluxes have been normalised to a distance of $184 \, \rm{Mpc}$, which corresponds to the distance of Mrk 231. We do not show the low-density run (nH1\_L46\_Z1), as it did not form any molecules. Note that the colour scale of the velocity-integrated map and the y-axis scale of the spectrum for the low-metallicity run (right-hand column) is lower than for the other two runs. 

In the velocity-integrated maps, we see that the CO emission is not smooth over the shell of swept up ISM, but rather is clumpy. This clumpiness was also seen in the 2D and 3D views of the gas density in Figs.~\ref{gasFig} and \ref{gasFig3D} respectively, and is triggered by the radiative cooling in the swept up material, as we see no clumpiness in the low-density run in Fig.~\ref{gasFig}, which does not cool. In the low-metallicity run (right hand column of Fig.~\ref{coEmissionFig}), the CO emission appears to be concentrated into much smaller clumps than in the other two runs. This is likely due to the lower dust abundance in this run, which results in weaker dust shielding. Since CO needs to be shielded from UV, it will therefore only trace the centres of dense clumps when the dust shielding is weaker. 

All three spectra in the bottom row of Fig.~\ref{coEmissionFig} show a significant dip at the systemic velocity ($\Delta v = 0$). However, this does not appear to be physical, and is likely due to the artifacts along the boundary between high- and low-resolution regions that we saw in Fig.~\ref{gasFig}. Since the CO emission is in a spherical, expanding shell, emission at $\Delta v = 0$ will come from the part of the shell moving perpendicular to the line of sight direction, i.e. along the boundary with the low-resolution region. However, we saw in Fig.~\ref{gasFig} that the spherical shell is disrupted near these boundaries. This is most evident in the top left and top right panels of Fig.~\ref{coEmissionFig}, where we see that the CO emission does not extend all the way to the left-hand and bottom edges of the maps. By symmetry, CO emission will also be missing along the third high resolution-low resolution boundary (perpendicular to the line of sight). This would then explain the dip in the CO spectra. We will therefore need to be careful that these artifacts do not affect our analysis below. 

In the two runs with an AGN luminosity of $10^{46} \, \rm{erg} \, \rm{s}^{-1}$, the CO spectra extend to $\pm 500 \, \rm{km} \, \rm{s}^{-1}$, while the low-luminosity run has a narrower CO spectrum, extending to $\pm 250 \, \rm{km} \, \rm{s}^{-1}$. Run nH10\_L46\_Z1 has the brightest CO flux, followed by nH10\_L45\_Z1 and then nH10\_L46\_Z0.1. This follows the trend in outflowing CO mass between these runs that we saw in Fig.~\ref{molecularMassFig}. Note that, unlike typical observations of AGN host galaxies, all of the CO emission in our simulations is from the outflow, and there is no pre-existing molecular disc in the host galaxy, which would appear as an additional narrow component in the CO spectrum \citep[e.g.][]{feruglio10, cicone14}. 

We can now use the CO spectra from our simulations to estimate the H$_{2}$ outflow rate in the same way as in the observations. \citet{cicone14} presented CO (1$-$0) observations of 7 ULIRGs and quasar hosts, of which 4 showed detections of fast molecular outflows. They also compared these with a further 12 sources from the literature. Not all of these galaxies are luminous AGN, so we only compare our simulations to the Seyfert 1 and 2 galaxies in the extended sample of \citet{cicone14}. This gives us a sample of 10 AGN host galaxies, of which 6 were observed by \citet{cicone14} and a further 4 were taken from the literature \citep{wiklind95, maiolino97, cicone12, feruglio13a, feruglio13b}. 

\citet{cicone14} measured the CO luminosity in the broad wings of the continuum-subtracted CO line. The velocity range that they integrated over depends on the source, but was typically from $\approx 300 \, \rm{km} \, \rm{s}^{-1}$ to the maximum velocity that the wing could be detected at (see their table 2). To match the method used for the observations, we integrate our simulated CO spectra in each wing from $300 \, \rm{km} \, \rm{s}^{-1}$ to a maximum $v_{\rm{max}}$, which we take to be the velocity at which the flux density falls to 1 per cent of its maximum value. This definition might not be equivalent to the $v_{\rm{max}}$ used in the observations, as we do not model the noise or detectability of CO in the simulated spectra, but it does not have a significant impact on the CO luminosity. For the low-luminosity run, we use a lower minimum velocity of $200 \, \rm{km} \, \rm{s}^{-1}$, as $v_{\rm{max}} < 300 \, \rm{km} \, \rm{s}^{-1}$. By restricting the CO luminosity to the wings of the line, we also avoid the artificial dip at $\Delta v = 0$. 

To obtain the H$_{2}$ mass from the CO luminosity, \citet{cicone14} assumed a conversion factor $\alpha_{\rm{CO} \, (1-0)} = M_{\rm{H_{2}}} / L_{\rm{CO} \, (1-0)} = 0.8 \, \rm{M}_{\odot} \, (\rm{K} \, \rm{km} \, \rm{s}^{-1} \, \rm{pc}^{2})^{-1}$. This value, which is $\approx 5$ times lower than the Milky Way value, is commonly used for ULIRGs. \citet{cicone12} showed that the molecular gas in Mrk 231 has similar excitation in the host galaxy and the outflow, and so the outflow plausibly has the same conversion factor as its ULIRG host. To compare with observations, we use the same conversion factor, although we will investigate this assumption further below. 

\citet{cicone14} then calculated the molecular mass outflow rate by assuming that the outflowing material is uniformly distributed over a radius $R$, measured from the CO maps. This gives an outflow rate of $\dot{M}_{\rm{H_{2}}} = 3 v M_{\rm{H_{2}}} / R$, where the velocity $v$ is the average velocity in the range used to integrate the wings of the CO line. In our simulations, the outflowing molecular gas is instead found in a shell, for which the outflow rate is $\dot{M}_{\rm{H_{2}}} = v M_{\rm{H_{2}}} / R$. We therefore divide the rates given in \citet{cicone14} by 3, for consistency with the simulations, although we note that this may underestimate the true outflow rate if the geometry of the observed systems is that of a uniform outflow rather than a shell. 

\begin{figure}
\centering
\mbox{
	\includegraphics[width=84mm]{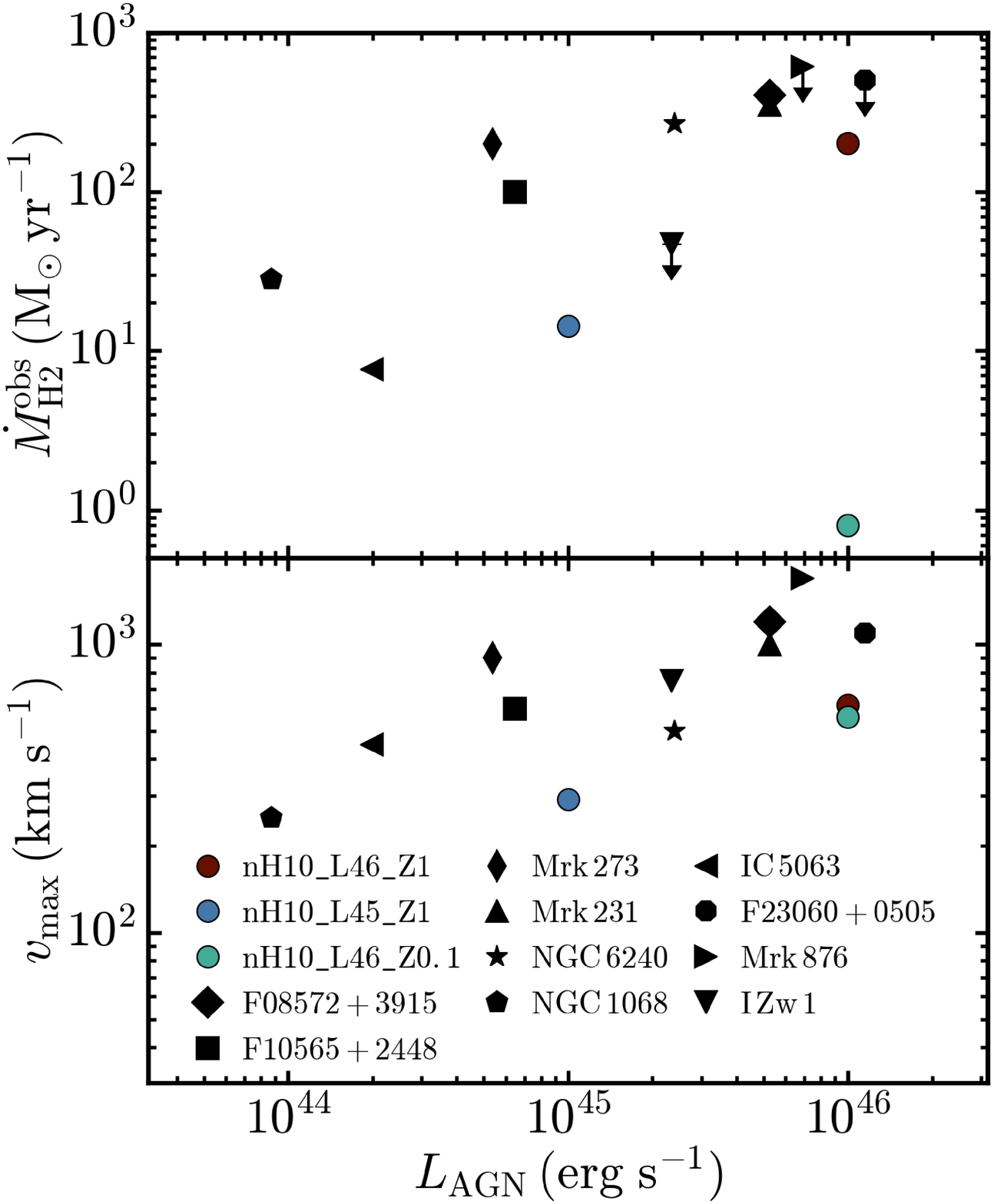}}
\caption{H$_{2}$ outflow rate (top panel) and maximum line of sight velocity (bottom panel), estimated from CO (1$-$0) spectra, plotted versus AGN luminosity. Coloured circles show the simulations: nH10\_L46\_Z1 (red), nH10\_L45\_Z1 (blue), and nH10\_L46\_Z0.1 (green). Black symbols show observed AGN host galaxies: IRAS F08572$+$3915, IRAS F10565$+$2448, IRAS F23060$+$0505, Mrk 273, Mrk 876, I Zw 1 \citep{cicone14}, IC 5063 \citep{wiklind95}, NGC 1068 \citep{maiolino97}, Mrk 231 \citep{cicone12}, and NGC 6240 \citep{feruglio13a, feruglio13b}. The H$_{2}$ outflow rates estimated from the simulated CO spectra are slightly lower than the observations, although they are still reasonably close, while the simulated line of sight velocities are also lower than in the observations, by a factor $\approx 2$.} 
\label{coObsFig}
\end{figure}

The top panel of Fig.~\ref{coObsFig} shows the molecular outflow rates derived from our simulated CO spectra, from run nH10\_L46\_Z1 (red circle), nH10\_L45\_Z1 (blue circle), and nH10\_L46\_Z0.1 (green circle), plotted versus AGN luminosity. The outflow rates from the high-resolution quadrant have been multiplied by 4 to give the outflow rate in the full spherical outflow. The black symbols show the observed sources from \citet{cicone14} and references therein. The molecular outflow rates in the two simulations at solar metallicity are slightly lower than the observations, although they are still reasonably close given the uncertainties in the geometry. The low-metallicity run is significantly below the observations, although this is unsurprising as these host galaxies are typically ULIRGs, with metallicities close to, or greater than, solar \citep[e.g.][]{rupke08, kilercieser14}. The molecular outflow rate increases with increasing AGN luminosity, as found by \citet{cicone14}. 

In the bottom panel of Fig.~\ref{coObsFig}, we show the maximum line of sight velocity of the CO wings versus AGN luminosity. In the simulations, this is defined as the velocity where the CO flux density falls to 1 per cent of its maximum value. These are generally lower than is seen in the observations, by a factor $\approx 2$. The definition of $v_{\rm{max}}$ in the simulations is not exactly equivalent to the observational definition, as we do not model noise or the detectability of CO in the simulations, but we can see that the simulated spectra in Fig.~\ref{coEmissionFig}  do not reach velocities up to $\sim 1000 \, \rm{km} \, \rm{s}^{-1}$ as in the observations. This may be due to the simplified geometry and the assumption of a uniform ambient ISM in the simulations. For example, if there are inhomogeneities in the ISM, there will be a range of outflow velocities, and the maximum velocity of the outflow will be determined by the lowest density channels, through which the outflow can escape more quickly. The velocity of the outflow increases with increasing AGN luminosity in the simulations and the observations. 

For the comparisons in Fig.~\ref{coObsFig}, we assumed a conversion factor between CO and H$_{2}$ of $\alpha_{\rm{CO} \, (1-0)} = 0.8 \, \rm{M}_{\odot} \, (\rm{K} \, \rm{km} \, \rm{s}^{-1} \, \rm{pc}^{2})^{-1}$, as used in the observations. However, we can also use our simulations to calculate this conversion factor, as well as conversion factors for the (2$-$1) and (3$-$2) lines. We first calculate the total CO luminosity in the high-resolution quadrant by integrating over the full CO line (not just the wings). We then calculate the total H$_{2}$ mass in the high-resolution quadrant (i.e. the high-resolution octant and mirrored in the line of sight direction). Since we integrate over the full CO line, the total CO luminosity will be affected by the artificial dip at $\Delta v = 0$. We therefore use the full high-resolution region to calculate the H$_{2}$ mass, rather than using only the high-resolution wedge and then scaling up. This way, the H$_{2}$ mass will also be affected by the artificial reduction in molecular gas along the boundaries with the low-resolution region. We find that, in the fiducial run, the masses of H$_{2}$ and CO in the high-resolution octant are 25 per cent lower than using the high-resolution wedge and scaling up (due to these artifacts), but the ratio of H$_{2}$ to CO mass changes by only 2 per cent. We therefore expect that the ratio of H$_{2}$ mass to CO luminosity (i.e. $\alpha_{\rm{CO}}$) calculated in this way should be unaffected by these artifacts. 

\begin{table}
\begin{minipage}{84mm}
\centering
\caption{Conversion factors from CO luminosity of the (1$-$0), (2$-$1) and (3$-$2) lines to H$_{2}$ mass after $1 \, \rm{Myr}$ in the three parameter variations runs that form molecules. For comparison, observations typically assume $\alpha_{\rm{CO} \, (1-0)} = 0.8 \, \rm{M}_{\odot} \, (\rm{K} \, \rm{km} \, \rm{s}^{-1} \, \rm{pc}^{2})^{-1}$ when inferring H$_{2}$ masses from CO (1$-$0) emission.}
\label{alphaCO_table}
\begin{tabular}{lccc}
\hline 
Simulation & \multicolumn{3}{c}{$\alpha_{\rm{CO}} = M_{\rm{H_{2}}} / L_{\rm{CO}}$\footnote{Units of $\alpha_{\rm{CO}}$ are $\rm{M}_{\odot} \, (\rm{K} \, \rm{km} \, \rm{s}^{-1} \, \rm{pc}^{2})^{-1}$.}} \\ 
 & (1$-$0) & (2$-$1) & (3$-$2) \\ 
\hline
nH10\_L46\_Z1 & $0.13$ & $0.08$ & $0.06$ \\ 
nH10\_L45\_Z1 & $0.15$ & $0.09$ & $0.07$ \\
nH10\_L46\_Z0.1 & $1.77$ & $0.82$ & $0.80$ \\ 
\hline
\end{tabular}
\vspace{-0.27in}
\end{minipage}
\end{table}

Table~\ref{alphaCO_table} shows the $\alpha_{\rm{CO}}$ conversion factors in our three simulations after $1 \, \rm{Myr}$, for the lines (1$-$0), (2$-$1) and (3$-$2). The value of $\alpha_{\rm{CO}}$ depends on the time at which we calculate it, but we quote values at $1 \, \rm{Myr}$ here as this corresponds to typical flow times, $r / v_{\rm{out}}$, of observed AGN-driven molecular outflows \citep[e.g.][]{gonzalezalfonso17}. In the fiducial and low-luminosity runs, at solar metallicity, we find $\alpha_{\rm{CO} \, (1-0)} = 0.13$ and $0.15 \, \rm{M}_{\odot} \, (\rm{K} \, \rm{km} \, \rm{s}^{-1} \, \rm{pc}^{2})^{-1}$, respectively, almost independent of the AGN luminosity. This is a factor of $5 - 6$ lower than is typically assumed in observations \citep[e.g.][]{cicone14}, which would result in a factor of $5 - 6$ lower H$_{2}$ outflow rates. For example, the H$_{2}$ outflow rates from \citet{cicone14} would then all be below 120 $\rm{M}_{\odot} \, \rm{yr}^{-1}$. 

At $0.1 \, \rm{Z}_{\odot}$, $\alpha_{\rm{CO} \, (1-0)}$ is more than a factor of 10 higher than in the other two runs. This is unsurprising, as there is less carbon and oxygen relative to hydrogen, and is also reflected in the lower CO abundances at low metallicity (see Fig.~\ref{molecularMassFig}). However, we caution that the $\alpha_{\rm{CO}}$ factors in the low-metallicity run are poorly converged with numerical resolution (see Appendix~\ref{resolution_appendix}). The metallicity dependence of the CO to H$_{2}$ conversion factor has also been noted in previous studies of interstellar gas across a range of galaxy properties \citep[e.g.][]{israel97, leroy11, feldmann12, genzel12, bolatto13a}. However, as the quasar host galaxies with observed fast molecular outflows are typically ULIRGs, with metallicities close to, or greater than, solar \citep[e.g.][]{rupke08, kilercieser14}, our solar metallicity estimate for $\alpha_{\rm{CO} \, (1-0)}$ is more relevant for these systems. 

The conversion factors for the (2$-$1) and (3$-$2) CO lines are a factor $\approx 2$ lower than for the (1$-$0) line in all three simulations. For comparison, \citet{cicone14} assume that the CO lines are thermalised and optically thick, such that the CO line luminosities, and hence the $\alpha_{\rm{CO}}$ conversion factors, of different rotational transitions are equal. 

In typical molecular clouds, the CO 1$-$0 line is optically thick. However, this line can become optically thin, for example, in a highly turbulent medium. In the optically thin regime, the $\alpha_{\rm{CO}}$ conversion factor reaches a lower limit. \citet{bolatto13a} showed that, for a CO to H$_{2}$ abundance $Z_{\rm{CO}} = n_{\rm{CO}} / n_{\rm{H}_{2}} = 10^{-4}$ and a CO excitation temperature $30 \, \rm{K}$, the optically thin limit is $\alpha_{\rm{CO}} = 0.34 \, \rm{M}_{\odot} \, (\rm{K} \, \rm{km} \, \rm{s}^{-1} \, \rm{pc}^{2})^{-1}$. While this appears to be inconsistent with the values that we find from our simulations, which are less than this at solar metallicity, we note that, for the solar abundances used in our simulations, $Z_{\rm{CO}} = 5 \times 10^{-4}$ when all carbon is in CO and all hydrogen is in H$_{2}$. This reduces the optically thin limit of $\alpha_{\rm{CO}}$ by a factor of 5, and so the values from the simulations are no longer inconsistent with this limit. Moreover, in Fig.~\ref{TrhoFig} we saw that CO is found across a wide range of temperatures ($\sim 10 - 10^{3} \, \rm{K}$) in our simulations, while the above optically thin limit assumes a single excitation temperature of 30 K and LTE. 

\citet{dasyra16} measured CO (4$-$3)/(2$-$1) flux ratios of 5 to 11 in the jet-driven winds in the galaxy IC 5063, which indicates that the CO gas in this outflow is highly excited and optically thin. Based on these flux ratios, they inferred optically thin $\alpha_{\rm{CO}}$ conversion factors of $0.27 - 0.44 \, \rm{M}_{\odot} \, (\rm{K} \, \rm{km} \, \rm{s}^{-1} \, \rm{pc}^{2})^{-1}$, which are reasonably close (within a factor $\approx 2 - 3$) to our simulations at solar metallicity. 

\citet{weiss01} demonstrated that the CO to H$_{2}$ conversion factor measured in different regions of the starburst galaxy M82 depends on the excitation of the molecular gas, with the conversion factor scaling with kinetic temperature, $T_{\rm{kin}}$, and H$_{2}$ density, $n_{\rm{H_{2}}}$, as $T_{\rm{kin}}^{-1} n_{\rm{H_{2}}}^{1/2}$. This scaling is also predicted by theoretical models of virialised clouds \citep[e.g.][]{maloney88}, although it is not clear whether the molecular clouds in an AGN wind will also be virialised. If they are not virialised, this will affect their line widths, which will affect the CO (1$-$0) luminosity (since this is typically an optically thick line), and hence the $\alpha_{\rm{CO}}$ factor, for a given $T_{\rm{kin}}$. \citet{cicone12} showed that the (2$-$1)/(1$-$0) CO line ratio in the core and the broad components in Mrk 231 are similar, and thus they suggested that the same conversion factor should be used for the outflow and the host galaxy. It is therefore instructive to compare the excitation of molecular gas in the simulations to these observations. 

\begin{figure}
\centering
\mbox{
	\includegraphics[width=84mm]{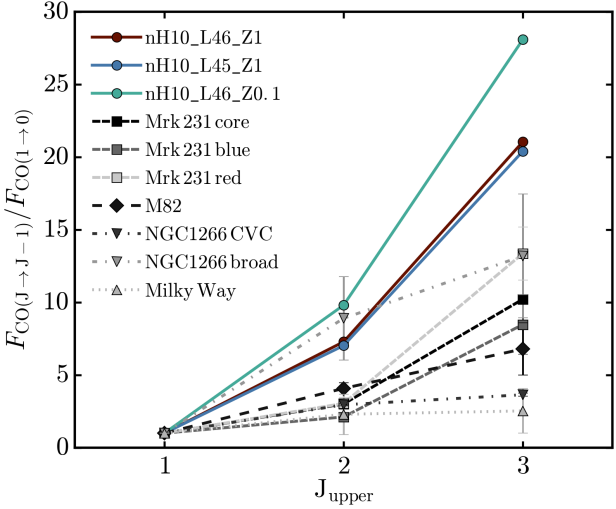}}
\caption{CO line ratios, relative to the (1$-$0) line, in the simulations nH10\_L46\_Z1 (red curve), nH10\_L45\_Z1 (blue curve), and nH10\_L46\_Z0.1 (green curve). Black and grey curves show observations of Mrk 231 \citep[from the core and the red- and blueshifted components;][]{cicone12, feruglio15}, the starburst galaxy M82 \citep{weiss05}, the S0 galaxy NGC 1266 with an AGN-driven outflow \citep[from the central velocity component and the broad component;][]{glenn15}, and the inner disc region of the Milky Way, with galactic longitude $2.5^{\circ} < |l| < 32.5^{\circ}$ \citep{fixsen99}. The line ratios use fluxes in units of $\rm{Jy} \, \rm{km} \, \rm{s}^{-1}$. The simulations generally show higher CO excitation than the observed systems, although the broad component of NGC 1266 has a similar (2$-$1)/(1$-$0) ratio.} 
\label{coRatiosFig}
\end{figure}

In Fig.~\ref{coRatiosFig} we show the CO line ratios for the (2$-$1) and (3$-$2) transitions, relative to the (1$-$0) line, from our three simulations after $1 \, \rm{Myr}$ (solid, coloured curves). We compare these to observed line ratios (black and grey curves) from Mrk 231 \citep{cicone12, feruglio15}, the starburst galaxy M82 \citep{weiss05}, the S0 galaxy NGC1266 with an AGN-driven molecular outflow \citep{glenn15}, and the inner disc of the Milky Way, with galactic longitude $2.5^{\circ} < |l| < 32.5^{\circ}$ \citep{fixsen99}. These line ratios are calculated with fluxes in units of $\rm{Jy} \, \rm{km} \, \rm{s}^{-1}$, for consistency with fig. 5 of \citet{cicone12}. Note that the value of the line ratios will depend on the units used, because converting between commonly used units (e.g. $\rm{Jy} \, \rm{km} \, \rm{s}^{-1}$, $\rm{W} \, \rm{m}^{-2}$, or $\rm{K} \, \rm{km} \, \rm{s}^{-1}$) will introduce factors of the rest wavelength, which will be different for different lines. 

\begin{figure*}
\centering
\mbox{
	\includegraphics[width=126mm]{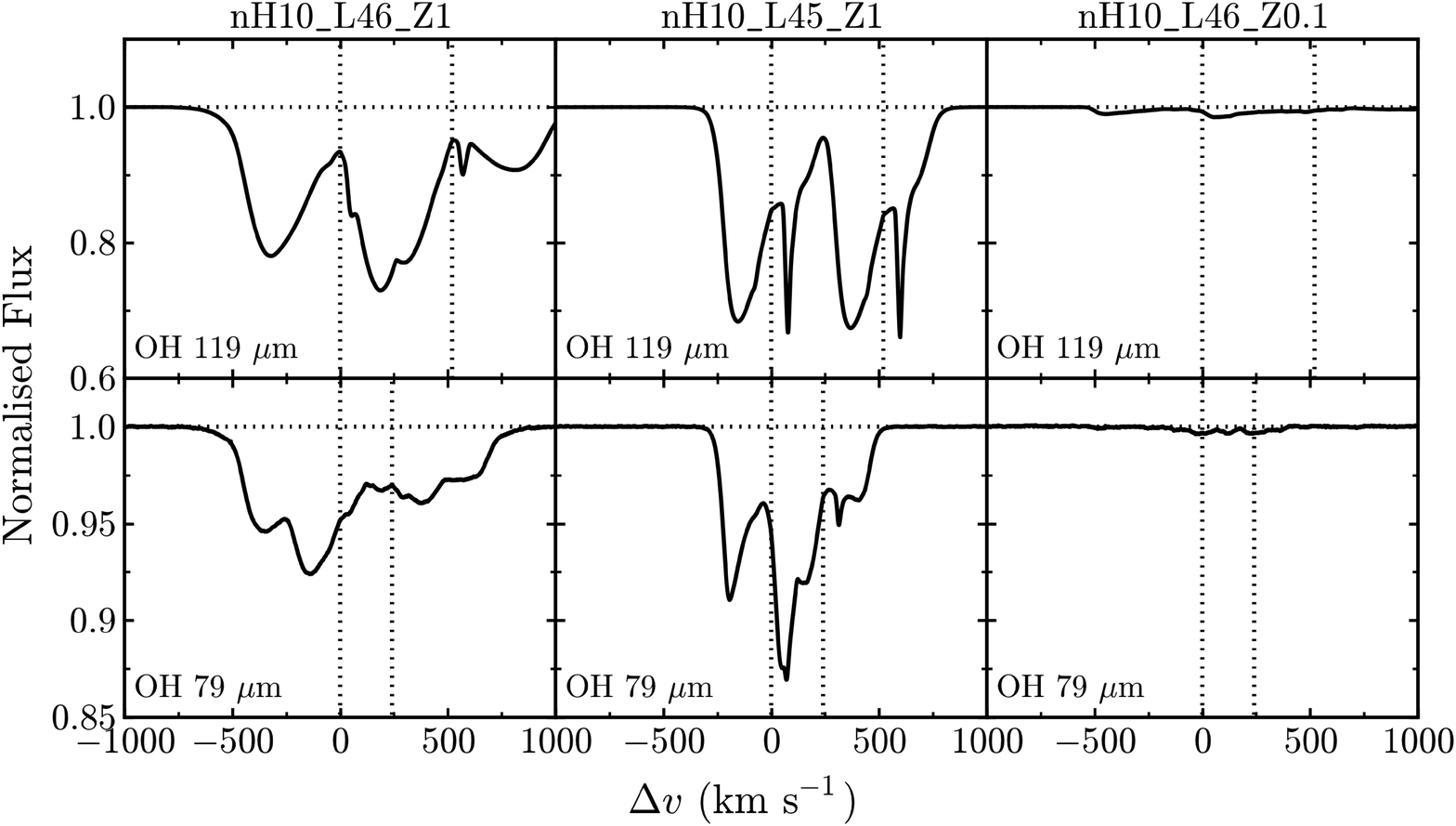}}
\caption{Continuum-normalised spectra of the OH doublets at $119 \, \rm{\mu m}$ (top row) and $79 \, \rm{\mu m}$ (bottom row) from the simulations nH10\_L46\_Z1 (left), nH10\_L45\_Z1 (centre), and nH10\_L46\_Z0.1 (right), after $1 \, \rm{Myr}$. The velocity scale is relative to the rest wavelength of the blueward component of the doublet. The positions of the two doublet components are indicated by vertical dotted lines.}  
\label{ohSpectraFig}
\end{figure*}

The molecular gas in our simulations shows significantly higher excitation than the core or broad components of Mrk 231 measured by \citet{cicone12} and \citet{feruglio15}. It is possible that the unusually high excitation in our simulations is at least partly due to the lack of inhomogeneities in the ambient ISM, which would lead to a range of shock velocities throughout the shell of swept up gas that would affect the shock heating, and hence excitation, of the molecular gas in the outflow. The higher excitation in the simulations suggests that the values of $\alpha_{\rm{CO} \, (1-0)}$ that we calculate in Table~\ref{alphaCO_table} may be lower than the true values in Mrk 231. We must however note that the comparison with Mrk 231 is for a single quasar and it is not clear whether the CO excitation in its wind is  representative of AGN-driven molecular outflows. Furthermore, the broad component in NGC 1266, with a width of $274 - 597 \, \rm{km} \, \rm{s}^{-1}$ (depending on the line), has a comparable (2$-$1)/(1$-$0) ratio to the simulations, although the (3$-$2)/(1$-$0) ratio is lower. This broad component has been attributed to an outflow powered by a buried AGN, as the star formation in this system is too weak to drive the observed outflow \citep{alatalo11, glenn15}. Note that NGC 1266 (a buried AGN in an S0 galaxy) is a very different system to Mrk 231 (a luminous quasar in a ULIRG). 

\subsection{OH absorption}\label{OH_lines_sect} 

Fast molecular outflows have been detected in several OH lines (see fig. 1 of \citealt{gonzalezalfonso14} for an energy level diagram of OH). We use \textsc{radmc-3d} to calculate the spectra of two commonly observed transitions to the ground state: the $^{2}\Pi_{3/2} \, \rm{J}=5/2-3/2$ doublet at $119 \, \rm{\mu m}$, and the $^{2}\Pi_{1/2}-^{2}\Pi_{3/2} \, \rm{J}=1/2-3/2$ doublet at $79 \, \rm{\mu m}$. Each of these transitions consists of a doublet, separated by $520 \, \rm{km} \, \rm{s}^{-1}$ and $240 \, \rm{km} \, \rm{s}^{-1}$, respectively. The molecular data for OH is taken from the \textsc{lamda} database. We compute the non-LTE level populations of OH including collisions from ortho- and para-H$_{2}$, using collisional rate coefficients from \citet{offer94}, assuming an ortho-to-para ratio of 3:1. As these rate coefficients are only tabulated up to $300 \, \rm{K}$ in the \textsc{lamda} database, we assume that they are constant above this temperature. We caution that, for OH, excitation by the far-infrared radiation field can be important. However, the implementation of the non-LTE LVG method in \textsc{radmc-3d} does not currently account for the radiation from the dust continuum or from the AGN when computing level populations. Nevertheless, as we discuss below, the OH $119 \, \mu\rm{m}$ line is typically optically thick, in which case the absorption strength of the $119 \, \mu\rm{m}$ line should depend primarily on the covering fraction of OH in the outflow, rather than the details of the OH level populations. 

Fig.~\ref{ohSpectraFig} shows the continuum-subtracted spectra of the OH doublets at $119 \, \rm{\mu m}$ (top row) and $79 \, \rm{\mu m}$ (bottom row) in nH10\_L46\_Z1 (left), nH10\_L45\_Z1 (centre), and nH10\_L46\_Z0.1 (right), after $1 \, \rm{Myr}$. The velocity scale is relative to the rest wavelength of the blueward component of each doublet. The rest positions of the two components are shown by the vertical dotted lines. We calculate the normalised spectra from spatial pixels within the radius of the outflow, as measured from the CO emission maps in Fig.~\ref{coEmissionFig}. 

The spectra from nH10\_L46\_Z1 and nH10\_L45\_Z1 show several distinct absorption features. These are most clearly illustrated in the OH $119 \, \rm{\mu m}$ spectrum of nH10\_L45\_Z1 (top centre panel of Fig.~\ref{ohSpectraFig}). Firstly, there is broad blueshifted absorption seen in both components of the doublet. This arises from outflowing material coming towards the observer. Secondly, there is a narrow, redshifted absorption feature, which is due to OH in the ambient ISM. This is redshifted because the ambient ISM is inflowing (due to the gravitational potential of the black hole and the host galaxy), so the ambient ISM that lies in front of the outflow is moving away from the observer. Note that the gas inflows in the simulations are different from a realistic galaxy due to our idealised setup (for example, we do not include rotation of the gas). Finally, there is a broad redshifted absorption feature (seen more clearly in the redward doublet component in the top left panel) due to outflowing material on the far side of the outflow. This is weaker than the broad blueshifted absorption because the receding side of the outflow lies behind most of the continuum-emitting dust, as viewed by the observer. 

In the $79 \, \rm{\mu m}$ spectra the two components of the doublet are closer together, so different absorption features overlap more. The absorption is weaker at $79 \, \rm{\mu m}$ than at $119 \, \rm{\mu m}$, as seen in observations \citep[e.g.][]{gonzalezalfonso17}. The low-metallicity run, nH10\_L46\_Z0.1, shows only negligible OH absorption, as we will discuss further below. 

It is likely that the artifacts along the boundaries between the high- and low-resolution regions will affect the OH spectra, similarly to the anomalous dips at $\Delta v = 0$ in the CO spectra in Fig.~\ref{coEmissionFig}. In the OH spectra, this will result in weaker absorption at $\Delta v = 0$. However, the different absorption features overlap, which makes it difficult to isolate the effects of these artifacts on the OH absorption. 

\begin{figure}
\centering
\mbox{
	\includegraphics[width=84mm]{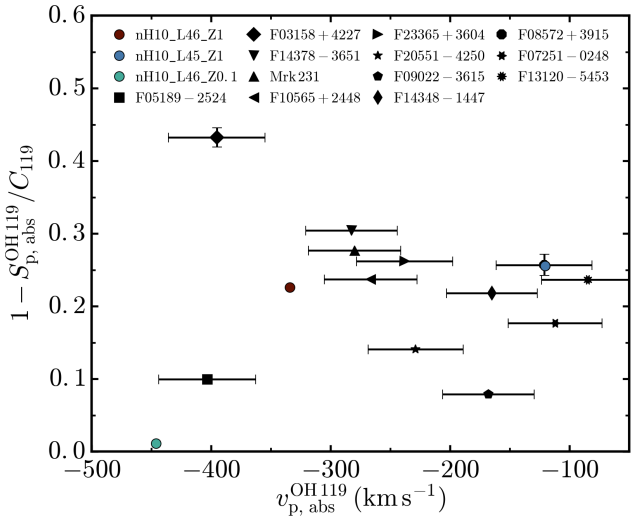}}
\caption{Peak blueshifted OH $119 \, \rm{\mu m}$ absorption strength ($1 - S_{\rm{p,} \, \rm{abs}}^{\rm{OH} \, 119} / C_{119}$) versus peak absorption velocity ($v_{\rm{p,} \, \rm{abs}}^{\rm{OH} \, 119}$) from the simulations nH10\_L46\_Z1 (red circle), nH10\_L45\_Z1 (blue circle) and nH10\_L46\_Z0.1 (green circle), and from observations of a sample of local ULIRGs (\citealt{gonzalezalfonso17}; black symbols). In the simulations, the peak absorption is defined as the minimum of the continuum-normalised spectrum at velocities $< -80 \, \rm{km} \, \rm{s}^{-1}$, while the observations measure the peak absorption strength from Gaussian fits, again limited to velocities $< -80 \, \rm{km} \, \rm{s}^{-1}$. The two simulations at solar metallicity are in good agreement with the observations, while the low-metallicity run has much weaker OH absorption, likely due to the lower OH covering factor.} 
\label{ohAbsorptionFig}
\end{figure}

To compare the strength of the OH absorption in our simulations to observations, we measure the strength of the peak blueshifted absorption feature from the OH $119 \, \rm{\mu m}$ transition, $1 - S_{\rm{p,} \, \rm{abs}}^{\rm{OH} \, 119} / C_{119}$. We take this to be the minimum in the continuum-normalised spectrum at velocities $< -80 \, \rm{km} \, \rm{s}^{-1}$, which will correspond to the broad outflow component, where $S_{\rm{p,} \, \rm{abs}}^{\rm{OH} \, 119}$ is the flux density of the spectrum at the minimum and $C_{119}$ is the flux density of the continuum. By looking at velocities $< -80 \, \rm{km} \, \rm{s}^{-1}$, we also limit the impact of the artifacts along the boundaries between the high- and low-resolution region, which primarily affect velocities close to $\Delta v = 0$, as in the CO spectra (see Fig.~\ref{coEmissionFig}). This OH absorption strength is shown in Fig.~\ref{ohAbsorptionFig}, plotted against the velocity of the absorption peak. The simulations are shown by the coloured circles, while the black symbols show the observed sample of local ULIRGs from \citet{gonzalezalfonso17} (see the top panel of their fig. 3). In the observations, the absorption peak was measured by fitting Gaussians to the absorption features. 

The two simulations at solar metallicity (nH10\_L46\_Z1 and nH10\_L45\_Z1) show OH $119 \, \rm{\mu m}$ absorption strengths and velocities that are comparable to those observed in the sample of local ULIRGs. However, the absorption strength in the low-metallicity run (nH10\_L46\_Z0.1) is lower than is observed. \citet{gonzalezalfonso17} highlighted that, since the OH $119 \, \rm{\mu m}$ doublet is optically thick, the absorption strength measures the covering fraction of OH in the outflow. In Fig.~\ref{coEmissionFig}, we saw that the covering fraction of dense, molecular clumps traced by CO emission is much lower in the low-metallicity run than in the other two simulations. Therefore, the weak OH absorption in this run is likely due to the low covering fraction of OH. It is unsurprising that this run does not match the observations, as the observed systems are ULIRGs, with higher metallicities. We also caution that the peak absorption velocity in the low-metallicity run is poorly converged with numerical resolution (see Appendix~\ref{resolution_appendix}). 

\begin{figure*}
\centering
\mbox{
	\includegraphics[width=126mm]{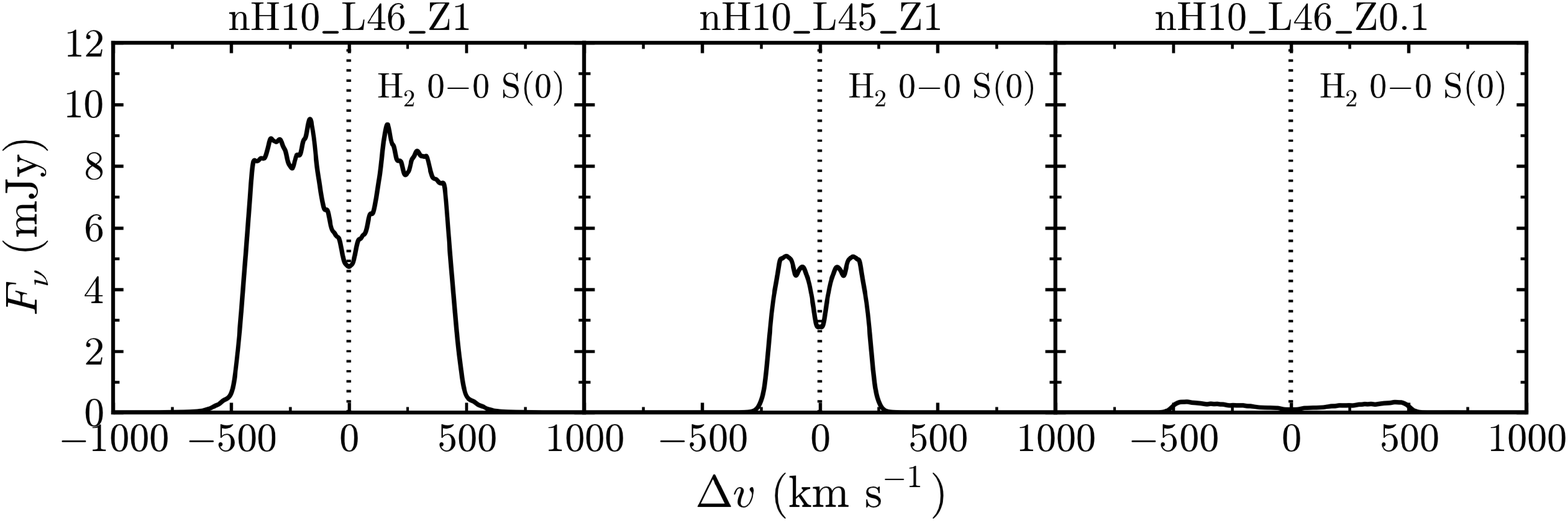}}
\caption{H$_{2}$ $0-0$ S(0) spectra from nH10\_L46\_Z1 (left), nH10\_L45\_Z1 (centre) and nH10\_L46\_Z0.1 (right) after $1 \, \rm{Myr}$, normalised to a distance of $184 \, \rm{Mpc}$. The dip at $\Delta v = 0 \, \rm{km} \, \rm{s}^{-1}$ is due to artifacts along the boundary between the high- and low-resolution regions. These artifacts will reduce the H$_{2}$ column densities calculated from the full line flux, but we find that the H$_{2}$ excitation temperatures calculated from line ratios are not strongly affected.} 
\label{H2spectraFig}
\end{figure*}

Several OH $119 \, \rm{\mu m}$ spectra in the sample of \citet{gonzalezalfonso17} show P-Cygni profiles, with redshifted emission in addition to blueshifted absorption. However, we do not find any OH emission in the simulations. It is possible that this is due to the simplified geometry of our simulations. For example, if most of the continuum emission comes from dusty material close to the AGN, inside the outflowing shell, then the side of the shell moving towards the observer will be seen in absorption, whereas the receding side will lie behind the continuum source and so will only be seen in emission, producing a P-Cygni profile. However, our simulations do not include a dusty torus or accretion disc around the AGN, and there is little dust inside the outflowing shell, as this gas is mostly hot ($\sim 10^{10} \, \rm{K}$), with zero dust abundance. Our simulations therefore represent AGN for which the continuum emission is primarily from the host galaxy itself, outside the outflowing shell, for which both sides of the shell are seen in absorption. See, for example, fig. 12 of \citet{gonzalezalfonso17} for schematic diagrams of the geometries that they assume in the radiative transfer models that they fit to their observations of OH lines in ULIRG outflows. 

\subsection{IR-traced warm H$_{2}$ emission}\label{H2_lines_sect} 

Cold H$_{2}$ at temperatures of a few tens of K, as found in typical giant molecular clouds, is difficult to observe directly, as the lowest rotational transition of the H$_{2}$ molecule has an energy of $E / k_{\rm{B}} = 510 \, \rm{K}$. However, warm H$_{2}$, at temperatures of a few hundred K or higher, exhibits emission lines in the near- and mid-infrared from ro-vibrational transitions. These lines have been observed in a variety of extragalactic systems, including colliding galaxies \citep{rieke85, sugai97, vanderwerf93, appleton06, appleton17}, starburst galaxies \citep{rosenberg13}, ULIRGs \citep{zakamska10, hill14}, local normal star-forming galaxies from the SINGS sample \citep{roussel07}, radio galaxies \citep{ogle07}, and AGN-driven outflows \citep{dasyra11, rupke13a}. 

We use \textsc{radmc-3d} to calculate the line emission in our simulations from the four lowest rotational transitions of the ground vibrational state of the H$_{2}$ molecule, i.e. the $0-0$~S(0) to S(3) transitions, with rest wavelengths of $28.2 - 9.7 \, \rm{\mu m}$. We use transition probabilities from \citet{wolniewicz98}, and we include collisional excitation of H$_{2}$ by H\textsc{i} \citep{wrathmall07}, H\textsc{ii} \citep{gerlich90}, para-H$_{2}$ \citep{flower98}, ortho-H$_{2}$ \citep{flower99}, He\textsc{i} \citep{floweretal98} and electrons \citep{draine83}. 

For the CO and OH lines, in Sections~\ref{CO_lines_sect} and \ref{OH_lines_sect} respectively, we included thermal emission, scattering and absorption from dust grains, and then subtracted the continuum emission from the line spectra (CO), or divided the total spectra by the continuum (OH). However, at the wavelengths of the warm H$_{2}$ lines, the dust continuum was stronger, and the resulting noise in the continuum (since \textsc{radmc-3d} uses a Monte Carlo method) became large compared to the lines themselves. It was too computationally expensive to increase the number of photon packets to sufficiently reduce this noise, so instead we simply excluded thermal emission and scattering from dust grains for the H$_{2}$ lines, and only included dust absorption. The lack of thermal dust emission will not affect our results, as we subtract the continuum. We found that dust scattering had very little effect on the CO spectra (not shown), although the H$_{2}$ lines are at a shorter wavelength.

Fig.~\ref{H2spectraFig} shows the S(0) spectra from nH10\_L46\_Z1 (left), nH10\_L45\_Z1 (centre) and nH10\_L46\_Z0.1 (right) after $1 \, \rm{Myr}$. The shape of this line is similar to that of the CO (1$-$0) line in these three simulations (see Fig.~\ref{coEmissionFig}). The dip at $\Delta v = 0 \, \rm{km} \, \rm{s}^{-1}$ is again due to artifacts along the boundary between the high- and low-resolution regions, as discussed in Section~\ref{CO_lines_sect}. 

Using the line fluxes of each rotational transition, we can calculate the column densities of the different energy levels, and hence deduce an excitation temperature of the IR-traced H$_{2}$ gas in a manner similar to what is done in observations. For a given transition, the column density $N_{\rm{u}}$ of the upper energy level is: 

\begin{equation}\label{Nu_eqn} 
N_{\rm{u}} = \frac{4 \pi \lambda}{h c} \frac{I}{A} \, \rm{cm}^{-2}, 
\end{equation} 
where $\lambda$ is the rest wavelength of the line, $A$ is the Einstein coefficient, and $I$ is the intensity of the line in units of $\rm{erg} \, \rm{s}^{-1} \, \rm{cm}^{-2} \, \rm{sr}^{-1}$. 

We can then compare the level populations to a thermal distribution with an excitation temperature $T_{\rm{exc}}$: 

\begin{equation}\label{thermal_eqn} 
N_{\rm{u}} / g_{\rm{u}} = \frac{N_{\rm{tot}} \exp(-E_{\rm{u}} / k_{\rm{B}} T_{\rm{exc}})}{Z(T_{\rm{exc}})}, 
\end{equation} 
where $N_{\rm{tot}}$ is the total H$_{2}$ column density, $E_{\rm{u}}$ is the energy of the given level, and $Z(T)$ is the partition function, which we take from \citet{herbst96}: 

\begin{equation} 
Z(T) = \frac{0.0247 (T / \rm{K})}{1 - \exp(-6000 \, \rm{K} / T)}. 
\end{equation} 
In equation~\ref{thermal_eqn}, $g_{\rm{u}}$ is the statistical weight of the upper energy level. For rotational state $J$, $g_{\rm{u}} = (2I + 1)(2J + 1)$, where the spin number $I$ is 0 for even $J$ and 1 for odd $J$. 

\begin{figure}
\centering
\mbox{
	\includegraphics[width=84mm]{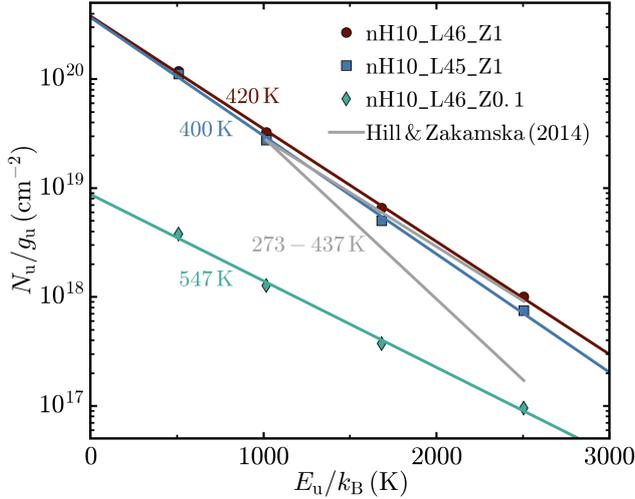}}
\caption{Level populations of H$_{2}$ derived from the line fluxes of the four lowest rotational transitions in the ground vibrational state, where $N_{\rm{u}}$ and $g_{\rm{u}}$ are the column density and statistical weight, respectively, of the upper energy level of each transition, plotted against the upper level energy. We show simulations nH10\_L46\_Z1 (red circles), nH10\_L45\_Z1 (blue squares) and nH10\_L46\_Z0.1 (green diamonds). The coloured lines show single-temperature thermal distributions fit to each simulation, with excitation temperatures as indicated. The slopes of the two grey lines show the range of excitation temperatures found in the 36 ULIRGs classified as AGN in the sample of \citet{hill14} (the normalisation of the grey lines is arbitrary). The excitation temperatures in the two simulations at solar metallicity are in good agreement with the observations of \citet{hill14}, although the low-metallicity run shows a higher excitation temperature.} 
\label{H2excitationFig}
\end{figure}

Fig.~\ref{H2excitationFig} shows the level populations, $N_{\rm{u}} / g_{\rm{u}}$, calculated from the four rotational transitions in the simulations (coloured symbols), plotted against the energy of the upper level. To calculate $N_{\rm{u}}$ from the simulations, we first calculate the intensity $I$ of each line by integrating the line over all velocity channels. This intensity is an average intensity over the area of the emitting region. We therefore only include spatial pixels within the projected radius of the outflow, as determined from the CO emission maps in Fig.~\ref{coEmissionFig}. We then use the intensity to calculate the column density $N_{\rm{u}}$, averaged over the projected area of the outflow, using equation~\ref{Nu_eqn}. 

The coloured lines in Fig.~\ref{H2excitationFig} show single-temperature thermal distributions fit to the level populations of each simulation. Both the excitation temperature, $T_{\rm{exc}}$, and the total H$_{2}$ column density of the thermal distribution, $N_{\rm{tot}}$, were fit to the values of $N_{\rm{u}} / g_{\rm{u}}$ from each simulation. We find excitation temperatures of $400 - 547 \, \rm{K}$. \citet{hill14} used archival \textit{Spitzer} data to measure the S(1) and S(3) lines in 115 ULIRGs from the \textit{IRAS} $1 \, \rm{Jy}$ sample, including 8 type 1 and 28 type 2 AGN. They derived excitation temperatures of $273 - 437 \, \rm{K}$, which is in good agreement with our two simulations at solar metallicity, although the low-metallicity run has a higher excitation temperature. The grey lines in Fig.~\ref{H2excitationFig} indicate the slopes of thermal distributions with this range of temperatures. The normalisation of the grey lines is arbitrary; they are intended to illustrate the slopes of the observed distributions. Although our solar metallicity wind simulations are consistent with the observations of \citet{hill14}, we note that the observations include all H$_{2}$ emission from the galaxies in their sample, which may include a significant non-wind component. 

The column densities, $N_{\rm{u}}$, will be affected by the artificial dips in the H$_{2}$ lines at $\Delta v = 0$ seen in Fig.~\ref{H2spectraFig}, which reduce the total line flux, and hence $N_{\rm{u}}$. However, the excitation temperatures are based on line ratios. We find that, if we calculate the excitation temperatures in the same way but using only the flux at velocities $|\Delta v| > 300 \, \rm{km} \, \rm{s}^{-1}$ (or $|\Delta v| > 200 \, \rm{km} \, \rm{s}^{-1}$ for nH10\_L45\_Z1), which avoids the artificial dips, then the excitation temperatures change by at most 2 per cent, compared to using the full lines. Thus the excitation temperatures are not strongly affected by these artificial dips in the spectra. 

\begin{table}
\begin{minipage}{84mm}
\centering
\caption{H$_{2}$ line luminosities from the simulations after $1 \, \rm{Myr}$. Note that these luminosities are from the high-resolution quadrant of the radiative transfer grid multiplied by 4, to represent the full sphere.}
\label{H2_table}
\begin{tabular}{lcccc}
\hline 
  & \multicolumn{4}{c}{Line luminosity ($\rm{erg} \, \rm{s}^{-1}$)} \\ 
Simulation & S(0) & S(1) & S(2) & S(3) \\ 
\hline
nH10\_L46\_Z1 \hspace{-0.18 in} & $4.0 \times 10^{40}$ \hspace{-0.15 in} & $1.2 \times 10^{42}$ \hspace{-0.15 in} & $8.5 \times 10^{41}$ \hspace{-0.15 in} & $2.2 \times 10^{42}$ \hspace{-0.15 in} \\
nH10\_L45\_Z1 \hspace{-0.18 in} & $1.1 \times 10^{40}$ \hspace{-0.15 in} & $3.1 \times 10^{41}$ \hspace{-0.15 in} & $1.9 \times 10^{41}$ \hspace{-0.15 in} & $4.8 \times 10^{41}$ \hspace{-0.15 in} \\ 
nH10\_L46\_Z0.1 \hspace{-0.18 in} & $1.4 \times 10^{39}$ \hspace{-0.15 in} & $5.3 \times 10^{40}$ \hspace{-0.15 in} & $5.3 \times 10^{40}$ \hspace{-0.15 in} & $2.3 \times 10^{41}$ \hspace{-0.15 in} \\ 
\hline
\end{tabular}
\end{minipage}
\end{table}

Using the total column density, $N_{\rm{tot}}$, from the best-fitting thermal distributions, we calculate the total mass of IR-traced H$_{2}$ that would be inferred from these mid-infrared emission lines. We find that the mass of IR-traced H$_{2}$ is approximately equal to the total H$_{2}$ mass in the simulations, to within 5 per cent. Thus nearly all of the molecular hydrogen in the outflow is warm and can be traced by the mid-infrared emission lines. Molecular hydrogen in the host galaxy -- not modelled here -- can have a substantial colder component. 

In Table~\ref{H2_table} we show the line luminosities for the four lowest rotational H$_{2}$ lines in the ground vibrational state from each simulation. These were calculated in the high-resolution quadrant of the radiative transfer grid and multiplied by 4 to represent the luminosity from the full spherical outflow. For comparison, the 36 AGN host galaxies included in \citet{hill14} show S(1) luminosities of $1.7 \times 10^{41} - 1.2 \times 10^{42} \, \rm{erg} \, \rm{s}^{-1}$ and S(3) luminosities of $7.0 \times 10^{40} - 5.8 \times 10^{41} \, \rm{erg} \, \rm{s}^{-1}$. This is comparable to our simulations, at least at solar metallicity, although the S(3) luminosity in the nH10\_L46\_Z1 run is a little high. We again note that the \citet{hill14} sample may include emission from the host galaxy, and not just the AGN outflow. 

\section{Conclusions}\label{conclusions_sect} 

We have run a suite of hydro-chemical simulations of an isotropic AGN wind interacting with a uniform medium to explore whether in-situ molecule formation within an AGN-driven wind can explain the fast molecular outflows that have been observed in quasars. This idealised setup matches the setup studied analytically by FGQ12. In this work, we focus on the chemistry of the gas to understand under what conditions molecules can form in the wind. 

We followed the non-equilibrium chemistry of the gas using a chemical network of 157 species, including 20 molecular species such as H$_{2}$, CO, OH and HCO$^{+}$. The simulations were run for $1 \, \rm{Myr}$, which corresponds to the typical flow times of observed molecular outflows \citep[e.g.][]{gonzalezalfonso17}, at a fiducial resolution of $30 \, \rm{M}_{\odot}$ per gas particle in the high-resolution octant for our main runs. We considered a range of initial ambient medium densities ($1 - 10 \, \rm{cm}^{-3}$), AGN luminosities ($10^{45} - 10^{46} \, \rm{erg} \, \rm{s}^{-1}$), and metallicities ($0.1 - 1 \, \rm{Z}_{\odot}$). We also varied certain aspects of the model (the cosmic ray ionization rate, AGN UV flux, shielding column density, dust abundance, and the presence of a Jeans limiter), to test how uncertainties in these affect our results. 

We then used the publicly available radiative transfer code \textsc{radmc-3d} \citep{dullemond12} to compute CO, OH, and IR-traced warm H$_{2}$ emission and absorption from our simulation snapshots in post-processing. This allowed us to compare the molecular outflows in our simulations to observations, and to make predictions for the conversion factors between CO emission and H$_{2}$ mass. 

We summarise our main results below. 

\begin{enumerate}
\item The hot wind bubble remains at high temperatures ($\ga 10^{10} \, \rm{K}$) throughout the simulations, and so the outflow is approximately energy conserving. However, in all simulations except the low-density run (nH1\_L46\_Z1), the gas swept up from the ambient ISM cools to below $10^{4} \, \rm{K}$ within $1 \, \rm{Myr}$ (Fig.~\ref{gasFig}). 
\item The fiducial simulation (nH10\_L46\_Z1) produces $2.7 \times 10^{8} \, \rm{M}_{\odot}$ of H$_{2}$ in outflowing gas after $1 \, \rm{Myr}$, with a molecular fraction of 0.24 and an H$_{2}$ outflow rate of $140 \, \rm{M}_{\odot} \, \rm{yr}^{-1}$ (Figs.~\ref{molecularMassFig} and \ref{outflowRateFig}). The molecular mass and outflow rate increase with increasing density, AGN luminosity and metallicity. 
\item The molecular gas is found at high densities ($n_{\rm{H}} \ga 10^{3} \, \rm{cm}^{-3}$) with a wide range of temperatures, from tens of K up to $\sim 10^{3} \, \rm{K}$. (Fig.~\ref{TrhoFig}). A significant fraction of the wind mass reaches densities $n_{\rm{H}} >10^{4} \, \rm{cm}^{-3}$ that may be observable with dense gas tracers such as HCN and HCO$^{+}$.
\item The H$_{2}$ outflow rate depends only weakly on the cosmic ray ionization rate, while it decreases strongly as we decrease the shielding column density or the dust-to-metals ratio (Fig.~\ref{modelVarsFig}). 
\item We used the CO line emission computed from our simulations in post-processing to calculate the H$_{2}$ outflow rates in the same way as in the observations. We find that the simulations at solar metallicity are slightly lower than, but still reasonably close to, the observed outflow rates from \citet{cicone14}, while the low-metallicity run (nH10\_L46\_Z0.1) is more than two orders of magnitude below the observations (Fig.~\ref{coObsFig}). The maximum line of sight velocities inferred from the CO (1$-$0) line in the simulations are a factor $\approx 2$ lower than in the observations. 
\item In our reference simulation (nH10\_L46\_Z1), the CO conversion factor for the (1$-$0) line is $\alpha_{\rm{CO} \, (1-0)} = 0.13 \, \rm{M}_{\odot} (\rm{K} \, \rm{km} \, \rm{s}^{-1} \, \rm{pc}^{2})^{-1}$ (Table~\ref{alphaCO_table}), a factor of 6 lower than is commonly assumed in observations. For the ($2-1$) and ($3-2$) lines, we find conversion factors of $0.08$ and $0.06 \, \rm{M}_{\odot} (\rm{K} \, \rm{km} \, \rm{s}^{-1} \, \rm{pc}^{2})^{-1}$, respectively. 
\item The simulations at solar metallicity show significant absorption from the OH doublets at $119$ and $79 \, \rm{\mu m}$, while the low-metallicity run shows only weak absorption (Fig.~\ref{ohSpectraFig}). The strength of the $119 \, \rm{\mu m}$ absorption in the solar metallicity runs is in good agreement with the observations of \citet{gonzalezalfonso17} (Fig.~\ref{ohAbsorptionFig}). 
\item We find strong mid-infrared emission from the four lowest rotational transitions of the ground vibrational state of H$_{2}$. The mass of H$_{2}$ inferred from these emission lines is within a few per cent of the total H$_{2}$ mass, i.e. nearly all of the molecular hydrogen in the outflow is traced by IR emission. The level populations derived from these lines indicate an excitation temperature of $400 - 547 \, \rm{K}$ (Fig.~\ref{H2excitationFig}), in good agreement with the observations of \citet{hill14}, at least at solar metallicity. 
\end{enumerate}

We have thus demonstrated that it is possible to form fast molecular outflows in-situ within an AGN-driven wind, provided that the ambient ISM density and metallicity are sufficiently high (at least $10 \, \rm{cm}^{-3}$ and solar metallicity, respectively). The molecular emission and absorption lines in our simulations are generally in good agreement with those seen in observations of AGN, although the maximum line of sight velocities from the CO (1$-$0) line are a factor $\approx 2$ lower than is observed at a given AGN luminosity, and the molecular gas in the simulations may be more highly excited than in observations. However, there are several caveats that we will need to address before we can fully understand the origin of fast molecular outflows in quasars. 

The biggest uncertainty is the role that dust grains play in AGN winds. Our simulations assume a constant dust-to-metals ratio equal to that in the Milky Way. If we reduce the dust-to-metals ratio by a factor of 10 or 100 in our reference model, the mass of outflowing H$_{2}$ after $1 \, \rm{Myr}$ decreases by a factor of $\approx 8$ or $\approx 150$, respectively (Fig.~\ref{modelVarsFig}). However, the simulations run by \citet{ferrara16} found that dust grains in an AGN wind will be rapidly destroyed by shocks and sputtering, within $10^{4} \, \rm{yr}$. As discussed in Section~\ref{model_vars_sect}, it may be possible for dust formation mechanisms, such as the accretion of metals from the gas phase onto grains, to re-form the grains after the swept up gas has cooled. To resolve these issues, we will need to develop a more rigorous treatment for the dust grains, including formation and destruction mechanisms and the resulting evolution of the grain size distribution. This will need to be coupled to the chemistry solver, along with alternative H$_{2}$ formation channels via PAH chemistry \citep[e.g.][]{boschman15}. 

Secondly, our simulations do not include thermal conduction. \citet{bruggen16} explored the effects of thermal conduction in simulations of cold clouds interacting with a hot, fast gas flow. They found that thermal conduction can evaporate the cold clouds, but it can also also compress the clouds, allowing them to survive longer. \citet{ferrara16} found that, while conduction can result in a more uniform medium in their simulations of an AGN wind, it does not have a significant impact on the final conditions of the gas, although their simulations did not include the hot wind bubble. 

Furthermore, thermal conduction can cause some of the cold gas in the shell of swept up material to expand into, and subsequently mix with, the hot wind bubble. \citet{castor75} showed that, in stellar wind bubbles, the mass of the hot, shocked wind region is dominated by gas that came from the swept up shell in such a conductive flow. The AGN-driven winds that we study here are analogous to the stellar wind bubbles of \citet{castor75}, although the latter are at much lower velocities and energies. In AGN-driven galactic winds, magnetic fields may act to limit the effects of thermal conduction. 

Thirdly, we show in Appendix~\ref{resolution_appendix} that the H$_{2}$ outflow rate is not well converged with resolution in our simulations. In the low-luminosity run (nH10\_L45\_Z1), the H$_{2}$ outflow rate increases by a factor $\approx 4$ as the mass resolution is increased from 240 to $10 \, \rm{M}_{\odot}$ per gas particle. It is encouraging that the H$_{2}$ outflow rate increases with increasing resolution, as this strengthens our main conclusion that molecules can form in an AGN wind. 

Finally, our simulations follow an idealised scenario of an isotropic AGN wind interacting with a uniform ambient medium. This allows us to focus on the details of the molecular chemistry, and makes it easier to understand how the chemistry is affected by the physical conditions such as density and AGN luminosity. However, we will need to explore the effects of more realistic geometries, such as a radially declining density profile, inhomogeneities in the ambient medium, and the presence of a galactic disc. It will therefore be interesting to apply our methods for injecting an AGN wind and following the non-equilibrium chemistry to more realistic galaxy simulations. 

In this work we have focussed on AGN-driven winds. However, molecular outflows have also been observed in starburst galaxies \citep[e.g.][]{nakai87, veilleux09, bolatto13b, hill14, beirao15}. It will thus be interesting to extend this study to the case of star formation-driven galactic winds. Indeed, we saw that some of our results, such as the H$_{2}$ outflow rate and velocity (Fig.~\ref{coObsFig}), scale with the AGN luminosity. Therefore, it might be possible to generalise our results to star formation-driven outflows by scaling down to the lower luminosities of starburst galaxies. 

Our simulations predict strong infrared emission from warm H$_{2}$, tracing nearly all of the mass of outflowing molecular gas. These H$_{2}$ lines may be observable by the upcoming James Webb Space Telescope (JWST). Future JWST observations could therefore provide valuable insight into the nature of these fast molecular outflows, and allow us to further constrain our models. 

\section*{Acknowledgements}

We thank the anonymous referee for their detailed report, which improved the quality of this manuscript. We also thank Chiara Feruglio and Eduardo Gonz\'{a}lez-Alfonso for their detailed comments, as well as Nadia Zakamska, Mike Shull, Desika Narayanan and Fabrizio Fiore for useful discussions We are grateful to Phil Hopkins and Paul Torrey for help with the \textsc{gizmo} code. AJR is supported by the Lindheimer fellowship at Northwestern University. CAFG was supported by NSF through grants AST-1412836, AST-1517491, and CAREER award AST-1652522, by NASA through grant NNX15AB22G, and by CXO through grant TM7-18007X. The simulations used in this work were run on the Stampede supercomputer at the Texas Advanced Computing Center (TACC) through allocations TG-AST160035 and TG-AST160059 granted by the Extreme Science and Engineering Discovery Environment (XSEDE), which is supported by NSF grant number ACI-154562; the Pleiades supercomputer through allocation s1480, provided through the NASA Advanced Supercomputing (NAS) Division at Ames Research Center; and the Quest computing cluster at Northwestern University, which is jointly supported by the Office of the Provost, the Office for Research, and Northwestern University Information Technology. 

{}

\appendix 

\section{Resolution tests}\label{resolution_appendix} 

We ran all four parameter variations runs (see Table~\ref{sims_table}) at a standard resolution of $30 \, \rm{M}_{\odot}$ per gas particle and a minimum gas gravitational softening of $0.1 \, \rm{pc}$. These four runs were also repeated with a factor 8 lower mass resolution ($240 \, \rm{M}_{\odot}$ per particle), with a minimum gravitational softening of $0.2 \, \rm{pc}$. We additionally ran the low-luminosity simulation (nH10\_L45\_Z1) at a factor 3 higher mass resolution of $10 \, \rm{M}_{\odot}$ per particle and a gravitational softening of $0.07 \, \rm{pc}$. We chose the low-luminosity run for the high-resolution test as it was computationally cheaper than the reference run ($240 \, 000$ and $840 \, 000$ CPU hours, respectively, at our fiducial resolution), while still forming a significant molecular outflow. Also, note that the increase in mass resolution is only a factor of 3 in the high-resolution run, rather than the factor of 8 from the low to standard resolution. Any higher resolution than this would have been prohibitively expensive. 

\begin{figure}
\centering
\mbox{
	\includegraphics[width=84mm]{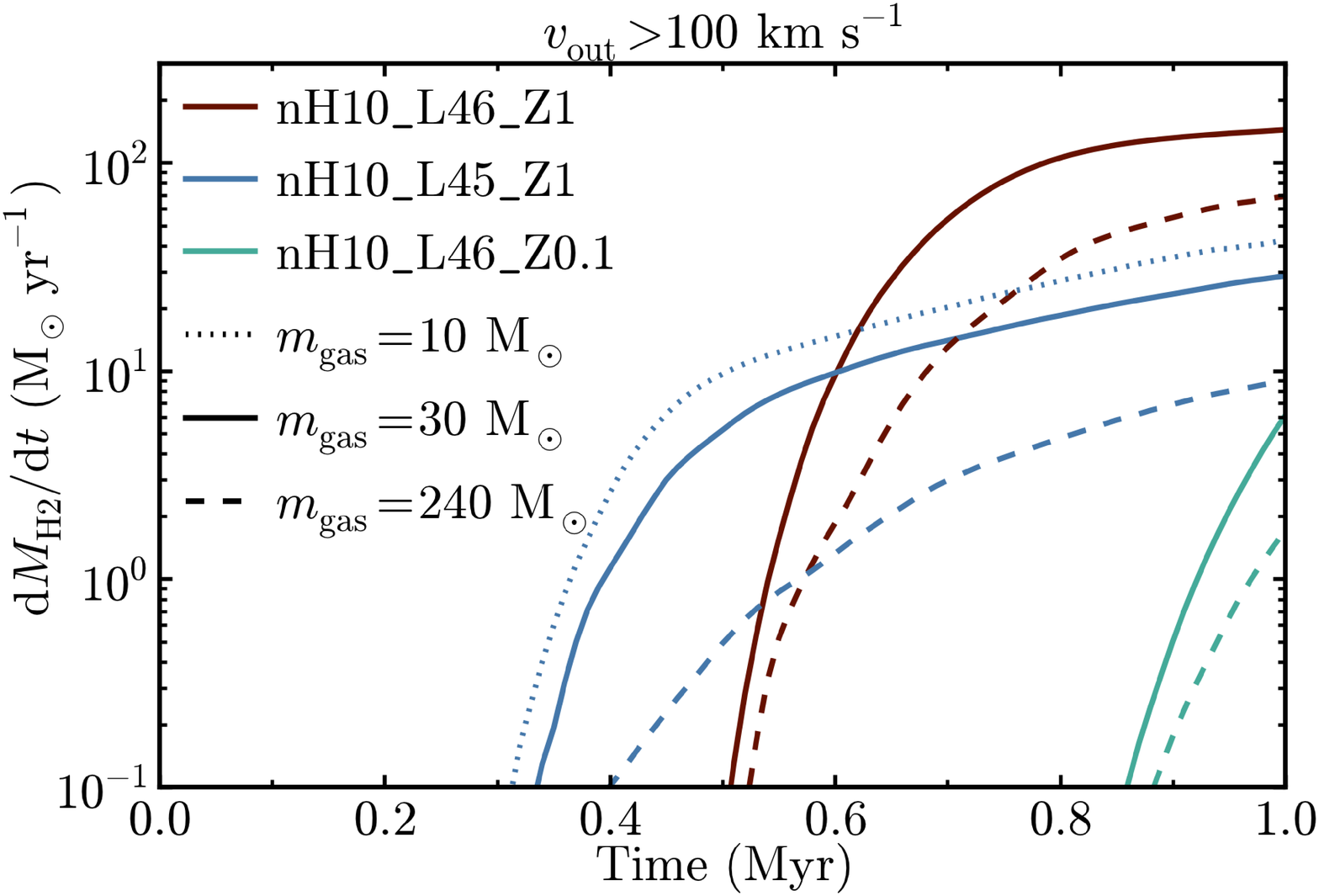}}
\caption{Mass outflow rates of H$_{2}$ versus time at low resolution (dashed curves), standard resolution (solid curves) and high resolution(dotted curve), for runs nH10\_L46\_Z1 (red curves), nH10\_L45\_Z1 (blue curves) and nH10\_L46\_Z0.1 (green curves). The H$_{2}$ outflow rate tends to increase with increasing resolution.} 
\label{resFig}
\end{figure}

To test whether the formation of fast molecular outflows in our simulations is sensitive to the resolution, we show in Fig.~\ref{resFig} the mass outflow rate of molecular hydrogen versus time, as in Fig.~\ref{outflowRateFig}, for the different resolution levels (low resolution, dashed curves; standard resolution, solid curves; and high resolution, dotted curve). The outflow rate was calculated in the high-resolution wedge, including only particles outflowing with $v_{\rm{out}} > 100 \, \rm{km} \, \rm{s}^{-1}$, then multiplied by 17 to give the outflow rate in the full spherical shell. We only show the three parameter variations that form molecules. 

The H$_{2}$ outflow rate increases with increasing resolution in all runs, by a factor $\approx 2 - 4$ from low to standard resolution, and by a factor $\approx 1.5$ from standard to high resolution. The H$_{2}$ outflow rate is thus not well converged with resolution. However, since it tends to increase with increasing resolution, this would further strengthen our main conclusion that fast molecular outflows can form in-situ within AGN-driven winds. 

The increase in H$_{2}$ outflow rate at high resolution is primarily due to an increase in the H$_{2}$ fraction of the outflowing material, rather than an increase in the total mass of gas in the outflow. The molecular fractions of CO, OH and HCO$^{+}$ also increase with increasing resolution, although the differences are largest for H$_{2}$. We note that the H$_{2}$ mass fraction $f_{\rm{H}_{2}} = M_{\rm{H}_{2}} / M_{\rm{H_{tot}}}$ at 1 Myr in the standard resolution run reaches 0.24 (see Fig.\ref{molecularMassFig}) and by definition cannot exceed 1.0. Thus, our standard resolution run cannot underestimate the H2 mass outflow rate by a large factor. 

\begin{figure}
\centering
\mbox{
	\includegraphics[width=84mm]{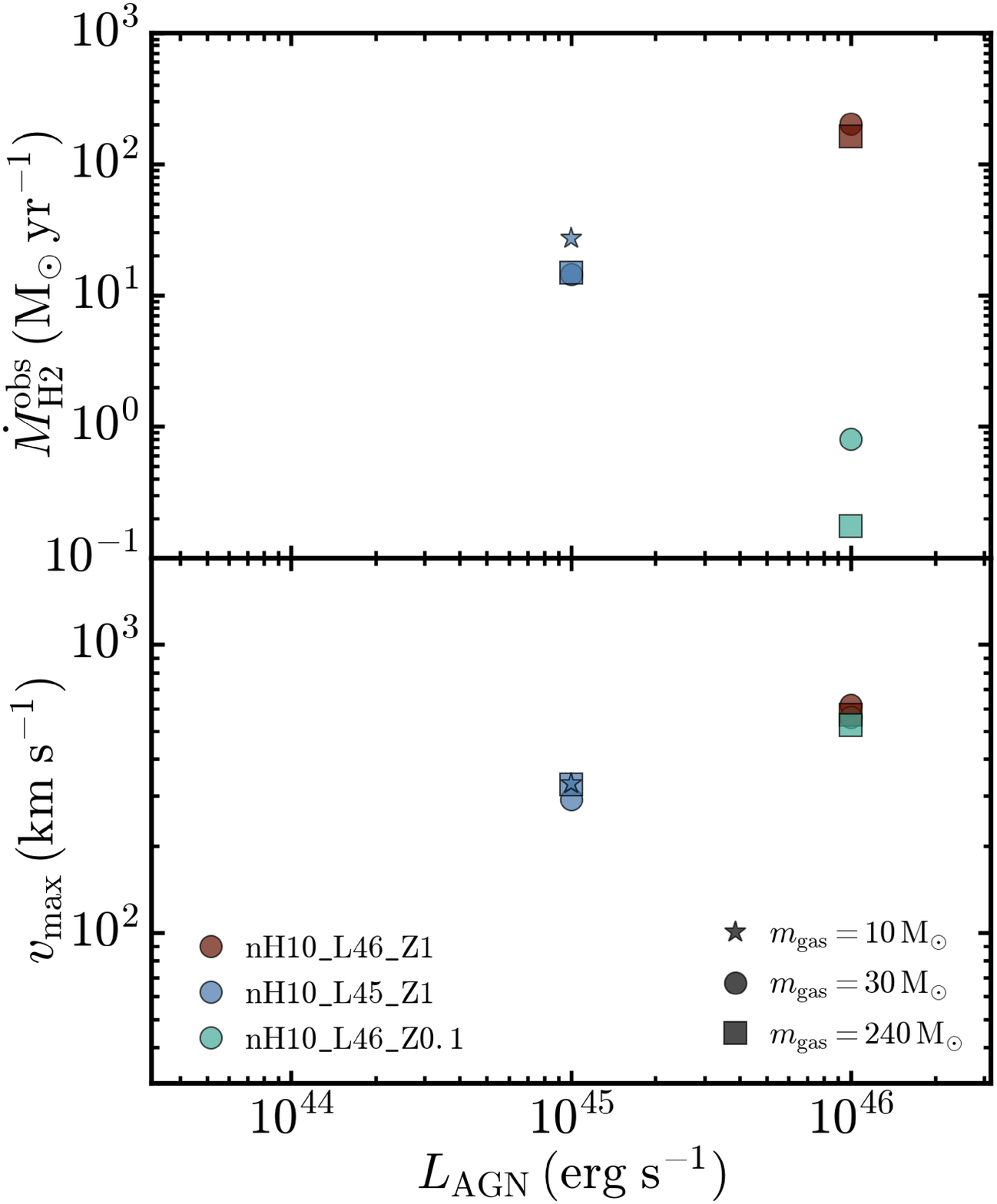}}
\caption{H$_{2}$ outflow rates calculated from CO emission (top panel) and maximum line of sight velocity from the CO spectra (bottom panel) plotted against AGN luminosity for runs nH10\_L46\_Z1 (red), nH10\_L45\_Z1 (blue) and nH10\_L46\_Z0.1 (green), at low (squares), standard (circles) and high (stars) resolution. The CO-derived H$_{2}$ outflow rate tends to increase with increasing resolution, especially in the low-metallicity run, while the maximum velocities are well converged.} 
\label{H2obsResFig}
\end{figure}

To test how our predictions for molecular emission and absorption lines are affected by numerical resolution, we repeated the radiative transfer calculations presented in Section~\ref{molecular_lines_sect} for the low- and high-resolution runs. Fig.~\ref{H2obsResFig} shows the H$_{2}$ outflow rate calculated from the CO 1$-$0 luminosity integrated over the wing of the CO line and assuming a standard ULIRG CO to H$_{2}$ conversion factor of $0.8 \, \rm{M}_{\odot} \, (\rm{K} \, \rm{km} \, \rm{s}^{-1} \rm{pc}^{2})^{-1}$ (top panel), and the maximum line of sight velocity from the CO spectra (bottom panel), plotted against AGN luminosity, as in Fig.~\ref{coObsFig}. We show the low (squares), standard (circles) and high (stars) resolution runs, for nH10\_L46\_Z1 (red), nH10\_L45\_Z1 (blue) and nH10\_L46\_Z0.1 (green). The H$_{2}$ outflow rates tend to increase with increasing resolution, particularly in the low-metallicity run, although the fiducial run is reasonably well converged. The maximum velocities show good convergence in all runs. 

\begin{figure}
\centering
\mbox{
	\includegraphics[width=84mm]{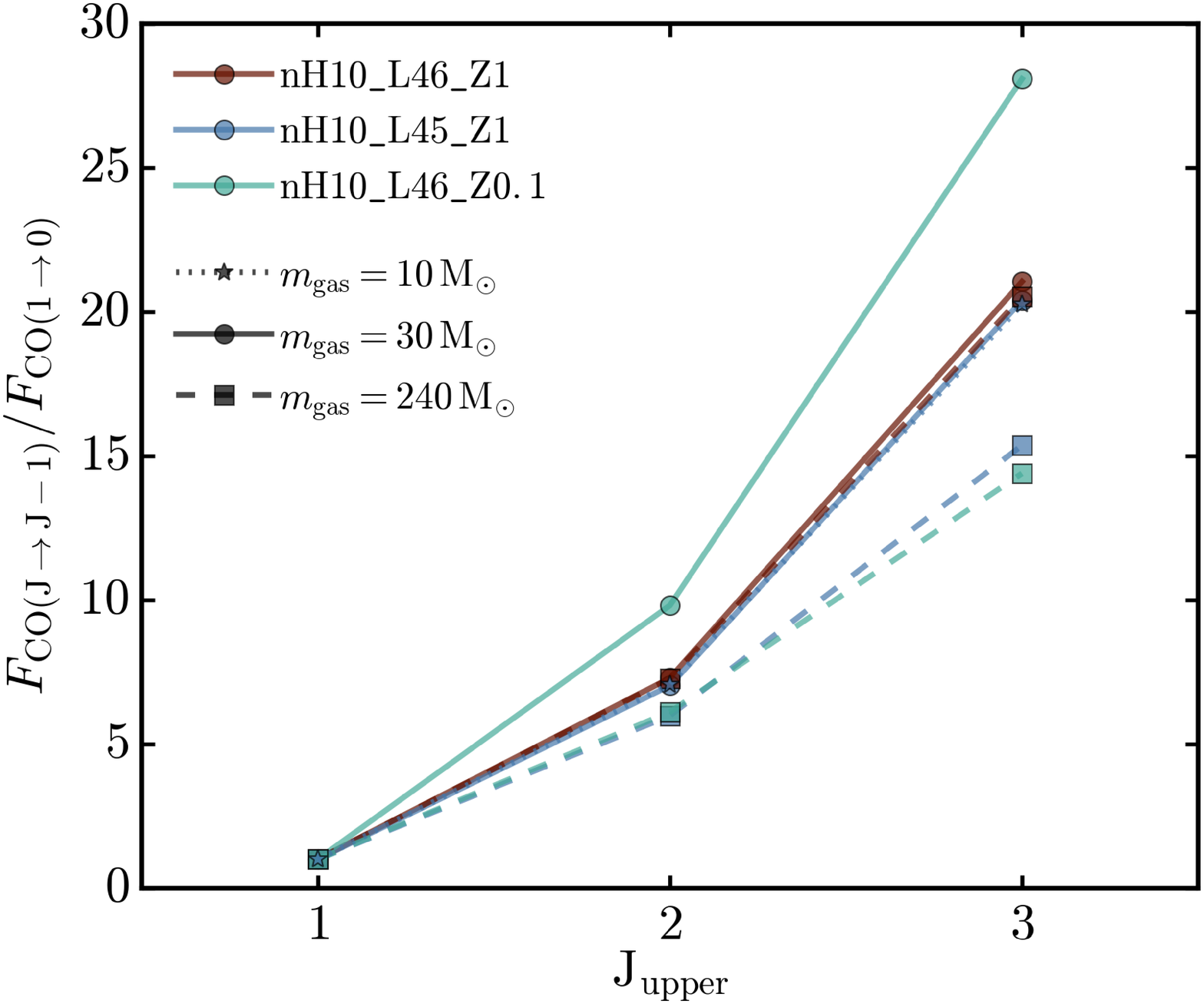}}
\caption{CO lines ratios measured from runs nH10\_L46\_Z1 (red), nH10\_L45\_Z1 (blue) and nH10\_L46\_Z0.1 (green), at low (squares and dashed curves), standard (circles and solid curves) and high (stars and dotted curves) resolution. The nH10\_L46\_Z1 runs are well converged, while the CO ratios in the low-luminosity and low-metallicity runs are lower at low-resolution. The standard and high-resolution runs of nH10\_L45\_Z1 are well converged.} 
\label{COratioResFig}
\end{figure}

Fig.~\ref{COratioResFig} shows CO line ratios relative to the 1$-$0 transition, where the line fluxes are measured in $\rm{Jy} \, \rm{km} \, \rm{s}^{-1}$, as in Fig.~\ref{coRatiosFig}. The nH10\_L46\_Z1 runs (red curves) are well converged, whereas in the low-luminosity and low-metallicity runs, the CO line ratios are lower in the low-resolution runs. The standard and high-resolution runs of nH10\_L45\_Z1 are well converged. 

\begin{figure}
\centering
\mbox{
	\includegraphics[width=84mm]{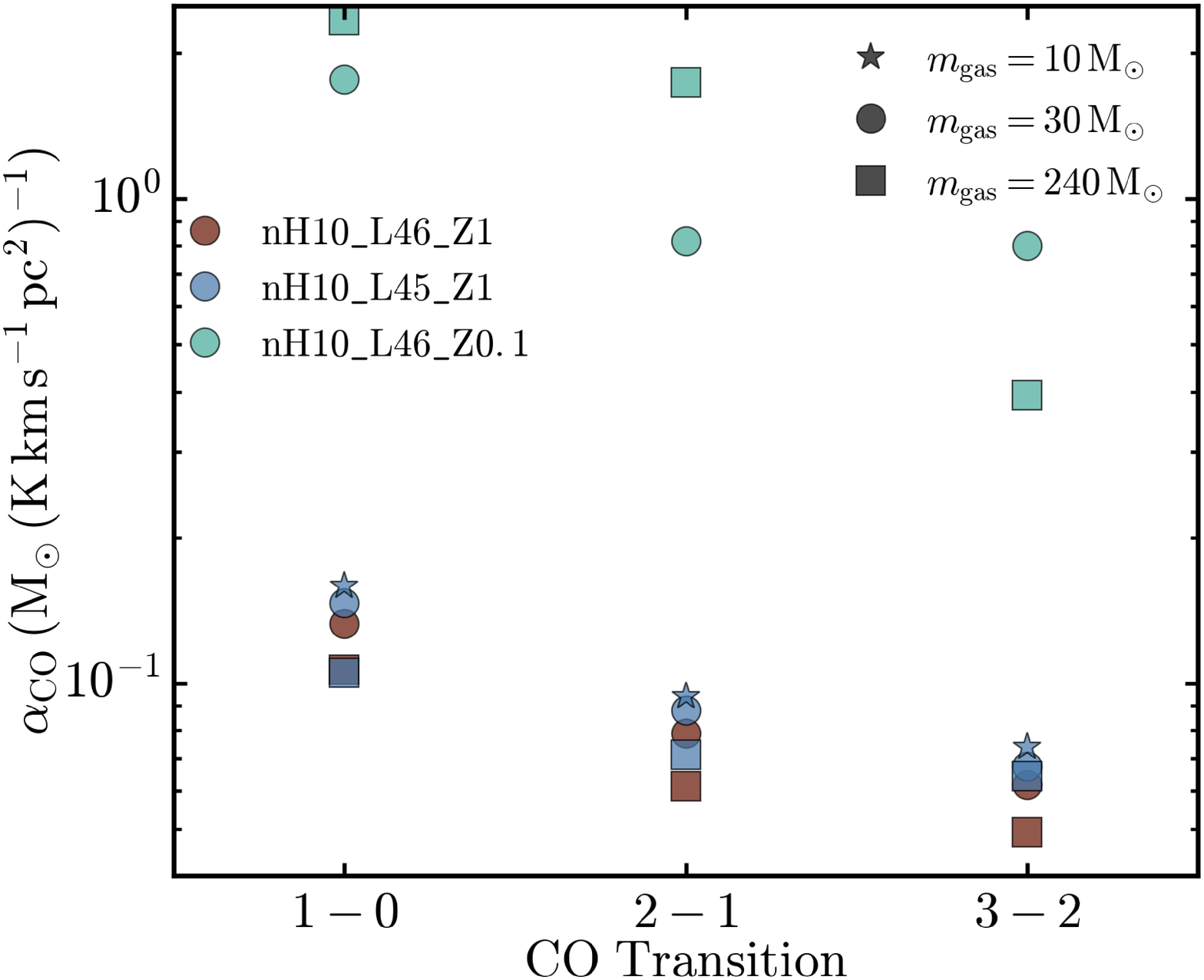}}
\caption{Resolution dependence of the CO to H$_{2}$ conversion factor, $\alpha_{\rm{CO}}$, for the three lowest rotational transitions of CO. In the two runs at solar metallicity, $\alpha_{\rm{CO}}$ increases with increasing resolution, by $\approx 25$ per cent from low to standard resolution. The low-metallicity run shows much poorer convergence in $\alpha_{\rm{CO}}$, which changes by up to a factor of 2 from low to standard resolution.} 
\label{COalphaResFig}
\end{figure}

Fig.~\ref{COalphaResFig} shows the CO to H$_{2}$ conversion factor, $\alpha_{\rm{CO}}$, calculated for the three lowest rotational CO lines. At solar metallicity (red and blue symbols), $\alpha_{\rm{CO}}$ tends to increase with increasing resolution, by $\approx 25$ per cent from low to standard resolution. The numerical convergence of $\alpha_{\rm{CO}}$ is much poorer in the low-metallicity run, varying by up to a factor of 2 from low to standard resolution. 

\begin{figure}
\centering
\mbox{
	\includegraphics[width=84mm]{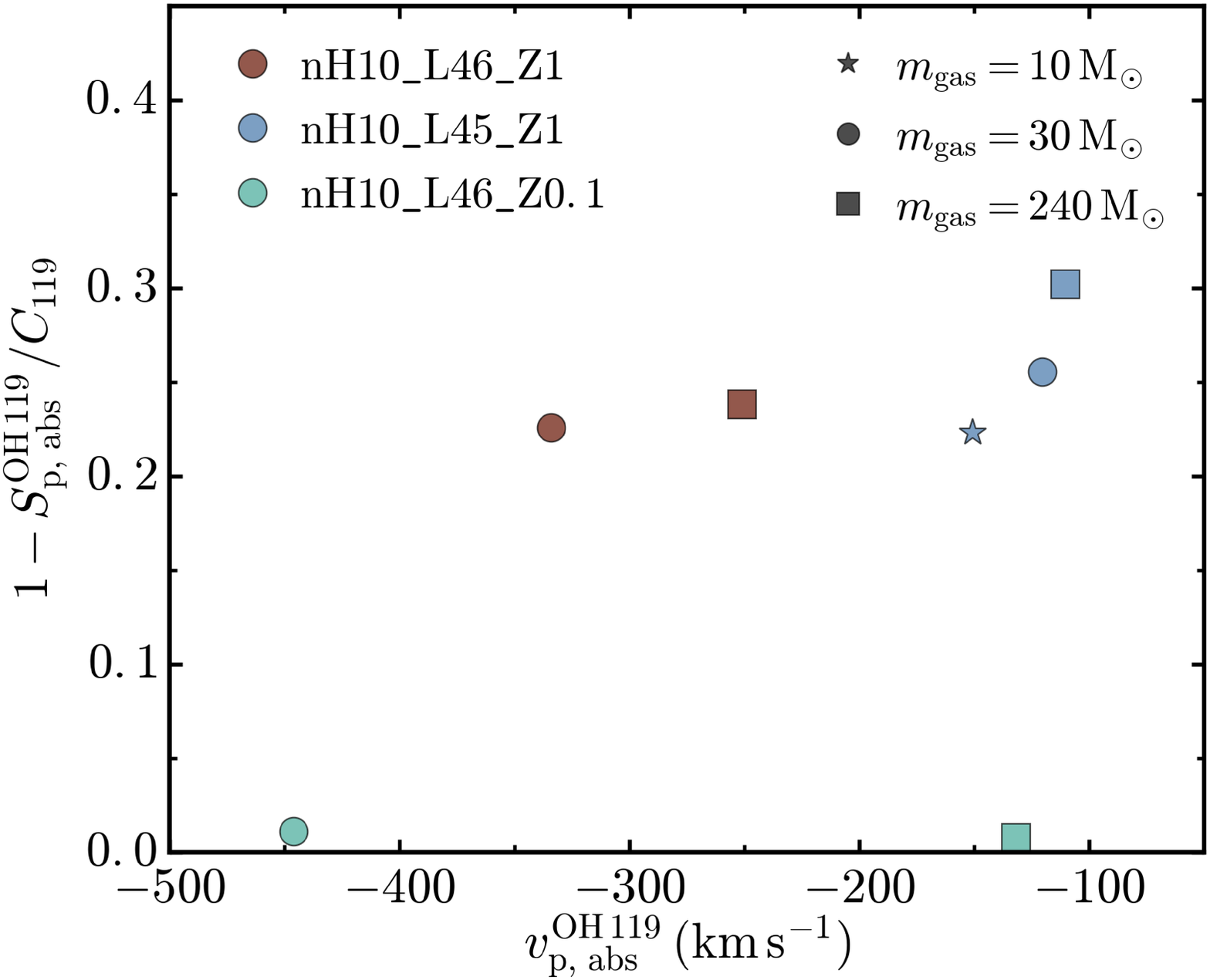}}
\caption{Resolution dependence of the absorption strength of the blue-shifted absorption feature of the $119 \, \mu\rm{m}$ OH line versus velocity of the absorption peak. The peak velocity of the OH $119 \, \mu\rm{m}$ absorption is poorly converged in the low-metallicity run (green), although this run only shows very weak OH absorption. In the two runs at solar metallicity, the OH absorption is weaker and more strongly blue-shifted at higher resolution, although, comparing to the black symbols in Fig.~\ref{ohAbsorptionFig}, we see that these two runs are still in agreement with the observations of \citet{gonzalezalfonso17} at all resolution levels.} 
\label{OHResFig}
\end{figure}

Fig.~\ref{OHResFig} shows the strength of the blue-shifted absorption peak of the OH $119 \, \mu\rm{m}$ line versus the peak absorption velocity, calculated as in Fig.~\ref{ohAbsorptionFig}. In the nH10\_L46\_Z1 run, the peak absorption velocity becomes more strongly blue-shifted as resolution increases, decreasing from $-250 \, \rm{km} \, \rm{s}^{-1}$ at low resolution to $-330 \, \rm{km} \, \rm{s}^{-1}$ at standard resolution. However, the strength of the absorption peak is well converged in this run. In the low-luminosity run, the absorption peak becomes weaker (from $(1 - S_{\rm{p,} \, \rm{abs}}^{\rm{OH119}}/C_{119}) = 0.30$ to 0.22) and more strongly blue-shifted (from $-110$ to $-150 \, \rm{km} \, \rm{s}^{-1}$) as the resolution increases from the low-resolution to the high-resolution run. However, if we compare to the black symbols in Fig.~\ref{ohAbsorptionFig}, we see that all runs at solar metallicity remain in good agreement with the observations of \citet{gonzalezalfonso17}. The low-metallicity run, on the other hand, shows much poorer convergence, as the peak absorption velocity decreases from $-130 \, \rm{km} \, \rm{s}^{-1}$ at low resolution to $-450 \, \rm{km} \, \rm{s}^{-1}$ at standard resolution. This is likely because this run shows very weak OH absorption, so the peak absorption velocity is poorly constrained. 

\begin{figure}
\centering
\mbox{
	\includegraphics[width=84mm]{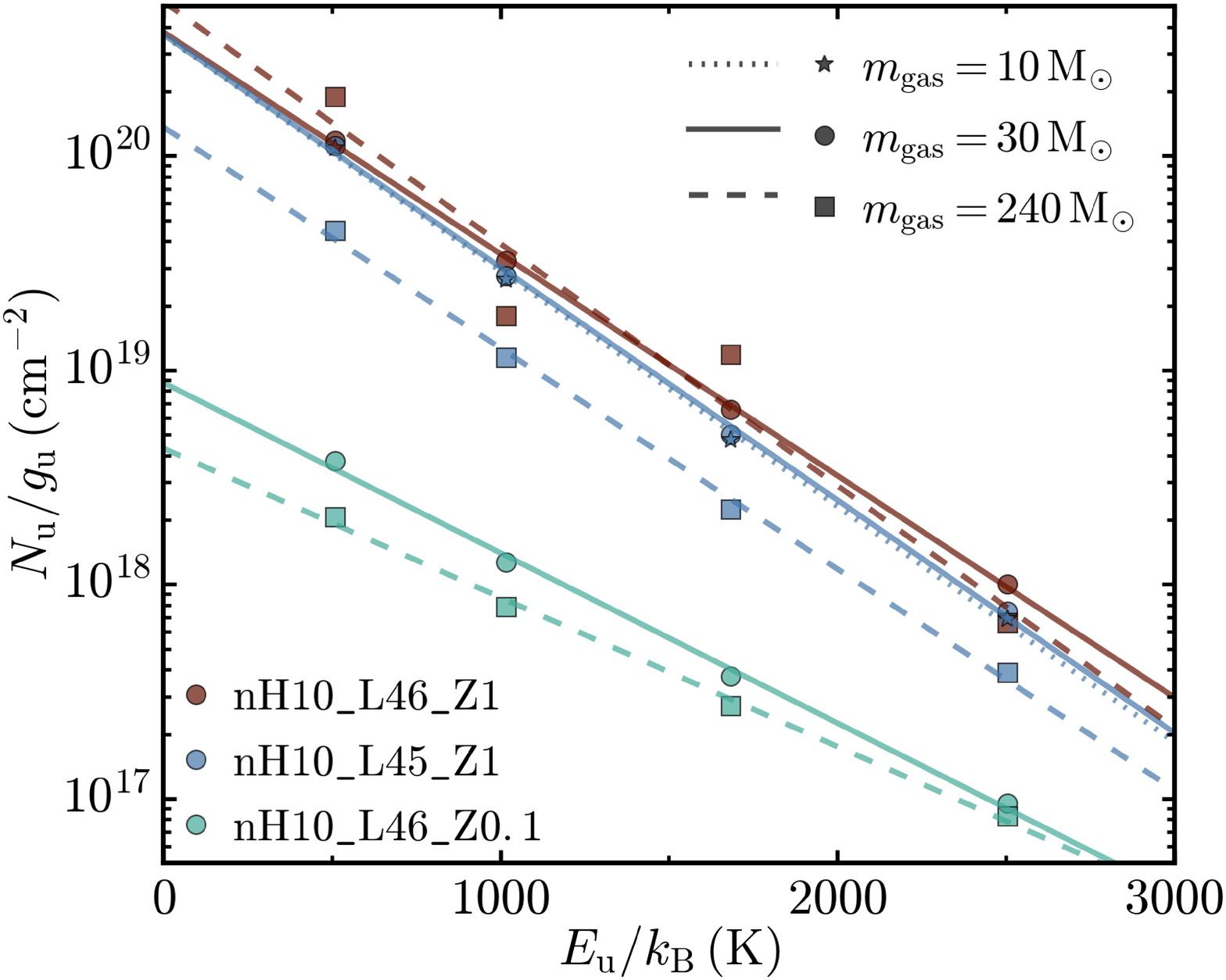}}
\caption{H$_{2}$ excitation diagram for different resolution levels. The symbols show H$_{2}$ level populations calculated from the line fluxes, plotted as the column density of the upper level, $N_{\rm{u}}$, divided by the statistical weight, $g_{\rm{u}}$, while the curves show single-temperature thermal distributions fitted to the level populations. The H$_{2}$ column densities increase with increasing resolution, consistent with the trend of increasing H$_{2}$ mass in the outflow with increasing resolution. However, the H$_{2}$ excitation temperatures, as determined from the slope of the best-fit thermal distributions, are well converged.} 
\label{H2ResFig}
\end{figure}

Fig.~\ref{H2ResFig} shows the level populations of H$_{2}$, calculated from the H$_{2}$ line fluxes, plotted against the energy of the upper level of each transition, as in Fig.~\ref{H2excitationFig}. The H$_{2}$ column densities increase with increasing resolution, which is consistent with the trend of increasing H$_{2}$ mass with increasing resolution. However, the H$_{2}$ excitation temperatures, as indicated by the slope of the best-fit single-temperature thermal distributions (curves), are well converged. In the fiducial, low-luminosity and low-metallicity runs, the H$_{2}$ excitation temperature varies by 9, 6 and 13 per cent, respectively, between resolution levels. 

\section{Local shielding approximation}\label{shielding_appendix} 

To correctly follow the attenuation of the photoionization and photodissociation rates in the simulations, we would need to include a treatment for the full 3D multi-frequency radiative transport of the UV radiation from the AGN, coupled to the non-equilibrium chemistry solver module. However, this is computationally expensive, so instead we use a local Sobolev-like approximation that assumes that each gas particle is only shielded locally, by gas within a shielding length $L_{\rm{sh}}$ that depends on the local density $\rho$ and density gradient (see Section~\ref{chemistry_sect}). 

In this section we test this approximation by comparing the column densities of individual particles in the final snapshot from simulation nH10\_L46\_Z1 after $1 \, \rm{Myr}$ to the true column densities calculated by ray tracing from the AGN to the gas particles in post-processing. 

\begin{figure}
\centering
\mbox{
	\includegraphics[width=84mm]{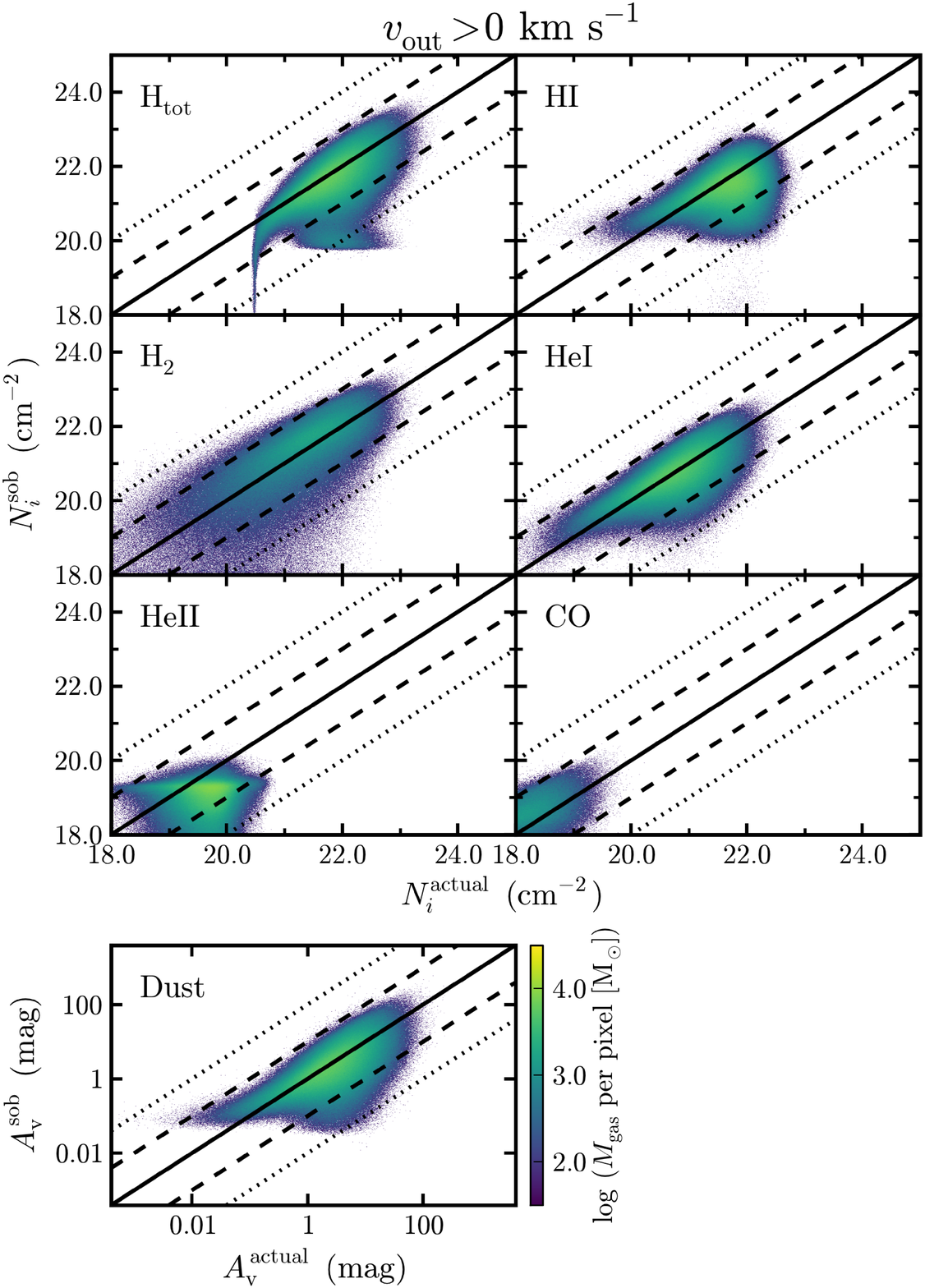}}
\caption{Column densities calculated using the Sobolev-like approximation (sob) and from ray tracing in post-processing (actual) in the simulation nH10\_L46\_Z1 after $1 \, \rm{Myr}$. The comparisons show the total hydrogen column density (top left), column densities of H\textsc{i} (top right), H$_{2}$ (second row, left), He\textsc{i} (second row, right), He\textsc{ii} (third row, left), CO (third row, right), and dust extinction (bottom left). Only particles within the high-resolution wedge and outflowing with $v_{\rm{out}} > 0 \, \rm{km} \, \rm{s}^{-1}$ are included. Solid lines indicate the one-to-one relation, while dashed and dotted lines indicate factors of 10 and 100, respectively, above and below the one-to-one relation. While the Sobolev-like approximation follows the actual column density on average, there is a large scatter of at least an order of magnitude.} 
\label{shieldFig}
\end{figure}

Fig.~\ref{shieldFig} shows this comparison between the Sobolev-like approximation column densities (sob) and the actual column densities from ray tracing (actual). The top six panels compare the total hydrogen column densities (top left), along with the column densities of H\textsc{i}, H$_{2}$, He\textsc{i}, He\textsc{ii} and CO. The bottom left panel compares the approximate and actual dust extinctions, $A_{\rm{v}} = 4.0 \times 10^{-22} N_{\rm{H_{tot}}} Z / \rm{Z}_{\odot}$, which include only gas particles below $10^{5} \, \rm{K}$. The solid lines indicate the one-to-one relation between the approximate and actual column densities, while the dashed and dotted lines show factors of 10 and 100, respectively, deviations from the one-to-one relation. 

We include only particles within the high-resolution wedge. We also exclude the ambient ISM gas by including only particles that are outflowing, with $v_{\rm{out}} > 0 \, \rm{km} \, \rm{s}^{-1}$. There is much poorer agreement between the Sobolev-like approximation and the actual column densities in the ambient ISM than in the outflowing gas. This is unsurprising, because the Sobolev-like approximation uses the density gradient, which would be zero (resulting in an infinite shielding length) if the ambient ISM were perfectly uniform. The shielding approximation that we use assumes that the gas is arranged in clumps that shield themselves, but this assumption is not valid in a uniform medium. However, we find that any molecules in the ambient ISM are destroyed when the forward shock sweeps up this gas into the outflow. Thus the molecular content of the outflow is independent of the chemical state of the ambient ISM. 

In the top left panel of Fig.~\ref{shieldFig} we see that, while the total hydrogen column density in the outflow computed using the Sobolev-like approximation on average follows the actual column density, there is a large scatter of an order of magnitude, with some particles underestimated by more than two orders of magnitude. To explore how these uncertainties might affect the resulting molecular outflows, we ran one simulation with column densities lowered by a factor of 10 (this run is labelled `lowShield' in Fig.~\ref{modelVarsFig}; see also the discussion in Section~\ref{model_vars_sect}). We found that this reduced the H$_{2}$ outflow rate after $1 \, \rm{Myr}$ by a factor of 4.3 compared to the reference model. However, reducing all column densities by a factor of 10 is a worst case scenario, and would correspond to all particles lying along the upper dashed lines in Fig.~\ref{shieldFig}. In practice, many particles lie below the one-to-one relation in this figure, i.e. the Sobolev-like approximation underestimates the actual column density for these particles. This would tend to underestimate the true H$_{2}$ fractions, and so this would tend to strengthen our main conclusion that molecules can form in-situ within AGN-driven outflows. 

Another uncertainty in the local shielding approximation is that the column densities of individual species are calculated by multiplying the total hydrogen column density by the abundance of that species. However, this assumes that the abundances are uniform over the shielding length. In Fig.~\ref{shieldFig} we also compare the local approximation to the actual column densities for individual species that are important for shielding. For most species, the scatter is similar to that for the total hydrogen column density. The H$_{2}$ column density does show broader scatter, exceeding two orders of magnitude. However, most particles still lie within a factor of 10 of the true value. Also, the largest scatter in $N_{\rm{H_{2}}}$ is below the one-to-one relation, so the true values, and hence the true H$_{2}$ fractions, would tend to be higher. 

\citet{safranekshrader17} compared several local shielding approximations to a full radiative transfer calculation in simulations of a section of a galactic disc in post-processing. They showed that a density gradient-based approach, as used in our work, performs poorly in terms of reproducing the correct column densities and equilibrium molecular fractions. Of the methods that use only local quantities (i.e. excluding their six-ray approximation), they found that using the Jeans length with a temperature capped at $40 \, \rm{K}$ gave the best agreement with the radiative transfer results. 

\begin{figure}
\centering
\mbox{
	\includegraphics[width=84mm]{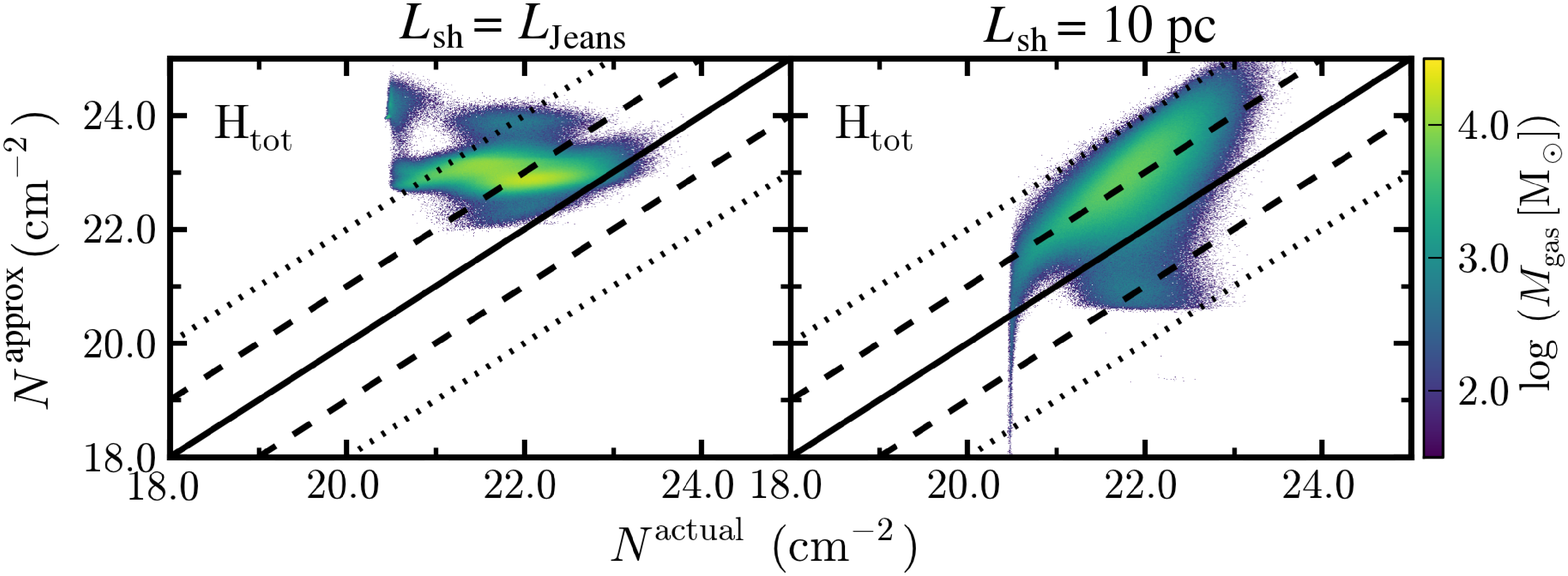}}
\caption{Total hydrogen column densities in simulation nH10\_L46\_Z1 calculated using the Jeans length (left-hand panel) and using a constant shielding length of $10 \, \rm{pc}$ (right-hand panel), versus the actual column density calculated from ray tracing in post-processing. We include particles within the high-resolution wedge that are outflowing with $v_{\rm{out}} > 0 \, \rm{km} \, \rm{s}^{-1}$. Both of these alternative approximations systematically overestimate the true column density by an order of magnitude or more.} 
\label{shieldAltFig}
\end{figure}

In the left-hand panel of Fig.~\ref{shieldAltFig} we compare the total hydrogen column density using the Jeans length in simulation nH10\_L46\_Z1 to the true column density from ray tracing. We do not cap the temperature to $40 \, \rm{K}$ as in \citet{safranekshrader17}, because we saw in Fig.~\ref{TrhoFig} that most of the molecular gas in our simulations is at temperatures of hundreds to thousands of K. When we use the Jeans length, the column density is systematically higher than from ray-tracing, typically by more than an order of magnitude. Comparing to the top left panel of Fig.~\ref{shieldFig}, we see that the density gradient-based method performs better than the Jeans length in our simulations. However, the physical conditions in these fast molecular outflows are very different from those in the galactic discs simulated by \citet{safranekshrader17}. Therefore, this comparison does not contradict the results of \citet{safranekshrader17}, as we are looking at a very different physical regime. 

If we impose a temperature cap of $40 \, \rm{K}$ (not shown), as in \citet{safranekshrader17}, we find that the column densities are lower and are closer to those from ray tracing. However, the slope of the $N_{\rm{H_{tot}}}^{\rm{approx}}$ versus $N_{\rm{H_{tot}}}^{\rm{actual}}$ relation is still flatter than the one-to-one relation, with the approximation tending to overestimate $N_{\rm{H_{tot}}}$ below $\sim 10^{22} \, \rm{cm}^{-2}$ by up to an order of magnitude, and underestimate it at higher column densities. 

Finally, the right-hand panel of Fig.~\ref{shieldAltFig} shows the total hydrogen column density assuming a constant shielding length of $10 \, \rm{pc}$, compared to the actual column density. This is the approach used by \citet{ferrara16}, as $10 \, \rm{pc}$ is the Field length at the mean density of their simulations, below which thermal conduction will suppress local thermal instabilities \citep{field65}. We see that this method systematically overestimates the column density, by an order of magnitude on average, although we do not include thermal conduction in our simulations. 

\label{lastpage}

\end{document}